\documentclass[3p,sort&compress,times]{elsarticle}
\pdfoutput=1 %This is for the ArXiv submission only
\usepackage{graphicx}
\usepackage{url}
\usepackage{import}
\usepackage{multirow}

\usepackage{color}
\usepackage{epstopdf}
\usepackage{comment}
\usepackage{amsmath} 

\usepackage{amssymb}
\usepackage{bm}
\usepackage{newcommand_yye}
\usepackage{tikz}
\usepackage{pgfplots}

\usepackage[breaklinks=true,colorlinks=true,linkcolor=blue,urlcolor=blue,citecolor=blue]{hyperref}

\def\n{\bm{n}}
\def\x{\bm{x}}
\def\X{\bm{X}}
\def\W{\bm{W}}

\def\d{\mathrm{d}}

\def\P{\mathbb{P}}
\def\R{\mathbb{R}}

\def\L{\mathcal{L}}

\def\N{\mathcal{N}}

\def\pa{\partial\Omega}

\def\ve{\varepsilon}

\def\ctanh{\mathrm{ctanh}}
\def\Var{\mathrm{Var}}

\journal{...}

\begin{document}
\begin{frontmatter}

\title{Escape-from-a-layer approach for simulating the boundary local time in Euclidean domains}

\author[inst1]{Yilin Ye}
\author[inst1]{Adrien Chaigneau}
\author[inst1]{Denis S. Grebenkov}

% \email{yilin.ye@polytechnique.edu}
%\affiliation[inst1]{Laboratoire de Physique de la Mati\`{e}re Condens\'{e}e, 
% \\ CNRS -- Ecole Polytechnique, Institut Polytechnique de Paris, 91120 Palaiseau, France}

% \email{adrien.chaigneau@polytechnique.edu}
%\affiliation[inst1]{Laboratoire de Physique de la Mati\`{e}re Condens\'{e}e, 
% \\ CNRS -- Ecole Polytechnique, Institut Polytechnique de Paris, 91120 Palaiseau, France}

% \email{denis.grebenkov@polytechnique.edu}
%\affiliation[inst1]{Laboratoire de Physique de la Mati\`{e}re Condens\'{e}e, 
% \\ CNRS -- Ecole Polytechnique, Institut Polytechnique de Paris, 91120 Palaiseau, France}

\address[inst1]{Laboratoire de Physique de la Mati\`{e}re Condens\'{e}e, 
\\ CNRS -- Ecole Polytechnique, Institut Polytechnique de Paris, 91120 Palaiseau, France}

\date{\today}

\begin{abstract}
We propose an efficient numerical approach to simulate the boundary
local time of reflected Brownian motion, as well as the time and
position of the associated reaction event on a smooth boundary of a
Euclidean domain.  This approach combines the standard walk-on-spheres
algorithm in the bulk with the approximate solution of the escape
problem in a boundary layer.  In this way, the most time-consuming
simulation of multiple reflections on the boundary is replaced by an
equivalent escape event.  We validate the proposed escape-from-a-layer
approach by comparing simulated statistics of the boundary local time
with exact results known for simple domains (a disk, a circular
annulus, a sphere, a spherical shell) and with the numerical results
obtained by a finite-element method in more sophisticated domains.
This approach offers a powerful tool for simulating reflected Brownian
motion in multi-scale confinements such as porous media or biological
environments, and for solving the related partial differential
equations. Its applications in the context of diffusion-controlled
reactions in chemical physics are discussed.
\end{abstract}

%\pacs{02.50.-r, 05.40.-a, 02.70.Rr, 05.10.Gg}

%02.50.-r       (Probability theory, stochastic processes, and statistics)
%05.40.-a 	Fluctuation phenomena, random processes, noise, and Brownian motion
%02.70.Rr       (General statistical methods)
%05.10.Gg 	Stochastic analysis methods (Fokker-Planck, Langevin, etc.) 

%02.50.Ey 	Stochastic processes  (Probability theory, stochastic processes, and statistics)

\begin{keyword}
boundary local time, reflected Brownian motion, escape problem,
walk-on-spheres, Monte Carlo simulation, diffusion-controlled
reaction, Robin and Neumann boundary conditions
\end{keyword}

\end{frontmatter} %\maketitle

\section{Introduction}
\label{sec:intro}

The boundary local time $\ell_t$ plays the central role in the theory
of stochastic processes \cite{Levy,Ito,Freidlin}.  For instance,
reflected Brownian motion $\X_t$ inside a given Euclidean domain
$\Omega \subset \R^d$ with a smooth boundary $\pa$ can be constructed
as the solution of the Skorokhod stochastic equation \cite{Ito,Freidlin}
\begin{equation} \label{eq:Skorokhod}
\d\X_t = \sqrt{2D}\, \d\W_t + \n(\X_t)\, \d\ell_t  ,
\qquad 
\X_0 = \x_0 ,
\qquad
\ell_0 = 0 , 
\end{equation}
where $\W_t$ is the standard Wiener process in $\R^d$, $D$ is the
diffusion coefficient, $\x_0 \in \Omega$ is the starting point,
$\n(\x)$ is the unit normal vector at a boundary point $\x\in\pa$
oriented inward the domain $\Omega$, and $\ell_t$ is a nondecreasing
stochastic process that increments at each encounter of $\X_t$ with
the boundary.  Qualitatively, the first term in
Eq. (\ref{eq:Skorokhod}) describes ordinary Brownian motion inside
$\Omega$, whereas the second term ensures that $\X_t$ is reflected
back normally into $\Omega$ at each encounter with the boundary.  From
the physical point of view, one can think of the second term as an
infinitely local force field that pushes the diffusing particle back
to the confining domain \cite{Grebenkov20}.  Curiously, the single
stochastic equation (\ref{eq:Skorokhod}) determines simultaneously two
tightly related stochastic processes: the position $\X_t$ and the
boundary local time $\ell_t$ (which has, despite its name, units of
length).  The latter should not be confused with a point local time,
which represents the residence time of Brownian motion in a vicinity
of a bulk point (see
\cite{mckean1975brownian,borodin2015handbook,majumdar2007brownian}).
Note that the boundary local time can be expressed as
\begin{equation}  \label{eq:ellt_def1}
\ell_t = \lim\limits_{\ve \to 0} \frac{D}{\ve} \int\limits_0^t \d t' \, \Theta(\ve - |\X_{t'} - \pa|),
\end{equation}
where $|\x - \pa|$ is the Euclidean distance between a point $\x$ and
the boundary $\pa$, and $\Theta(z)$ is the Heaviside step function:
$\Theta(z) = 1$ for $z > 0$ and $0$ otherwise.  The integral
represents the residence time of the process $\X_t$ in a thin layer
$\pa_\ve = \{ \x\in\Omg: \llv \x - \pa \rrv < \ve \}$ of width $\ve$
near the boundary $\pa$, which is rescaled by $\ve$ to get a
nontrivial limit, as $\ve \to 0$.  Alternatively, one has
\begin{equation}  \label{eq:ellt_def2}
\ell_t = \lim\limits_{\ve\to 0} \ve \, \N_t^{(\ve)} ,
\end{equation}
where $\N_t^{(\ve)}$ is the number of crossings of the layer $\pa_\ve$ up
to time $t$.  In other words, the boundary local time $\ell_t$ can be
understood either as the rescaled residence time in a thin boundary
layer, or as the rescaled number of encounters with that boundary.

Apart from its major role in the theory of stochastic processes, the
boundary local time was recently employed as the conceptual pillar to
build the encounter-based approach to diffusion-controlled reactions
\cite{Grebenkov20}.  This approach laid a theoretical ground for
assessing the statistics of encounters between the diffusing particle
and the boundary, provided an intuitively clear probabilistic
interpretation of partial reactivity and allowed one to consider much
more general surface reaction mechanisms
\cite{Grebenkov21,Grebenkov22a,Bressloff22,Grebenkov22,Bressloff22d,Benkhadaj22,Grebenkov22b,Grebenkov23b}.  
In fact, the reaction event of a diffusing particle on the
boundary can be initiated when the boundary local time $\ell_t$
exceeds a chosen random threshold $\hat{\ell}$.  In this way, the
probability distribution of the threshold determines the surface
reaction mechanism {\it independently} of the diffusive dynamics.  So,
the exponential distribution of $\hat{\ell}$ corresponds to the
standard constant reactivity implemented via Robin boundary condition,
whereas other distributions can describe, for instance, progressive
activation or passivation of the boundary by the diffusing particle
\cite{Grebenkov23b,Grebenkov24f}.  While the former setting can be
simulated by conventional methods, implementation of more
sophisticated surface reaction mechanisms requires the knowledge of
the boundary local time.  Moreover, these concepts were further
extended to describe permeation across membranes
\cite{Bressloff22c,Bressloff23a,Bressloff23b}, diffusive exchange
between adjacent compartments \cite{fieremans2010monte}, and the
related snapping out Brownian motion \cite{Lejay16,lejay2018monte}.

Despite these advances, the fundamental relation between the
statistics of the boundary local time $\ell_t$ and the geometrical
structure of the confining domain $\Omega$ remains poorly understood.
For instance, the distribution of $\ell_t$ is known explicitly only
for two simple domains: a half-line and the exterior of a sphere
\cite{Grebenkov21,Grebenkov19,Grebenkov20b}.  Indeed, the
symmetries of these domains allow for the separation of variables and
thus reduction to a one-dimensional problem that can be solved
exactly.  In contrast, most diffusive processes in nature and
industrial applications occur in multi-scale media with irregular,
sophisticated boundaries, for which numerical tools are needed to
access the statistics of encounters.  Even though the spectral
expansions developed in \cite{Grebenkov20} are formally still
applicable in such domains, their numerical implementation (e.g., by a
finite-element method) can be very costly and time-consuming,
especially in three dimensions.  In this light, one may prefer Monte
Carlo techniques that offer great flexibility and moderate
computational costs \cite{Sabelfeld,Sabelfeld2,Milshtein}.  While
these techniques are broadly used for simulating stochastic processes
and diffusion-controlled reactions in complex domains
\cite{litwin1980monte,Torquato89,Zheng89,Lee89,Ossadnik91,batsilas2003stochastic,dagdug2003diffusion,Grebenkov05a,Grebenkov05b,Grebenkov06c,Levitz06,Deaconu06,opplestrup2006first,hall2009convergence,Zein10,berezhkovskii2013trapping,ghosh2015non,ghosh2015anomalous,bernoff2018boundary,palombo2019generative,ianus2021mapping,le2022first,cherry2022trapping},
their adaptation for studying the boundary local time is not
straightforward.

The most basic way to simulate the boundary local time consists in
approximating reflected Brownian motion by a random walk on a lattice
with spacing $\ve$ and counting the number $\N_t^{(\ve)}$ of its
encounters with the discretized boundary $\pa$.  According to
Eq. (\ref{eq:ellt_def2}), $\ve \N_t^{(\ve)}$ is an approximation of
$\ell_t$, if $\ve$ is small enough.  The need for boundary
discretization and excessively long trajectories (with $2dDt/\ve^2$
steps) are major drawbacks of this method.  The first drawback can be
relaxed by performing off-lattice random walks, i.e., a sequence of
centered Gaussian jumps of variance $\sigma^2$ with normal reflections
on the boundary.  In this case, one can either count the number
$\N_t^{(\sigma)}$ of reflections and use again
Eq. (\ref{eq:ellt_def2}) with $\sigma$ instead of $\ve$, or calculate
the residence time of this random walk inside a boundary layer of
width $\ve \sim \sigma$ and use Eq. (\ref{eq:ellt_def1}) to
approximate $\ell_t$ (see also \cite{bossy2004symmetrized} for the
discussion of Euler schemes).  Since a typical one-jump displacement
$\sigma$ should be the smallest length scale of the problem, this
method can be efficient for simple domains (see, e.g.,
\cite{Grebenkov07}) but it becomes too time-consuming in multi-scale
media.  This issue was partly resolved by Zhou {\it et al.} who
combined the standard walk-on-spheres (WOS) algorithm by Muller
\cite{Muller56} and the fixed-length displacements in a thin layer
near boundary \cite{Zhou17}.  Their hybrid method exploits the
well-known advantages of the WOS algorithm for fast simulations of
large-scale displacements far from the boundary (see details in
Sec. \ref{sec:WOS}).  As a consequence, the major remaining
computational limitation is caused by modeling small displacements
inside the boundary layer of small width $\ve$.  This method was
further improved and adapted to simulating snapping out Brownian
motion in planar bounded domains by Schumm and Bressloff \cite{Schumm23}.
In particular, they suggested using walk-on-spheres even inside the
boundary layer of width $\ve$, with the jump distance being fixed to
be $2 \ve$, in order to speed up simulations within that layer.
Despite this progress, an accurate estimation of the boundary local
time $\ell_t$ within a boundary layer requires taking $\ve$ sufficient
small so that modeling of reflected Brownian motion within this layer
remains the critical bottleneck of these multi-scale Monte Carlo
simulations.

\begin{figure}[t!]
\begin{center}
\begin{tikzpicture}
\node (fig1) at (0,0) {\includegraphics[height=40mm]{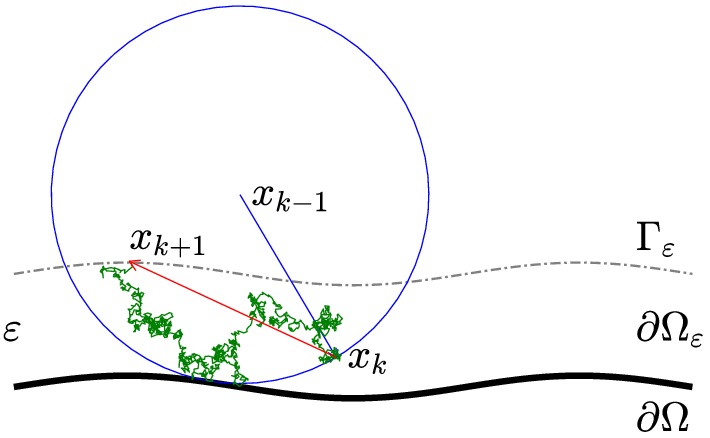}};
\node[] at (fig1.south west) {(a)};
\node (fig2) at (8,-0.25) {\includegraphics[height=35mm]{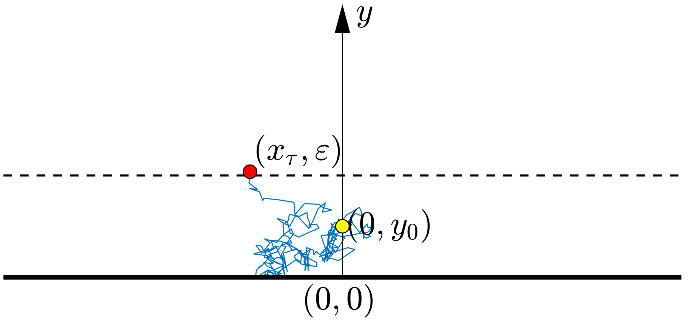}};
\node[] at (fig2.south west) {(b)};
\end{tikzpicture}
\caption{
\tbf{(a)} 
A zoom near the boundary $\partial \Omg$: once the WOS arrives inside
the boundary layer $\partial \Omg_\ve$ of width $\ve$, one aims at
replacing a costly detailed simulation of the random trajectory (in
green) from the entrance point $\x_k$ to the exit point $\x_{k+1}$ by
a single escape event.  For this purpose, one needs to generate the
escape position $\x_{k+1}$, the escape time $\tau$, and the acquired
boundary local time $\ell^\prime_\tau$.
\tbf{(b)} 
Flat layer approximation: starting from a point $(0,y_0)$ (yellow
circle), a random trajectory (in blue) experiences numerous
reflections on the flat boundary before escaping an infinite stripe of
width $\ve$ at a random escape time $\tau$ in a random escape position
$(x_\tau,\ve)$ (red circle).  These reflections are characterized by
the acquired boundary local time $\ell^\prime_\tau$.  The random
trajectory is thus replaced by a simple jump from $(0, y_0)$ to
$(x_\tau, \ve)$.  }
\label{fig:scheme_1D}
% A_Yilin_simu_scheme;
\end{center}
\end{figure}

In this paper, we propose a different approach to simulate the
stochastic process $(\X_t,\ell_t)$ inside a confining domain.  We
still employ the WOS algorithm for fast simulations of large-scale
displacements far from the boundary.  In turn, we replace the detailed
time-consuming simulation of the stochastic process within a boundary
layer by a single escape event from that layer.  A similar approach
was recently implemented for simulating the escape of a sticky particle
\cite{Scher23,Scher24}.  In fact, once the process arrives inside the
layer, we introduce the random escape time $\tau$ from that layer,
i.e., the first-passage time to the set $\Gamma_\ve = \{ \x
\in\Omega~:~ |\x - \pa| = \ve\}$ of points in $\Omega$ that are
equidistant to the boundary (Fig. \ref{fig:scheme_1D}a).

To realize an escape event, the central quantity of interest is the
joint probability density of the escape time $\tau$, the escape
position $\X_\tau$, and the acquired boundary local time $\ell_\tau$.
Its exact spectral expansion was recently derived by means of the
encounter-based approach \cite{Grebenkov23}.  As a smooth boundary is
locally flat, the spectral expansion takes a simple form that allows
its direct implementation in Monte Carlo simulations.  Moreover, when
the boundary is composed of flat elements, the simulation of the
escape event becomes almost exact.  In addition, we propose an
improvement to account for the curvature of the boundary in order to
enable using a larger width $\ve$ of the boundary layer.  In this way,
there is no need for simulating multiple reflections on the boundary
that was the most time-consuming step and the main source of
accumulating errors in former techniques.

The paper is organized as follows.  In Sec. \ref{sec:method}, we
describe the escape-from-a-layer (EFL) approach.  We start with a
simpler approximation of a flat boundary layer and then discuss its
improvement to curved boundary layers.  Section \ref{sec:algorithm}
summarizes an algorithmic implementation of the EFL approach.  Section
\ref{sec:validation} validates this method for several confining
domains by comparing simulated results to either exact values, or
numerical values obtained by a finite-element method.  We also compare
the performance of our method to the state-of-the-art method by Schumm
and Bressloff \cite{Schumm23}.  In Sec. \ref{sec:discon}, we discuss
further improvements, extensions, applications and future perspectives
of this method.

\section{Escape-from-a-layer approach}
\label{sec:method}

\subsection{Walk-on-spheres algorithm}
\label{sec:WOS}

We start by recalling the walk-on-spheres algorithm for simulating
Brownian motion in a confining domain $\Omega \subset \R^d$ with an
{\it absorbing} boundary $\pa$ \cite{Muller56}.  From a prescribed
starting point $\x_0 \in \Omega$, one determines the Euclidean
distance $\rho_0 = |\x_0 - \pa|$ to the boundary $\pa$ and draws a
ball $B_{\rho_0}(\x_0)$ of radius $\rho_0$ centered at $\x_0$.  The
continuity of Brownian motion implies that any trajectory must cross
the boundary of this ball, $\partial B_{\rho_0}(\x_0)$, before hitting
the boundary $\pa$.  One can therefore replace lengthy simulations of
the trajectory inside the ball by an escape event, i.e., a jump to a
uniformly distributed point $\x_1$ on $\partial B_{\rho_0}(\x_0)$.
The probability distribution of the time $\tau_1$ needed to escape
from the ball is known explicitly (see \ref{app:exittime}) that allows
one to easily generate the random variable $\tau_1$.  From the point
$\x_1$, the previous step is repeated, i.e., one determines the
distance $\rho_1 = |\x_1 -\pa|$ and generates the escape time $\tau_2$
and the escape position $\x_2$ as a uniformly distributed point on the
sphere $\partial B_{\rho_1}(\x_1)$, and so on.  The algorithm is
stopped when the distance to the boundary becomes smaller than a
prescribed threshold $\ve$.  This procedure samples a sequence of
points $\X_{t_1}, \X_{t_2}, \ldots, \X_{t_k}$ of a random trajectory
$\X_t$ at random times $t_1 = \tau_1,~t_2 = t_1 + \tau_2, \ldots , t_k
= t_{k-1} + \tau_k$.  As the jump length is progressively adapted
according to the proximity to the boundary, large jumps can be
performed far from the boundary.  This variability ensures remarkably
fast convergence of the method \cite{Binder12} and allows for
efficient simulations of Brownian motion in multi-scale environments
\cite{Torquato89,Zheng89,Lee89,Ossadnik91,Grebenkov05a,Grebenkov05b,Grebenkov06c,Levitz06}.

The threshold $\ve$ determines the width of a boundary layer, which is
considered as absorbing: the process is stopped once it arrives to
this layer.  The value of $\ve$ can be either assigned from the
underlying microscopic model (e.g., the interaction range between the
diffusing particle and the boundary), or treated as a numerical
parameter of the method.  As shown by Binder and Braverman
\cite{Binder12}, the computational time scales linearly with $\ln
(1/\ve)$, i.e., one can choose an extremely small $\ve$ with almost no
computational cost.  This remarkably fast convergence explains the
efficiency and popularity of the WOS algorithm in domains with
absorbing boundaries.  However, if the boundary is not absorbing, one
loses such fast convergence.  Indeed, if the process is not stopped
upon the first arrival into the boundary layer, one needs to simulate
reflections on the boundary that forms the critical bottleneck for an
extension of the WOS method to reflecting boundaries.  In the next
subsections, we propose an efficient way to solve this problem.

\subsection{The layer escape problem}
 
One can run the WOS algorithm until some time $t_k$ when the process
arrives to a point $\x_k$ inside the boundary layer $\Omega_\ve$ of
width $\ve$: $\Omega_\ve = \{ \x\in\Omega~:~ |\x - \pa| < \ve\}$.  Our
goal is to substitute a lengthy simulation of the trajectory
$\X^\prime_t = \X_{t + t_k}$ inside this layer by a single
displacement out of $\Omega_\ve$ (as reflected Brownian motion is
Markovian, the trajectory $\X_t$ with $t < t_k$ does not matter so
that we can consider $t_k$ as a new origin of time for the shifted in
time process $\X^\prime_t$, denoted by prime).  For this purpose, we
aim at generating three random variables: (i) the escape time $\tau =
\inf\{ t > 0 ~:~ \X^\prime_t \notin \Omega_\ve ~|~ \X^\prime_0 =
\x_k\}$ from the layer, i.e., the first instance when the process
leaves $\Omega_\ve$ by crossing the surface $\Gamma_\ve = \{
\x\in\Omega~:~ |\x - \pa| = \ve\}$; (ii) the position $\X^\prime_\tau$
at the escape time, and (iii) the boundary local time
$\ell^\prime_\tau$ acquired during this escape event
(Fig. \ref{fig:scheme_1D}a).  If we manage to generate these random
variables, one can execute a single escape event from the layer,
without simulating the trajectory inside it.  After the escape, the
new position is $\X_{t_{k+1}} = \x_{k+1} = \X^\prime_\tau$, the time
is incremented by $\tau$, $t_{k+1} = t_k + \tau$, and the boundary
local time is incremented by $\ell^\prime_\tau$: $\ell_{t_{k+1}} =
\ell_{t_k} + \ell^\prime_\tau$.  From this position, the WOS algorithm
is resumed, until the particle arrives again inside the layer, and so
on. The simulation is stopped according to a prescribed stopping
condition (see Sec. \ref{sec:stopcond}).

This layer escape problem was solved in \cite{Grebenkov23}.  In fact,
for any bounded domain with a smooth boundary, a formal spectral
expansion was derived for the probability density flux
$j(\x,\ell,t|\x_k)$, which is the joint probability density of the
escape position $\X^\prime_\tau \in \Gamma_\ve$, the acquired boundary
local time $\ell^\prime_\tau$, and the escape time $\tau$.  In the
Laplace domain, this spectral expansion reads for any $\x \in
\Gamma_\ve$:
\begin{align}   \nonumber
 \tilde{j}(\x,\ell,p|\x_k) & = \int\limits_0^\infty \d t \, e^{-pt} \, j(\x,\ell,t|\x_k) \\   \label{eq:jp}
& = \tilde{j}_0(\x,p|\x_k) \delta(\ell) - \sum\limits_{k=0}^\infty [V_k^{(p)}(\x_k)]^* \, (\partial_n V_k^{(p)}(\x)) e^{-\mu_k^{(p)}\ell} ,
\end{align}  
where $\partial_n$ is the normal derivative on the boundary
$\Gamma_\ve$ oriented outwards the domain $\Omega_\ve$, asterisk
denotes complex conjugate, $p \geq 0$ is a scalar,
$\tilde{j}_0(\x,p|\x_k)$ is the Laplace transform of the probability
density flux onto $\Gamma_\ve$ in the case of the absorbing boundary
$\pa$, while $\mu_k^{(p)}$ and $V_k^{(p)}(\x)$ with $k = 0, 1, 2,
\ldots$ are the eigenvalues and eigenfunctions of the
generalized Steklov problem in the boundary layer $\Omega_\ve$:
\begin{subequations}
\begin{align}
(p - D \Delta) V_k^{(p)}(\x) & = 0 \quad (\x\in\Omega_\ve),  \\
\partial_n V_k^{(p)}(\x) & = \mu_k^{(p)} V_k^{(p)}(\x) \quad (\x\in \pa), \\
V_k^{(p)}(\x) & = 0 \quad (\x\in \Gamma_\ve)
\end{align}
\end{subequations}
(for mathematical details about the Steklov problem, see
\cite{Levitin} and references therein).  The first term in
Eq. (\ref{eq:jp}) represents the contribution of trajectories that do
not hit the boundary $\pa$ until escaping the layer $\Omega_\ve$.  As
a consequence, the acquired boundary local time $\ell^\prime_\tau$ is
zero, as represented by the Dirac distribution $\delta(\ell)$.  In
turn, the second term accounts for other trajectories that arrived
onto $\pa$ and thus yielded positive $\ell^\prime_\tau$.  Note that
all the ``ingredients'' of Eq. (\ref{eq:jp}) depend on the shape of
the boundary layer $\Omega_\ve$, so that advanced numerical tools are
in general needed to access these quantities.  In practice, their
computation is usually more challenging than that of the boundary
local time $\ell_t$ in the original domain $\Omega$.  In other words,
a direct application of this spectral expansion for Monte Carlo
simulations is pointless.  In the next subsection, we propose an
approximate scheme that relies on the smallness of the width $\ve$.

\begin{comment}
\begin{figure}[t!]
\begin{center}
\includegraphics[width=70mm]{scheme_1D.eps}
\end{center}
\caption{
Starting from a point $(0,y_0)$ (yellow circle), a random trajectory
(in blue) experiences numerous reflections on the flat boundary before
escaping a thin layer of width $\ve$ at the random escape time $\tau$
in a random escape position $(x_\tau,\ve)$ (red circle). These
reflections are characterized by the acquired boundary local time
$\ell^\prime_\tau$.  }
\label{fig:scheme_1D}
% load('scheme1D_XY.mat');  A_Yilin_1D_scheme(X,Y);
\end{figure}
\end{comment}

\subsection{Flat layer approximation (FLA)}
\label{sec:FLA}

For clarity, we focus on the planar case ($d = 2$) but the method is
immediately applicable in any dimension $d \geq 2$ (see below).  We
assume that the boundary $\pa$ is smooth and can thus be considered as
flat at a small scale $\ve$.  In other words, the boundary layer
$\Omega_\ve$ in the vicinity of $\x_k$ can be approximated by an
infinite stripe of width $\ve$: $\R \times (0,\ve)$.
Let $\x_b \in \pa $ be the boundary point, which is the closest to
$\x_k$.  For simplicity of notations, we introduce local coordinates
with the origin at $\x_b$ and with $y$ axis being perpendicular to the
boundary (Fig. \ref{fig:scheme_1D}b).  In other words, $\x_b = (0,0)$
and $\x_k = (0, y_0)$, where $y_0 = |\x_k - \pa| = |\x_k - \x_b|$ is
the distance to the boundary from $\x_k$.  As lateral and transverse
displacements in the stripe are independent, one can focus on
diffusion on the interval $(0,\ve)$ with reflecting endpoint $0$ and
absorbing endpoint $\ve$, so that the general spectral expansion can
be considerably simplified.  In fact, one has
\begin{equation}  \label{eq:j_factored}
j(x,\ve,\ell,t|x_0,y_0) = \frac{e^{-(x-x_0)^2/(4Dt)}}{\sqrt{4\pi Dt}} \, J(\ell,t|y_0),
\end{equation}
where the first factor is the Gaussian probability density of the
lateral displacement $x_\tau$, while the second factor describes the
joint probability density of $\ell^\prime_\tau$ and $\tau$ for the transverse
displacement.  According to Eq. (\ref{eq:jp}), the latter admits the
following representation in the Laplace domain:
\begin{align} 
\tilde{J}(\ell,p|y_0) 
 & = \tilde{J}_0(p|y_0) \delta(\ell) + \sum\limits_{k=0}^\infty [V_k^{(p)}(y_0)]^* e^{-\mu_k^{(p)}\ell} C_k^{(p)} ,
\end{align}
where 
\begin{equation}
C_k^{(p)} = - \bigl(\partial_n V_k^{(p)}\bigr)\bigr|_{y = \ve} ,
\end{equation}
$\mu_k^{(p)}$ and $V_k^{(p)}(y)$ are the eigenvalues and
eigenfunctions of the Steklov problem on the interval $(0,\ve)$, and
$\tilde{J}_0(p|\x_0)$ is the Laplace transform of the probability flux
onto the absorbing endpoint $\ve$ of the interval under the condition
of not hitting the endpoint $0$.  In the case of an interval $(0,\ve)$
with one absorbing endpoint $\ve$, the above spectral expansion
includes only one eigenmode ($k = 0$), and all the ingredients are
known explicitly:
\begin{subequations}
\begin{align}
V_0^{(p)}(y) &= \frac{\sinh(\sqrt{p/D}(\ve - y))}{\sinh(\sqrt{p/D} \ve)},  \\
\mu_0^{(p)} &= \sqrt{p/D} \, \ctanh(\sqrt{p/D} \ve), \\
C_0^{(p)} & = \frac{\sqrt{p/D}}{\sinh(\sqrt{p/D} \ve)} \,, \\
\tilde{J}_0(p|y_0) & = \frac{\sinh(\sqrt{p/D}\, y_0)}{\sinh(\sqrt{p/D} \ve)} \,.
\end{align}
\end{subequations}
As a consequence, one has
\begin{equation}  \label{eq:j1tilde}
 \tilde{J}(\ell,p|y_0) = \frac{\sinh(\sqrt{p/D}\, y_0)}{\sinh(\sqrt{p/D} \ve)} \delta(\ell) 
+ \frac{\sqrt{p/D} \sinh(\sqrt{p/D}(\ve - y_0))}{\sinh^2(\sqrt{p/D} \ve)}  e^{-\ell  \sqrt{p/D} \, \ctanh(\sqrt{p/D} \ve)}  .
\end{equation}

In order to generate two random variables $\ell^\prime_\tau$ and $\tau$, one
can first generate $\ell^\prime_\tau$ and then employ the conditional
distribution of $\tau$.  The marginal probability density
$\rho(\ell|y_0)$ of the random variable $\ell^\prime_\tau$ is obtained by
integrating the joint probability density $J(\ell,t|y_0)$ over the
second variable $t$, i.e., by setting $p = 0$ in the Laplace
transform:
\begin{equation} \label{eq:rhoD}
\rho(\ell|y_0) = \int\limits_0^\infty \d t \, J(\ell,t|y_0) = \tilde{J}(\ell,0|y_0) 
 = \pi_0 \delta(\ell) + (1-\pi_0) \frac{e^{-\ell/\ve}}{\ve}   ,
\end{equation}
where $\pi_0 = \tilde{J}_0(0|y_0) = y_0/\ve$ is the splitting
probability of reaching the endpoint $\ve$ first.  This is a mixture
of the Dirac distribution at $0$ and the exponential distribution with
mean $\ve$.  In other words, the random variable $\ell^\prime_\tau$
admits a simple form:
\begin{equation} \label{eq:ell_tau_plate}
\ell^\prime_\tau = \begin{cases}  0, \hskip 5mm \textrm{with probability~} \pi_0 , \cr
\ve \eta  , \quad \textrm{with probability~} 1 - \pi_0, \end{cases}
\end{equation}
where $\eta$ is the standard exponentially distributed random
variable (with mean $1$).

Even though the conditional distribution of the escape time $\tau$ is
found exactly in the Laplace domain (see below), the inversion of the
Laplace transform and the consequent generation of $\tau$ can be
time-consuming (see \ref{app:condproden}). To overcome this
limitation, we propose a simple approximation, in which the random
escape time $\tau$ is replaced by its mean value, which can be found
explicitly (see \ref{sec:justify} for discussion on its validity).

In practice, one can first generate a binary random variable that
determines which endpoint of the interval $(0,\ve)$ is reached first:
the endpoint $\ve$ (with probability $\pi_0$) or the endpoint $0$
(with probability $1 - \pi_0$).

\begin{enumerate}[(i)]
\item %(i) 
If the endpoint $\ve$ is reached first, then the boundary local time
$\ell^\prime_\tau$ is zero, while the escape time (denoted here as
$\tau_0$) is determined by the conditional probability density
$H_0(t|y_0)$ (see \ref{app:condproden}), whose Laplace transform is
\begin{equation}  \label{eq:H0_int}
\langle e^{-p\tau_0} \rangle = \tilde{H}_0(p|y_0) = \frac{\ve}{y_0} \, \frac{\sinh(\sqrt{p/D}\, y_0)}{\sinh(\sqrt{p/D}\, \ve)} \,.
\end{equation}
The mean escape time is then 
\begin{equation}   \label{eq:tau_mean_calc_plate1}
\langle \tau_0 \rangle = - \llp \partial_p \tilde{H}_0 (p | y_0) \rrp_{p=0} 
= \frac{\ve^2 - y_0^2}{6D} \,. 
\end{equation}

\item %(ii) 
If the endpoint $0$ is reached first, then the boundary local time
$\ell^\prime_\tau$ is nonzero and can be generated from its marginal
distribution (\ref{eq:rhoD}), i.e., by setting $\ell^\prime_\tau = \ve \,
\eta$, where $\eta$ is the standard exponential random variable.
In turn, the Laplace transform of the conditional probability density
of the escape time (denoted here as $\tau_\ell$) is determined by the
second term in Eq. (\ref{eq:j1tilde}), which is divided by the second
term in Eq. (\ref{eq:rhoD}):
\begin{equation}  \label{eq:Hcond2} 
\langle e^{-p\tau_\ell} \rangle = \tilde{H}(p|y_0,\ell) =  \frac{1}{(1 - \pi_0) e^{-\ell/\ve}/\ve } 
\frac{\sqrt{p/D} \sinh(\sqrt{p/D}(\ve - y_0))}{\sinh^2(\sqrt{p/D} \ve)} 
 e^{-\ell  \sqrt{p/D} \, \ctanh(\sqrt{p/D} \ve)} \,.
\end{equation}
The mean escape time is then
\begin{equation}   \label{eq:tau_mean_calc_plate2} 
\langle \tau_\ell \rangle = - \llp \partial_p \tilde{H} (p | y_0, \ell) \rrp_{p=0} 
= \frac{2\ell \ve + 2\ve^2 - (\ve - y_0)^2}{6D} \,. 
\end{equation}

\end{enumerate}
Finally, the lateral displacement $x_\tau$ can be generated as the
Gaussian random variable with mean zero and variance $2D \langle
\tau_{0} \rangle$ or $2D \langle
\tau_{\ell} \rangle$.

In higher dimensions ($d > 2$), the boundary layer near $\x_k$ is
approximated by a slab $\R^{d-1} \times (0,\ve)$ between two parallel
(hyper)planes.  As the lateral and transverse displacements are still
independent, the probability density flux $j(\x,\ell,t|\x_k)$ is again
factored into a Gaussian density for the lateral displacement and the
same joint probability density $J(\ell,t|y_0)$ for the interval
$(0,\ve)$, like in Eq. (\ref{eq:j_factored}).  As a consequence, the
generation of $\ell^\prime_\tau$ and $\tau$ remains unchanged, whereas
the displacement along each lateral coordinate is still an independent
Gaussian random variable with mean $0$ and variance $2D\langle \tau_0
\rangle$ or $2D\langle \tau_\ell \rangle$.

\subsection{Curved layer approximation (CLA)}
\label{sec:CLA}

The above flat layer approximation ignores the curvature of the
boundary, that is only valid for small enough $\ve$.  Such a choice
may slow down simulations, as the process stays too long in the
vicinity of the boundary.  This approximation can be further improved
by approximating the boundary layer by a circular annulus or a
spherical shell, with the same radius of curvature.  We start by the
two-dimensional case and then extend it to three dimensions.

\subsubsection{Convex boundary in two dimensions}

\begin{figure}[t!]
\begin{center}
\includegraphics[trim={0cm 0cm 0cm 1.215cm}, clip, width=70mm]{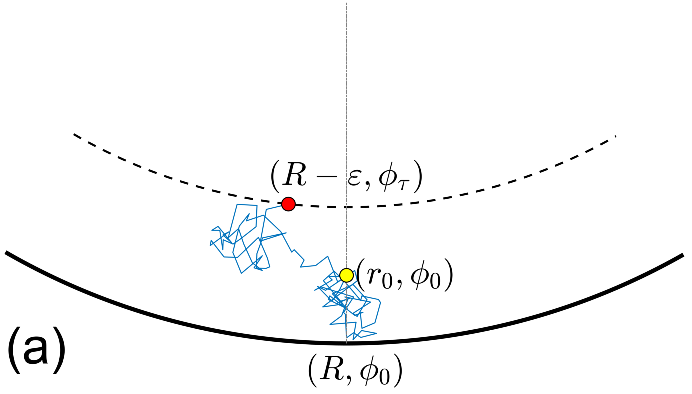}
\hskip 5mm \includegraphics[width=70mm]{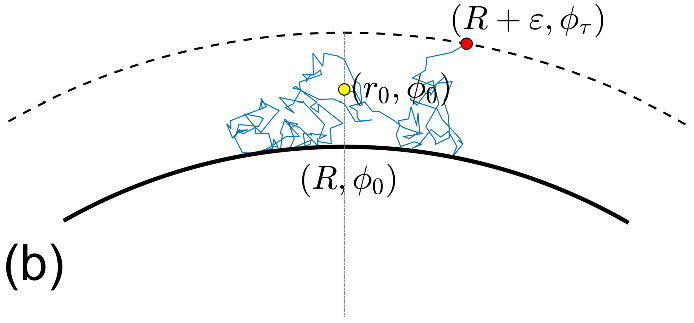}
\end{center}
\caption{
Curved layer approximation: starting from a point $(r_0,\phi_0)$ in
polar coordinates (yellow circle), a random trajectory (in blue)
experiences numerous reflections on the circle of radius $R$ before
escaping a thin layer of width $\ve$ at a random escape time $\tau$ in
a random escape position $(R\pm \ve,\phi_\tau)$ (red circle), with
sign minus corresponding to a convex boundary {\bf (a)} and sign plus
corresponding to a concave boundary {\bf (b)}.  These reflections are
characterized by the acquired boundary local time $\ell^\prime_\tau$.}
\label{fig:scheme_1D_curved}
% load('scheme1D_convex_XY.mat');  A_Yilin_1D_scheme_convex(X,Y);
% load('scheme1D_concave_XY.mat');  A_Yilin_1D_scheme_concave(X,Y);
\end{figure}

We first consider the case of a convex boundary
(Fig. \ref{fig:scheme_1D_curved}a).
Let $R$ be the radius of curvature of the boundary at the boundary
point $\x_b$, which is the closest to $\x_k$.  The boundary layer can
be approximated by a circular annulus between two circles of radii
$R-\ve$ and $R$, separated by distance $\ve$, i.e., $\Omega_\ve = \{
\x \in\R^2 ~:~ R - \ve < |\x| < R\}$.  We are interested in the escape
event from that annulus through the inner circle $\Gamma_\ve = \{
\x\in\R^2 ~:~ |\x| = R-\ve\}$ of radius $R-\ve$, which is thus
absorbing.  The joint probability density $J(\ell,t|\x_k)$ of the
escape time $\tau$ and the boundary local time $\ell^\prime_\tau$,
acquired on the outer circle up to $\tau$, is determined by the radial
displacements.  This can also be seen by integrating the probability
density flux $j(\x,\ell,t|\x_k)$ over the escape position
$\X^\prime_\tau$ on the inner circle, in which case all terms in the
sum of Eq. (\ref{eq:jp}) with $k \ne 0$ vanish due to the periodic
dependence of the Steklov eigenfunctions $V_k^{(p)}$ on the angle
\cite{Grebenkov20b}.  In other words, we get
\begin{align}  \label{eq:Jp_curved}
\tilde{J}(\ell,p|r_0) & = \tilde{J}_0(p|r_0) \delta(\ell) + V_0^{(p)}(r_0) C_0^{(p)} e^{-\mu_0^{(p)}\ell}  ,
\end{align}
where $r_0 = |\x_k|$ is the radial coordinate of the starting point
(here, we do not employ local coordinates centered at $\x_b$ so that
$r_0 \in (R - \ve,R]$), and
\begin{equation}
C_0^{(p)} = - \int\limits_{\Gamma_\ve} \d\x \, (\partial_n V_0^{(p)}).
\end{equation}
For a circular annulus with an absorbing inner boundary, one has
\cite{Grebenkov20b}
\begin{subequations}
\begin{align}
V_0^{(p)}(r_0) & = \frac{1}{\sqrt{2\pi R}} \, \frac{K_0(\alpha (R-\ve)) I_0(\alpha r_0) - I_0(\alpha (R-\ve)) K_0(\alpha r_0)}
{K_0(\alpha (R-\ve)) I_0(\alpha R) - I_0(\alpha (R-\ve)) K_0(\alpha R)} \,, \\
\mu_0^{(p)} & = \alpha \frac{K_0(\alpha (R-\ve)) I_1(\alpha R) + I_0(\alpha (R-\ve)) K_1(\alpha R)}
{K_0(\alpha (R-\ve)) I_0(\alpha R) - I_0(\alpha (R-\ve)) K_0(\alpha R)} \,, \\
C_0^{(p)} & = \frac{\sqrt{2\pi}}{\sqrt{R}} \, \frac{1}{K_0(\alpha (R-\ve)) I_0(\alpha R) - I_0(\alpha (R-\ve)) K_0(\alpha R)} \,, \\
\tilde{J}_0(p|r_0) & = \frac{K_0(\alpha R) I_0(\alpha r_0) - I_0(\alpha R) K_0(\alpha r_0)}
{K_0(\alpha R) I_0(\alpha (R-\ve)) - I_0(\alpha R) K_0(\alpha (R-\ve))} \,,
\end{align}
\end{subequations}
where $\alpha = \sqrt{p/D}$, $I_\nu(z)$ and $K_\nu(z)$ are the
modified Bessel functions of the first and second kind, and we used
$I'_0(z) = I_1(z)$, $K'_0(z) = - K_1(z)$, and the Wronskian $K_\nu(z)
I'_\nu(z) - I_\nu(z) K'_\nu(z) = 1/z$.

In the limit $p\to 0$, we get the marginal probability density of the
boundary local time $\ell^\prime_\tau$:
\begin{align}  \label{eq:rho_curved}
\rho(\ell|r_0) & = \tilde{J}(\ell,0|r_0) = \tilde{J}_0(0|r_0) \delta(\ell) + V_0^{(0)}(r_0) \, C_0^{(0)} \, e^{-\mu_0^{(0)}\ell} ,
\end{align}
where
\begin{subequations}
\label{eq:pi0convex2d}
\begin{align}
V_0^{(0)}(r_0) & = \frac{1}{\sqrt{2\pi R}} \, \frac{\ln(r_0/(R-\ve))}{\ln(R/(R-\ve))} \,, \\
\mu_0^{(0)} & = \frac{1}{R \ln (R/(R-\ve))} \,, \\
C_0^{(0)} & = \frac{\sqrt{2\pi}}{\sqrt{R}} \, \frac{1}{\ln (R/(R-\ve))} \,, \\
\pi_0 & = \tilde{J}_0(0|r_0) = \frac{\ln(R/r_0)}{\ln(R/(R-\ve))} \,,
\end{align}
\end{subequations}
and we used the asymptotic behavior of the modified Bessel
functions.  As a consequence, we find
\begin{align} 
\rho(\ell|r_0) & = \pi_0 \delta(\ell) + (1 - \pi_0) \mu_0^{(0)} e^{-\mu_0^{(0)}\ell}  .
\end{align}
As in the case of the flat layer approximation, we get a mixture of
the Dirac distribution at $0$ and the exponential distribution with
the mean $1/\mu_0^{(0)}$, where $\pi_0$ is the splitting probability
of the first arrival onto the absorbing (inner) circle, in which case
the boundary local time is zero.  Setting $r_0 = R - y_0$, one sees
that $\pi_0(r_0) \approx y_0/\ve$ in the limit $\ve \ll 1$ and $y_0
\ll 1$, as in the flat case.

By generating a binary random variable, one can choose between two
options:

(i) with probability $\pi_0$, one has $\ell^\prime_\tau = 0$, and the escape
time $\tau_0$ is determined by
\begin{equation}
\langle e^{-p\tau_0} \rangle = \tilde{H}_0(p|r_0) = \frac{\tilde{J}_0(p|r_0)}{\pi_0} \,.
\end{equation}
In particular, the mean escape time is
\begin{equation} \label{eq:tau_0_2Dconvex}
\langle \tau_0 \rangle = - \llp \partial_p \tilde{H}_0(p|r_0) \rrp_{p=0}
= \frac{1}{4D} \biggl( (R-\ve)^2 - r_0^2  
 + \frac{R^2 - r_0^2}{\ln (R/r_0)} - \frac{R^2- (R-\ve)^2}{\ln(R/(R-\ve))} \biggr).
\end{equation}

(ii) with probability $1-\pi_0$, one has $\ell^\prime_\tau = \eta /\mu_0^{(0)}$
(where $\eta$ is the standard exponential random variable), while the
escape time $\tau_\ell$ is determined by
\begin{equation}
\langle e^{-p\tau_\ell} \rangle =
\tilde{H}(p|r_0,\ell) = \frac{V_0^{(p)}(r_0) C_0^{(p)} e^{-\mu_0^{(p)}\ell} }{(1 - \pi_0) \mu_0^{(0)} e^{-\mu_0^{(0)}\ell}} .
\end{equation}
The mean escape time is then 
\begin{equation}  \label{eq:tau_ell_2Dconvex}
\langle \tau_\ell \rangle = - \llp \partial_p \tilde{H}(p|r_0,\ell) \rrp_{p=0} 
= \frac{\ell \, V_0^{(0)} C_0^{(0)} \bigl(\partial_p \mu_0^{(p)}\bigr)_{p=0} 
- \bigl(\partial_p V_0^{(p)}(r_0) C_0^{(p)}\bigr)_{p=0}}{(1-\pi_0) \mu_0^{(0)}} \,,
\end{equation}
where
\begin{equation}
\llp \partial_p \mu_0^{(p)} \rrp_{p=0} = R \frac{1 - (1-\ve/R)^2 + 2\ln(1-\ve/R) \bigl(\ln(1-\ve/R)+1\bigr)}{4D \ln^2(1-\ve/R)} \,, 
\end{equation}
and
\begin{align}
\llp \partial_p V_0^{(p)}(r_0) C_0^{(p)} \rrp_{p=0} 
= & -\frac{1}{4D R \ln^3(R/(R-\ve))} 
\biggl((2R^2+(R-\ve)^2-r_0^2)\ln(r_0/(R-\ve))\ln(R/(R-\ve)) \nonumber \\ 
& + 2(R^2-(R-\ve)^2)\ln(R/r_0)  
-(2R^2-(R-\ve)^2-r_0^2)\ln(R/(R-\ve))\biggr).
\end{align}
Substituting these expressions into Eq. (\ref{eq:tau_ell_2Dconvex}),
one can evaluate the mean escape time $\langle \tau_\ell \rangle$.

%%%  See A_Yilin_1Dconcave_H_p_fig();

The last step consists in generating the escape position
$\X^\prime_\tau$ on the inner circle.  In contrast to the flat case,
``lateral'' displacements along the tangential direction are not
independent from the transverse (radial) displacements due to the
curvature.  The exact distribution of the escape position can be found
from the probability density flux $j(\x,\ell,t|\x_0)$, but its
practical computation could be too sophisticated.  In turn, two
approximations can be useful.

(i) The simplest approximation consists in replacing the random
variable $\X_\tau^\prime$ by its mean value.  By symmetry, one has
$\langle \X_\tau^\prime \rangle = (R-\ve,\phi_0)$ in polar
coordinates, where $\phi_0$ is the angle of the starting position
$\x_k = (r_0,\phi_0)$.  For a flat boundary, this approximation would
correspond to replacing $x_\tau$ by $\langle x_\tau\rangle = 0$.

(ii) This approximation can be improved for small $\ve$, for which the
lateral motion can be treated as lying on the inner circle of radius
$R-\ve$ (instead of displacements inside the thin annulus of width
$\ve$).  In this approximation, the displacements on the circle are
characterized by the angle $\phi_t$, which is independent of radial
displacements, and can thus be generated from known distributions (see
\cite{Carlsson10,Burrage22}).

\subsubsection{Concave boundary in two dimensions}

The analysis for a locally concave boundary
(Fig. \ref{fig:scheme_1D_curved}b) is very similar.  In this case, one
considers a thin circular annulus $\Omega_\ve = \{ \x\in\R^2 ~:~ R <
|\x| < R+\ve\}$ with the absorbing outer boundary $\Gamma_\ve = \{
\x\in\R^2 ~:~ |\x| = R+\ve\}$.  As the absorbing boundary is now
located at $R+\ve$ instead of $R-\ve$, one mainly needs to change the
sign of $\ve$:
\begin{subequations}
\begin{align}
V_0^{(p)}(r_0) & = \frac{1}{\sqrt{2\pi R}} \, \frac{K_0(\alpha (R+\ve)) I_0(\alpha r_0) - I_0(\alpha (R+\ve)) K_0(\alpha r_0)}
{K_0(\alpha (R+\ve)) I_0(\alpha R) - I_0(\alpha (R+\ve)) K_0(\alpha R)} \,, \\
\mu_0^{(p)} & = - \alpha \frac{K_0(\alpha (R+\ve)) I_1(\alpha R) + I_0(\alpha (R+\ve)) K_1(\alpha R)}
{K_0(\alpha (R+\ve)) I_0(\alpha R) - I_0(\alpha (R+\ve)) K_0(\alpha R)} \,, \\
C_0^{(p)} & = - \frac{\sqrt{2\pi}}{\sqrt{R}} \, \frac{1}{K_0(\alpha (R+\ve)) I_0(\alpha R) - I_0(\alpha (R+\ve)) K_0(\alpha R)} \,, \\
\tilde{J}_0(p|r_0) & = \frac{K_0(\alpha R) I_0(\alpha r_0) - I_0(\alpha R) K_0(\alpha r_0)}
{K_0(\alpha R) I_0(\alpha (R+\ve)) - I_0(\alpha R) K_0(\alpha (R+\ve))} \,.
\end{align}
\end{subequations}
Note also that the signs in front of $\mu_0^{(p)}$ and $C_0^{(p)}$ are
also changed due to the opposite direction of the normal derivative.
These expressions determine $\tilde{J}(\ell,p|r_0)$ via
Eq. (\ref{eq:Jp_curved}). One also gets
\begin{subequations}
\label{eq:pi0concave2d}
\begin{align}
V_0^{(0)}(r_0) & = \frac{1}{\sqrt{2\pi R}} \, \frac{\ln(r_0/(R+\ve))}{\ln(R/(R+\ve))} \,, \\
\mu_0^{(0)} & = -\frac{1}{R \ln (R/(R+\ve))} \,, \\
C_0^{(0)} & = -\frac{\sqrt{2\pi}}{\sqrt{R}} \, \frac{1}{\ln (R/(R+\ve))} \,, \\
\pi_0 & = \tilde{J}_0(0|r_0) = \frac{\ln(R/r_0)}{\ln(R/(R+\ve))} \,,
\end{align}
\end{subequations}
that determine $\rho(\ell|r_0)$ via Eq. (\ref{eq:rho_curved}).
Finally, the mean escape time is given by
\begin{align} \label{eq:tau_0_2Dconcave}
\langle \tau_0 \rangle & = \frac{1}{4D} \biggl( (R+\ve)^2 - r_0^2  
 + \frac{R^2 - r_0^2}{\ln (R/r_0)} - \frac{(R+\ve)^2-R^2}{\ln(1 + \ve/R)} \biggr),
\end{align}
whereas $\langle \tau_\ell \rangle$ is still determined by
Eq. (\ref{eq:tau_ell_2Dconvex}), in which
\begin{equation}
\bigl(\partial_p \mu_0^{(p)}\bigr)_{p=0} = - R \frac{1 - (1+\ve/R)^2 + 2\ln(1+\ve/R) \bigl(\ln(1+\ve/R)+1\bigr)}{4D \ln^2(1+\ve/R)} \,,
\end{equation}
and
\begin{align}
 (\partial_p V_0^{(p)}(r_0) C_0^{(p)})_{p=0} 
= &\frac{1}{4D R \ln^3(R/(R+\ve))} 
\biggl((2R^2+(R+\ve)^2-r_0^2)\ln(r_0/(R+\ve))\ln(R/(R+\ve)) \nonumber \\ 
& + 2(R^2-(R+\ve)^2)\ln(R/r_0) -(2R^2-(R+\ve)^2-r_0^2)\ln(R/(R+\ve))\biggr).
\end{align}
%%%  See A_Yilin_1Dconcave_H_p_fig();

\subsubsection{Convex boundary in three dimensions}

A natural extension of the above analysis to the three-dimensional
case would consist in approximating a smooth boundary with two radii
of curvature by an ellipsoidal or hyperboloidal layer.  Unfortunately,
there is no explicit representation of the joint probability density
of $\ell^\prime_\tau$ and $\tau$ in these cases.  We therefore
restrict the discussion to the particular setting, in which the smooth
boundary can be locally approximated by a sphere of radius $R$
(i.e. two radii of curvature are equal).  In this case, the extension
is straightforward, while the resulting formulas are actually simpler
than in two dimensions.

A convex boundary is approximated by a spherical shell $\Omega_\ve =
\{ \x\in\R^3 ~:~ R-\ve < |\x| < R\}$ with the absorbing inner sphere 
$\Gamma_\ve = \{ \x\in\R^3 ~:~ |\x| = R - \ve\}$.  In this case, the
modified Bessel functions $I_\nu(z)$ and $K_\nu(z)$ are replaced by
the modified spherical Bessel functions $i_\nu(z)$ and $k_\nu(z)$,
which admit simple expressions.  Skipping technical steps, we get
\begin{subequations}
\begin{align}
V_0^{(p)}(r_0) & = \frac{\sinh(\alpha (r_0- R +\ve))}{\sqrt{4\pi} \, r_0 \sinh(\alpha \ve)} \,, \\
\mu_0^{(p)} & = \alpha\, \ctanh(\alpha \ve) - \frac{1}{R} \,, \\
C_0^{(p)} & = \frac{\sqrt{4\pi}\, \alpha (R-\ve)}{\sinh(\alpha \ve)} \,, \\
\tilde{J}_0(p|r_0) & = \frac{(R-\ve) \sinh(\alpha (R-r_0))}{r_0 \sinh(\alpha \ve)} \,,
\end{align}
\end{subequations}
while their limits as $p\to 0$ are
\begin{subequations}
\label{eq:pi0convex3d}
\begin{align}
V_0^{(0)}(r_0) & = \frac{r_0 - R + \ve}{\sqrt{4\pi} r_0 \ve}  \,, \\
\mu_0^{(0)} & = \frac{1}{\ve} - \frac{1}{R} \,, \\
C_0^{(0)} & = \frac{\sqrt{4\pi} (R-\ve)}{\ve}  \,, \\
\pi_0 & = \tilde{J}_0(0|r_0) = \frac{(R-\ve)(R-r_0)}{\ve r_0}  \,.
\end{align}
\end{subequations}
As a consequence, one can easily evaluate the mean escape times:
\begin{subequations} \label{eq:tau_3Dconvex}
\begin{align}
\langle \tau_0 \rangle & = \frac{\ve^2 - (R-r_0)^2}{6D} \,, \\
\langle \tau_\ell \rangle &= \frac{2\ell \ve + 2\ve^2 - (r_0 - R + \ve)^2}{6D} \,.
\end{align}
\end{subequations}
Setting $y_0 = R - r_0$ as the distance to the boundary, one
retrieves Eqs. (\ref{eq:tau_mean_calc_plate1},
\ref{eq:tau_mean_calc_plate2}) for the flat case.
As previously, one first employs the splitting probability $\pi_0$ to
choose between two options, generates the boundary local time
$\ell^\prime_\tau$, and uses the mean escape time.  The escape
position can be either generated, or replaced by its mean value, which
by symmetry is $(R-\ve, \theta_0,\phi_0)$ in spherical coordinates,
where $\theta_0$ and $\phi_0$ are the angular coordinates of the
starting position $\x_k$.

\subsubsection{Concave boundary in three dimensions}

A concave boundary is approximated by a spherical shell $\Omega_\ve =
\{ \x\in\R^3 ~:~ R < |\x| < R + \ve\}$ with the absorbing outer sphere 
$\Gamma_\ve = \{ \x\in\R^3 ~:~ |\x| = R + \ve\}$.  One has
\begin{subequations}
\begin{align}
V_0^{(p)}(r_0) & = \frac{\sinh(\alpha (R-r_0+\ve)))}{\sqrt{4\pi} \, r_0 \sinh(\alpha \ve)} \,, \\
\mu_0^{(p)} & = \alpha\, \ctanh(\alpha \ve) + \frac{1}{R} \,, \\
C_0^{(p)} & = \frac{\sqrt{4\pi}\, \alpha (R+\ve)}{\sinh(\alpha \ve)} \,, \\
\tilde{J}_0(p|r_0) & = \frac{(R+\ve) \sinh(\alpha (r_0-R))}{r_0 \sinh(\alpha \ve)} \,,
\end{align}
\end{subequations}
and
\begin{subequations}
\label{eq:pi0concave3d}
\begin{align}
V_0^{(0)}(r_0) & = \frac{R-r_0+\ve}{\sqrt{4\pi} r_0 \ve}  \,, \\
\mu_0^{(0)} & = \frac{1}{\ve} + \frac{1}{R} \,, \\
C_0^{(0)} & = \frac{\sqrt{4\pi} (R+\ve)}{\ve}  \,, \\
\pi_0 & = \tilde{J}_0(0|r_0) = \frac{(R+\ve)(r_0-R)}{\ve r_0} \,.
\end{align}
\end{subequations}
As a consequence, one can easily evaluate the mean escape
times:
\begin{subequations} \label{eq:tau_3Dconcave}
\begin{align}
\langle \tau_0 \rangle & = \frac{\ve^2 - (r_0 - R)^2}{6D} \,, \\
\langle \tau_\ell \rangle & = \frac{2\ell \ve + 2\ve^2 - (R-r_0 +\ve)^2}{6D} \,.
\end{align}
\end{subequations}
Setting $y_0 = r_0 - R$ as the distance to the boundary, one
retrieves Eqs. (\ref{eq:tau_mean_calc_plate1},
\ref{eq:tau_mean_calc_plate2}) for the flat case.

\subsection{Stopping conditions}
\label{sec:stopcond}

In the conventional setting of domains with an absorbing boundary, the
simulation was stopped upon the first arrival into the boundary layer
of width $\ve$. In turn, as we deal with reflected Brownian motion,
the simulation would never stop without imposing a suitable stopping
condition. In many applications, one employs one of the three
conditions (or their combinations):
\begin{enumerate}[(i)]
\item 
at a fixed time $T$; 

\item 
at a random stopping time $\dlt$ with a prescribed probability distribution; 

\item 
at a random time ${\mathcal T}$ when the boundary local time $\ell_t$
exceeds a fixed or random threshold $\hat\ell$ with a prescribed
probability distribution.
\end{enumerate}

The first stopping condition allows one to characterize the diffusive
dynamics in a confining domain $\Omg$, i.e., the position $\X_T$ and
the boundary local time $\ell_T$ at a given time $T$. In particular,
one can obtain the empirical distributions of these quantities, their
moments and correlations. The second stopping condition is often
employed to simulate ``mortal'' particles with a finite random
lifetime with a given distribution
\cite{Yuste13,Meerson15,Grebenkov17d}.  The most common choice is the
exponential distribution of the stopping time $\dlt$, $\mathbb{P} \{
\dlt > t \} = e^{-pt}$, with $p$ being the decay rate or,
equivalently, $1/p$ being the mean lifetime. Basic examples are
nuclear radioactive disintegration, spin relaxation, photobleaching,
first-order bulk reactions, etc. In this case, one generates the
statistics of the position $\X_\dlt$ and the acquired boundary local
time $\ell_\dlt$ until the ``death'' time $\dlt$.  The third stopping
condition was introduced in \cite{Grebenkov20,Grebenkov07} to describe
various surface reactions which may occur after a sufficient number
(represented by the threshold $\hat\ell$) of failed reaction attempts.
In this way, one can generate the statistics of the reaction times
$\mathcal{T}$ and of the reaction locations $\X_\mathcal{T}$ on the
boundary.  If the threshold $\hat\ell$ is fixed, the surface reaction
is triggered after $\hat\ell/\ve$ encounters of the diffusing particle
with the reactive layer of width $\ve$. More generally, the number of
required encounters can be drawn from a given distribution. For
instance, the most common setting of a constant partial reactivity
$\kpa$ in the Robin boundary condition corresponds to the exponential
distribution of the threshold, $\mathbb{P} \{ \hat\ell > \ell \} =
e^{-q \ell}$, with $q = \kpa / D$ \cite{Grebenkov20}.  In turn, other
distributions can account for more sophisticated surface reactions
such as activation or passivation of the reactive boundary, or
non-Markovian binding \cite{Grebenkov23b,Grebenkov24f}.  The third
stopping condition was also applied to describe permeation across the
boundary \cite{Bressloff22c,Bressloff23a,Bressloff23b}.

An accurate implementation of the stopping condition is known to be
delicate even for the absorbing boundary, i.e., for the classical WOS
algorithm. For instance, if the stopping time $T$ is fixed, the
simulation does not actually stop at $T$ but when $t_{k+1}
>T$. However, if the last jump between $t_k$ and $t_{k+1}$ was big,
the position $\X_T$ lying inside the large ball $B_{\rho_k}(\x_k)$ can
significantly differ from the last generated position $\X_{t_{k+1}}$
on the sphere $\partial B_{\rho_k}(\x_k)$. To amend this problem, one
often employs specific treatment of the last step such as, e.g.,
generating the random position $\X_T$ for the intermediate time $T$
between $t_k$ and $t_{k+1}$ from the conditional probability density,
an interpolation between $\X_{t_k}$ and $\X_{t_{k+1}}$, the inclusion
of the upper bound on the possible escape time, etc
\cite{maire2013monte,Grebenkov11}.  As this issue is classical, we do
not discuss these improvements.  In our work, we focus on the
statistics of the boundary local time, for which the last step is less
critical. Indeed, if the last step (after which the simulation is
stopped) is a big jump to the sphere $\partial B_{\rho_k}(\x_k)$, the
bulk trajectory inside the ball $B_{\rho_k}(\x_k)$ cannot touch the
boundary, so that the boundary local time has not changed during this
jump: $\ell_{t_{k+1}} = \ell_T = \ell_{t_k}$.
%%%
In turn, if the last jump is an escape from the boundary layer
$\Omega_\varepsilon$ of width $\varepsilon$, the boundary local time
can change, but its random increment $\ell^\prime_\tau$ is typically
of the order of $\varepsilon$. If the width $\ve$ is small enough, the
induced error of $\ell_T$ is small due to $\ell_{t_k} \leqslant \ell_T
\leqslant \ell_{t_{k+1}}$. As a consequence, either of these two
bounds can be used to approximate $\ell_T$ (moreover, one can record
both $\ell_{t_k}$ and $\ell_{t_{k+1}}$ to get lower and upper bounds).
Alternatively, one can employ a suitable interpolation between
$\ell_{t_k}$ and $\ell_{t_{k+1}}$, e.g., the Skorokhod integral
representation of the boundary local time (see \ref{app:Skorokhod})
suggests to set
\begin{equation}
\ell_T = \ell_{t_k} + \llp \ell_{t_{k+1}} - \ell_{t_k} \rrp \frac{\sqrt{T} - \sqrt{t_k}}{\sqrt{t_{k+1}} - \sqrt{t_k}} \, .
\end{equation}
Finally, one can also attempt to derive the conditional probability
density of $\ell_T$.  In the following numerical examples, we employ
the simplest approximation by setting $\ell_T = \ell_{t_{k+1}}$.

\section{Implementation of the algorithm}
\label{sec:algorithm}

We summarize the main steps of our method by focusing explicitly on
domains in two and three dimensions.  An extension to higher
dimensions is immediately available for FLA; in turn, additional
computations would be needed for implementing CLA (not present).

\begin{enumerate}
  
\item 
Consider a domain $\Omg$ defined by a function $\rho(\x)$, which
determines the distance from any point $\x$ to the boundary $\partial
\Omg$.  Initialize all the parameters: initial position
$\x_\mathrm{ini}$, diffusion coefficient $D$, number of particles $N$,
width $\ve$ of the boundary layer, stopping condition, etc.  Load the
pre-computed data to generate the escape times for the unit disk in
2d, or the unit sphere in 3d (see \ref{app:exittime}).  For each
particle ($i = 1, 2, \ldots, N$), run the following simulation.

\item 
Initiate $\x_0 = \x_\mathrm{ini}$, $t_0 = 0$, $\ell_0 = 0$; for the
second stopping condition, generate the random stopping time
$\dlt$ (e.g., if $\dlt$ obeys the exponentially distribution, set
$\dlt = - \ln(u) / p$, where $u$ is the uniform random variable on
$(0,1)$, and $p > 0$ is a given parameter); for the third stopping
condition, generate the random threshold $\hat{\ell}$ (e.g., if
$\hat{\ell}$ obeys the exponentially distribution, set $\hat\ell = -
\ln(u) / q$, where $u$ is the uniform random variable on $(0,1)$,
and $q > 0$ is a given parameter).

\item 
At each step $k = 0, 1, 2, \ldots$, the radius $\rho_k = \rho(\x_k)$
is determined by the current position $\x_k$.

\begin{itemize}
\item 
If $\rho_k > \ve / 2$, the new position $\x_{k+1}$ is generated to be
uniformly distributed on $\partial B_{\rho_k} (\x_k)$. In two
dimensions, it is the circle of radius $\rho_k$, centered at $\x_k$,
so that
\begin{equation} 
\left\{ \begin{aligned} x_{k+1} &= x_k +\rho_k \cos\varphi \,, \\ y_{k+1} &= y_k + \rho_k \sin\varphi \,, \end{aligned} \right. 
\end{equation}
where $\varphi$ is a random variable uniformly distributed in
$(0,2\pi)$.  In three dimensions, it is the sphere of radius $\rho_k$,
centered at $\x_k$, so that
\begin{equation}
\left\{ 
      \begin{aligned} x_{k+1} &= x_k +\rho_k \sin\theta \cos\varphi \,, \\ y_{k+1} &= y_k + \rho_k \sin\theta \sin\varphi \,, \\ z_{k+1} &= z_k + \rho_k \cos\theta  \,, \end{aligned} \right. 
\end{equation}
where $\varphi$ is a random variable uniformly distributed in
$(0,2\pi)$, and $\theta$ is a random variable such that $\cos\theta$
is uniformly distributed in $(-1,1)$. \\
The diffusion time is incremented by
\begin{equation}
t_{k+1} = t_{k} + \frac{\rho_k^2}{D} \tau_c \,, 
\end{equation}
where $\tau_c$ is the escape time either from the unit disk (in 2d) or
from the unit ball (in 3d).  This random variable is generated in a
standard way (see \ref{app:exittime}).

\item 
If $\rho < \ve/2$, initiate the escape-from-a-layer event: generate a
random variable $w$ uniformly distributed in $(0,1)$.
    
\begin{itemize}
\item 
If $w \in (0,\pi_0)$, the particle escapes the layer without touching
the boundary, so that $\ell^\prime_\tau = 0$.  We set therefore
$t_{k+1} = t_k + \langle \tau_0 \rangle$ and $\ell_{t_{k+1}} =
\ell_{t_k}$.

\item 
If $w \in (\pi_0,1)$, the particle escapes the layer after reflections
on the boundary. The boundary local time increases as $\ell_{t_{k+1}}
= \ell_{t_k} + \eta / \mu_0^{(0)}$, where $\eta$ is a standard
exponentially random variable, and $t_{k+1} = t_k + \langle \tau_\ell
\rangle$.
\end{itemize}

According to different approximations (FLA or CLA), the curvature sign
(convex or concave), and space dimension ($d=2$ or $d=3$), $\pi_0$,
$\langle \tau_0 \rangle$, $\langle \tau_\ell \rangle$, and
$\mu^{(0)}_0$ are given by equations listed in Table
\ref{table:eqs_alg}, with $R$ being the radius of curvature at the
boundary point $\x_b$ closest to $\x_k$, and $r_0 = \llv R - \rho_k
\rrv$.  The new position $\x_{k+1}$ is located at distance $\ve$ from
the boundary point $\x_b$.
        
\begin{table}[t!]
    \centering
    \begin{tabular}{c|c|c|c|c|c}
    \hline\hline
   \multirow{2}{*}{} & \multirow{2}{*}{FLA} & \multicolumn{4}{c}{CLA} \\
   \cline{3-6} 
& & convex 2d & concave 2d & convex 3d & concave 3d \\
     \hline
    $\pi_0$ & Eq. (\ref{eq:rhoD}) &  Eq. (\ref{eq:pi0convex2d}d) & Eq. (\ref{eq:pi0concave2d}d) & Eq. (\ref{eq:pi0convex3d}d) & Eq. (\ref{eq:pi0concave3d}d) \\ \hline
    $\langle \tau_0 \rangle$ & Eq. (\ref{eq:tau_mean_calc_plate1}) & Eq. (\ref{eq:tau_0_2Dconvex}) & Eq. (\ref{eq:tau_0_2Dconcave}) & Eq. (\ref{eq:tau_3Dconvex}a) & Eq. (\ref{eq:tau_3Dconcave}a) \\ \hline
    $\langle \tau_\ell \rangle$ & Eq. (\ref{eq:tau_mean_calc_plate2}) & Eq. (\ref{eq:tau_ell_2Dconvex}) & Eq. (\ref{eq:tau_ell_2Dconvex}) & Eq. (\ref{eq:tau_3Dconvex}b) & Eq. (\ref{eq:tau_3Dconcave}b)  \\ \hline
    $\mu^{(0)}_0$ & Eq. (\ref{eq:ell_tau_plate}) & Eq. (\ref{eq:pi0convex2d}b) & Eq. (\ref{eq:pi0concave2d}b) & Eq. (\ref{eq:pi0convex3d}b) & Eq. (\ref{eq:pi0concave3d}b) \\
    \hline\hline
    \end{tabular}
\caption{
Summary of approximations for the quantities that determine the
escape-from-a-layer event.}  
\label{table:eqs_alg}
\end{table}

\end{itemize}

\item 
Check for the stopping condition ($t_{k+1} > T$, or $t_{k+1} > \dlt$,
or $\ell_{t_{k+1}} > \hat\ell$): if it is not satisfied, return to
step 3; otherwise, stop the simulation and record the boundary local
time $\ell_{t_{k+1}}$ for this particle.

\item 
Repeat the steps 2-4 for $N$ particles to get an empirical statistics
of boundary local times.
\end{enumerate}

%%%%%

In the above implementation, the escape-from-a-layer event is
triggered when the distance to the boundary $\rho_k$ is below $\ve/2$
(instead of the earlier discussed condition $\rho_k < \ve$).  This
``trick'' allows one to eliminate too often immediate returns to the
layer after the escape. The choice of the fraction $1/2$ in front of
$\ve$ is somewhat arbitrary, it can be replaced by any number between
0 and 1.

\section{Validation}
\label{sec:validation}

In order to check the accuracy of the proposed algorithm and the role
of its parameters (such as $\ve$), we consider reflected Brownian
motion in a bounded domain $\Omega \subset \R^d$ started from a fixed
point $\x_0\in\Omega$.  The probability density $\rho(\ell,t|\x_0)$ of
the boundary local time $\ell_t$ reads in the Laplace domain as
\cite{Grebenkov19}
\begin{equation}  \label{eq:PDF_general}
\tilde{\rho}(\ell,p|\x_0) = \int\limits_0^\infty \md t \, e^{-pt} \, \rho(\ell,t|\x_0) 
= \tilde{S}_0(p|\x_0) \delta(\ell) + \sum\limits_{k=0}^\infty
e^{-\mu_k^{(p)} \ell} c_k^{(p)}(\x_0) \,,
\end{equation}
where
\begin{equation}
c_k^{(p)}(\x_0) = \frac{[V_k^{(p)}(\x_0)]^*}{D} \int\limits_\Omega \d\x \, V_k^{(p)}(\x) \,,
\end{equation}
and $\tilde{S}_0(p|\x_0)$ is the Laplace-transformed survival
probability in the presence of an absorbing boundary $\pa$.  While
$\tilde{S}_0(p|\x_0)$, $\mu_k^{(p)}$ and $V_k^{(p)}(\x)$ can be found
numerically for a given domain $\Omega$ (e.g., by a finite-element
method), the need for the Laplace transform inversion makes the above
spectral expansion less suitable for validating Monte Carlo
simulations.

To overcome this difficulty, we focus on the boundary local time
$\ell_\delta$, stopped at an exponentially distributed random time
$\delta$, i.e., $\P\{\delta > t\} = e^{-p t}$, with a given rate $p >
0$ (the second stopping condition).  The probability density of
$\ell_\delta$ can be obtained by averaging $\ell_t$ as
\begin{equation}
\Psi(\ell,p|\x_0) = \int\limits_0^\infty \d t \, \underbrace{p\, e^{-pt}}_{=\ \textrm{pdf of}~\delta} \rho(\ell,t|\x_0) 
= p \, \tilde{\rho}(\ell,p|\x_0).
\end{equation}
In other words, the statistics of $\ell_\delta$ is directly accessible
in the Laplace domain and can thus be obtained without Laplace
transform inversion.

%%%%%%

\begin{figure}[t!]
\begin{center}
\begin{tikzpicture}[scale=0.83]
\node (2d1) at (0,0) {\includegraphics[width=1.0in]{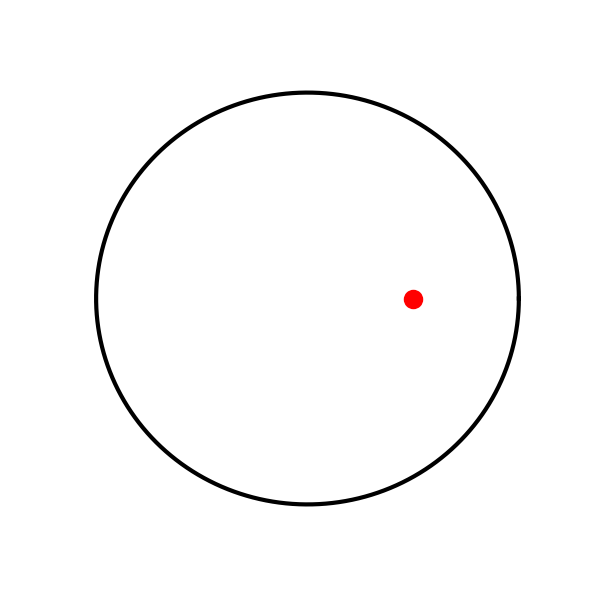}};
\node[] at (2d1.south west) {(a)};
\node (2d2) at (5,0) {\includegraphics[width=1.0in]{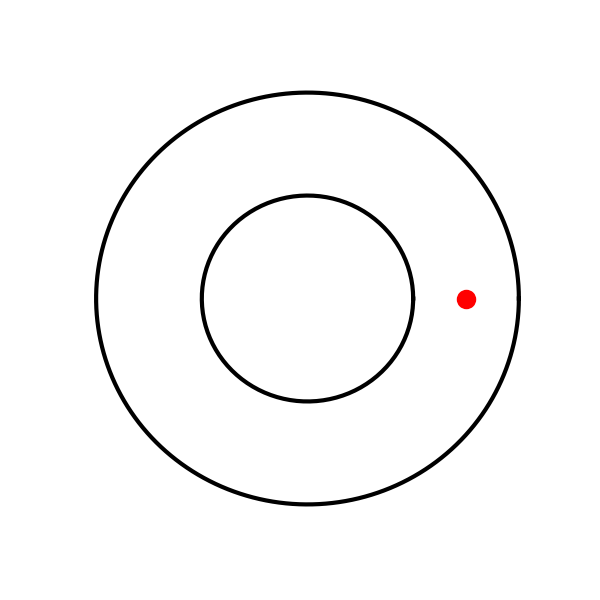}};
\node[] at (2d2.south west) {(b)};
\node (2d3) at (10,0) {\includegraphics[width=1.0in]{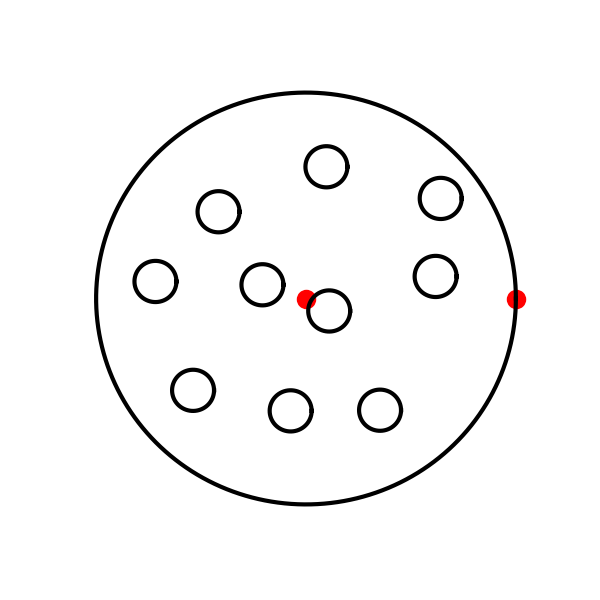}};
\node[] at (2d3.south west) {(c)};
\node (3d1) at (0,-3.5) {\includegraphics[trim={2.2cm 2.2cm 2cm 2cm}, clip, height=1.0in]{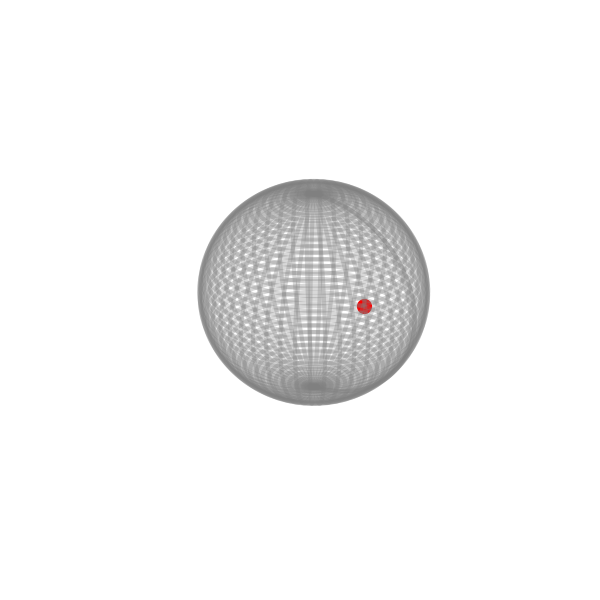}};
\node[] at (3d1.south west) {(d)};
\node (3d2) at (5,-3.5) {\includegraphics[trim={2.2cm 2.2cm 2cm 2cm}, clip, height=1.0in]{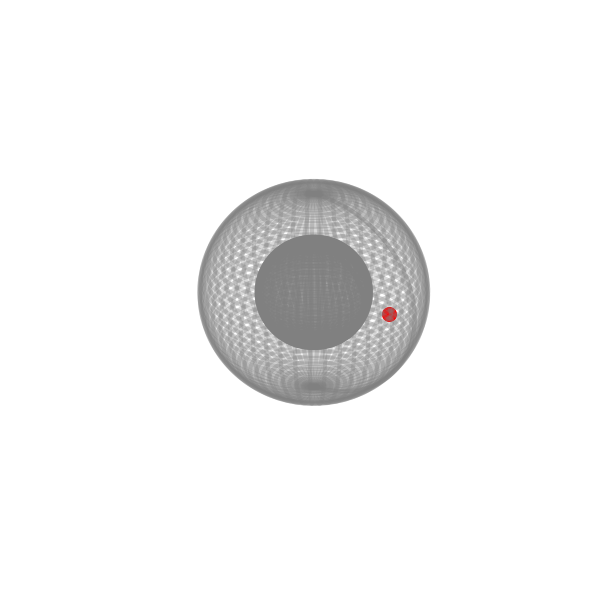}};
\node[] at (3d2.south west) {(e)};
\node (3d3) at (10,-3.5) {\includegraphics[trim={4.5cm 2.4cm 4cm 2cm}, clip, height=1.0in]{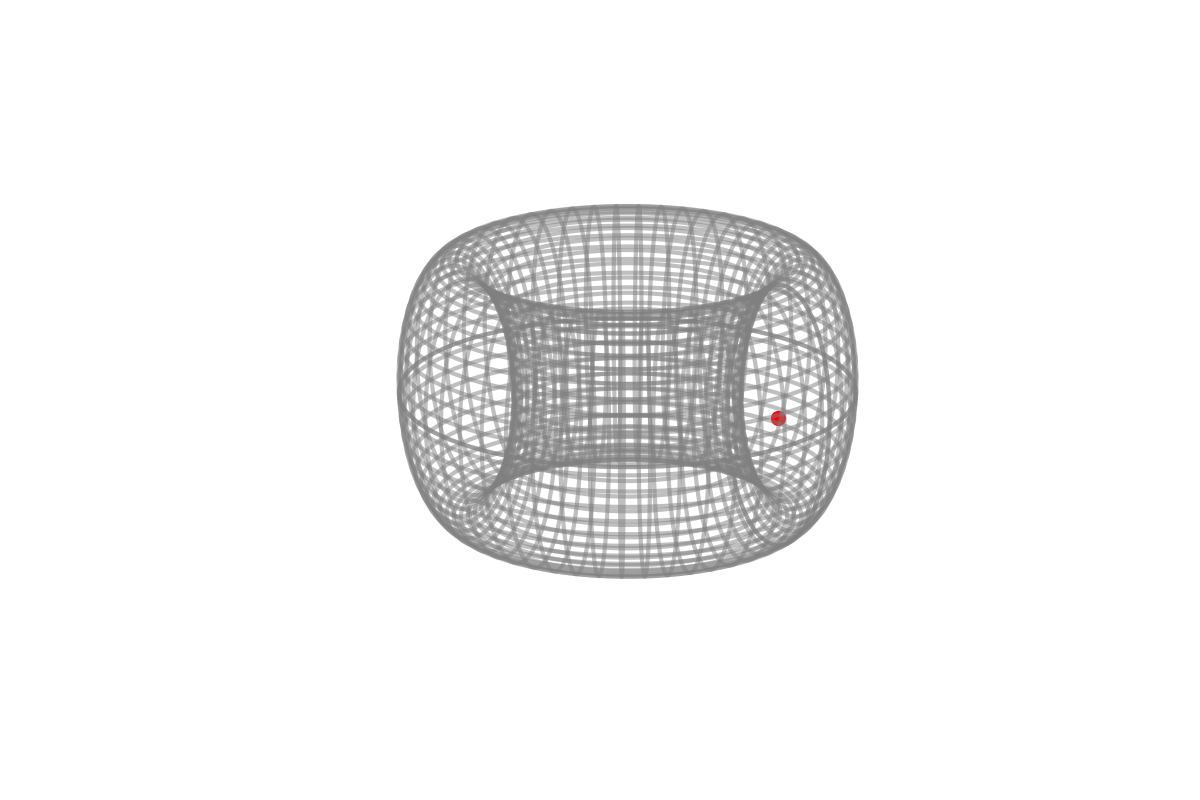}};
\node[] at (3d3.south west) {(f)};
\end{tikzpicture}
\caption{Considered examples. %Examples of 2D domains. 
\tbf{(a)} 
Unit disk centered at the origin; the initial point (in red) is
located at $(r_0=0.5, \vphi_0=0)$.
\tbf{(b)} 
Annulus between two concentric circles of radii $R=1$, $L=2$; the
initial point (in red) is located at $(r_0=1.5, \vphi_0=0)$.
\tbf{(c)} 
Unit disk ($R=1$) centered at the origin and perforated by ten small
circular holes of the same radius $R_0=0.1$ located randomly (their
centers are at $(-0.4167, 0.4213)$, $(0.0973, 0.6397)$, $(0.6418,
0.4860)$, $(-0.7173, 0.0827)$, $(-0.2075, 0.0666)$, $(0.1105,
-0.0605)$, $(0.6176, 0.1071)$, $(-0.5381, -0.4460)$, $(-0.0731,
-0.5457)$, $(0.3528, -0.5425)$); the initial points are: $(R, 0)$,
$(0,0)$. %\\
%Examples of 3D domains. 
\tbf{(d)} 
Unit sphere ($R=1$) centered at the origin; the initial point (in red)
is located at $r_0=0.5$.
\tbf{(e)} 
Spherical shell between two concentric spheres of radii $R=1$ and
$L=2$; the initial point (in red) is located at $r_0=1.5$.
\tbf{(f)} 
Torus with $R = 1.5$, $R_0 = 0.5$; the initial point (in red) is
located at $(1.5, 0, 0)$.  }
\label{fig:illustration}
\end{center}
\end{figure}

We test our approach for six domains shown in
Fig. \ref{fig:illustration}.  For these domains, we obtain the
empirical statistics of $\ell_\dlt$ by using FLA and CLA with
different $\ve$ and compare to either exactly known results, or
numerical results obtained by a finite-element method
\cite{Chaigneau24}. We also implemented the Skorokhod integral method
(SIM) considered as a benchmark state-of-the-art Monte Carlo technique
\cite{Schumm23}.  In \ref{app:Skorokhod}, we briefly describe the
original method as being introduced by Schumm and Bressloff (and
referred to as SIM1), and a minor variation of this method (referred
to as SIM2).  Throughout all cases, we fix $D=1$.  The discussion of
cases shown in Fig. \ref{fig:illustration}b, \ref{fig:illustration}d,
\ref{fig:illustration}e is re-delegated to \ref{app:valid}.

\subsection{Disk}
\label{sec:2dvalid}

For a disk of radius $R$ (Fig. \ref{fig:illustration}a), the
probability density $\Psi(\ell,p|\x_0)$ is known explicitly
\cite{Grebenkov19,Grebenkov20b}:
\begin{equation}  \label{eq:PDF_elldelta}
\Psi(\ell,p|\x_0) = \pi_0 \delta(\ell) 
+ (1 - \pi_0)\, \mu_0^{(p)}\, e^{-\mu_0^{(p)}\ell}  ,
\end{equation}
where $\alpha = \sqrt{p/D}$ and 
\begin{subequations}
\begin{align}
\mu_0^{(p)} & = \alpha \frac{I_1(\alpha R)}{I_0(\alpha R)} \,, \\
\pi_0 & = 1 - \frac{I_0(\alpha r_0)}{I_0(\alpha R)} \,.
\end{align}
\end{subequations}
Once again, we have a mixture of the Dirac distribution and the
exponential distribution.  For this distribution, it is easy to
compute all the moments of the $\ell_\delta$, in particular,
\begin{equation}
\langle \ell_\delta\rangle = (1-\pi_0)/\mu_0^{(p)}, \qquad 
\sigma^2 = \mathrm{Var}\{ \ell_\delta\} = (1-\pi_0^2)/(\mu_0^{(p)})^2 \,. 
\label{eq:ellmean}
\end{equation}

%%% [A]
We fix $p=1$ and the initial point $r_0 = 0.5$ inside the unit disk
$(R=1)$.  As the first test, we compute the empirical mean $\lla
\ell_\dlt \rra_\mrm{emp}$ by using $N=10^6$ particles and evaluate its
relative error with respect to $\lla \ell_\dlt \rra$ by three methods
(SIM, FLA, CLA) and three values of the layer width $\ve$.  Table
\ref{table:2Ddisk} presents these results, as well as the CPU time for
each simulation (all simulations were run on the same MacBook Pro with
1.4GHz quad-core Intel Core i5 processor and 8GB of 2133MHz LPDDR3
onboard memory).  First of all, the three methods have similar CPU
times for a given $\ve$, the SIM being slightly faster. Importantly,
the CPU time grows linearly with $1/\ve$ for the three methods. This
is in sharp contrast to the $\ln(1/\ve)$ scaling of the CPU time for
absorbing boundaries. Indeed, according to Eq. (\ref{eq:ellt_def2}),
the number of encounters with the boundary layer of width $\ve$
behaves as $\ell_\dlt/\ve$, requiring on average $\lla \ell_\dlt \rra
/ \ve$ escape events for each simulated trajectory.  The expected
linear scaling of the CPU time with $1/\ve$ was the main motivation of
our work, which aims at computing the boundary local time faster and
more accurately by allowing larger $\ve$.

Let us now compare the accuracy of the empirical mean computed by
three methods. The relative error of the benchmark SIM1 changes from
$-7.43 \%$ to $-0.91 \%$ and then to $1.80 \%$ as $\ve$ goes down from
$10^{-1}$ to $10^{-2}$ and then to $10^{-3}$.  It seems that there is
an ``optimal'' width layer $\ve$ that minimizes the error, whereas
further decreases of $\ve$ is counter-productive. Note that a similar
trend can be noticed in Fig. 3b of \cite{Schumm23}, when the relative
error of the mean first-passage time is minimal at $\ve \simeq 0.01$.
Even though such accuracy can be sufficient for some applications, it
significantly exceeds the relative statistical error, estimated as
$\frac{\sigma}{\lla \ell_\dlt \rra \sqrt{N}} \simeq 0.1 \%$. In other
words, the relative error cannot be explained by statistical
fluctuations (which are present in any finite sample) and thus
originates from the employed approximations (e.g., finite $\ve$).  The
accuracy of our modification (SIM2) is worse for this example.

The situation is very different for the FLA, for which the relative
error drops from 4.8\% at $\ve = 10^{-1}$ to 0.1\% at $\ve =
10^{-3}$. In other words, we managed to reduce the relative error by
factor $\sim 20$ and to achieve the level of statistical errors by
using almost the same CPU time as for the SIM1.  The accuracy is
further improved by CLA. Even for $\ve=10^{-1}$, the relative error of
the empirical mean is $0.5\%$, i.e., one achieves a two-fold
improvement in accuracy by a 7 times faster algorithm, as compared to
the SIM1 with $\ve = 10^{-2}$.  We conclude that the CLA dramatically
outperforms two other methods.

\begin{figure}[t!]
\begin{center}
\includegraphics[width=0.33\linewidth]{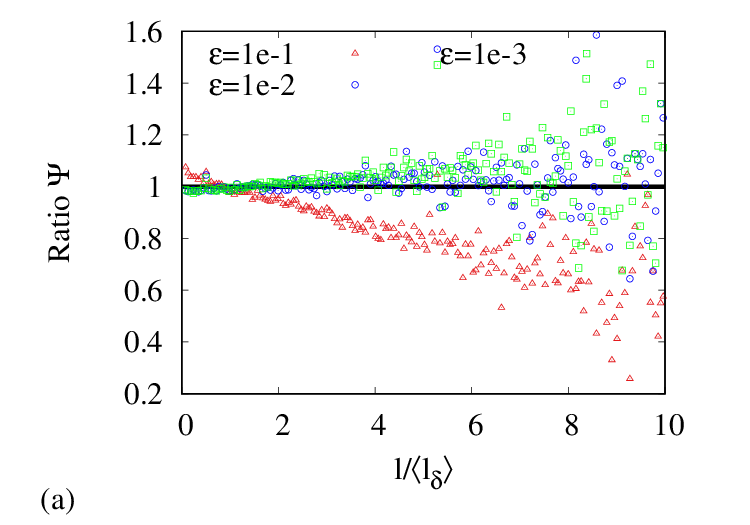}
\includegraphics[width=0.33\linewidth]{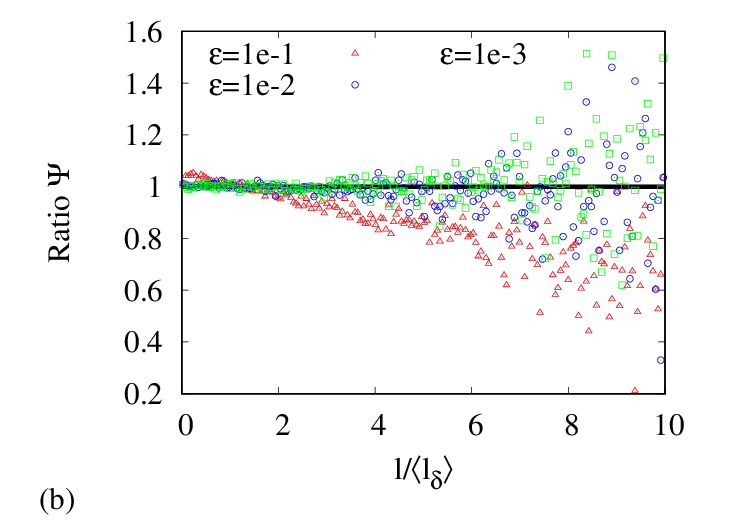}
\includegraphics[width=0.33\linewidth]{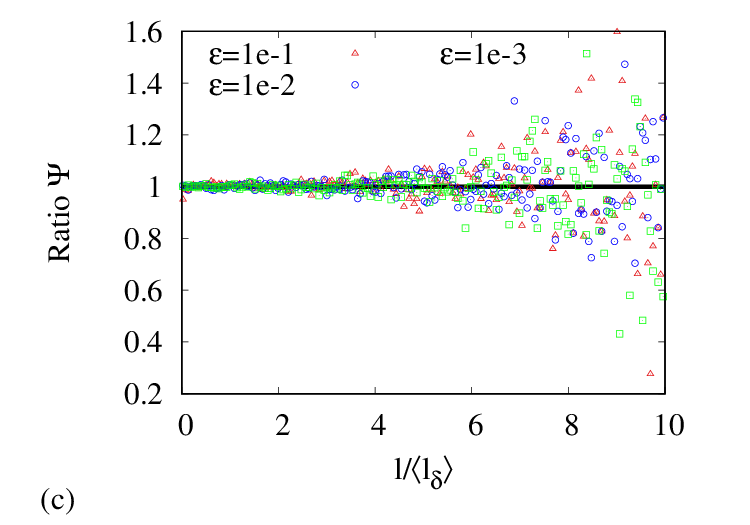}
\end{center}
\caption{
Ratio $\Psi_\mrm{emp}(\ell,p|\x_0)$ over $\Psi(\ell,p|\x_0)$ as a
function of $\ell/\lla\ell_\dlt\rra$ for the unit disk
(Fig. \ref{fig:illustration}a), with $p=1$ and $r_0 = 0.5$.  The
empirical PDF $\Psi_\mrm{emp}$ was estimated from $N=10^6$ simulated
values of $\ell_\dlt$ by \tbf{(a)} SIM1, \tbf{(b)} FLA, \tbf{(c)} CLA,
with different choices of the boundary layer width: $\ve=10^{-1}$
(triangles), $\ve=10^{-2}$ (circles), and $\ve=10^{-3}$ (squares).
The trajectories that never touched the boundary and thus yielded
$\ell_\delta = 0$, were excluded. }
\label{fig:disk_comp}
\end{figure}

The comparison between three methods carries on for the probability
density $\Psi(\ell,p|\x_0)$ of the $\ell_\dlt$.  Figure
\ref{fig:disk_comp} shows the ratio between the empirical probability
density obtained from the simulated sample of size $N=10^6$, and the
exact result in Eq. (\ref{eq:PDF_elldelta}); in both cases, we exclude
the trajectories that never touched the boundary, for which
$\ell_\delta = 0$; in other words, we compare only regular parts of
the probability density.
The SIM1 yields a systematic bias in the probability density at large
$\ell$, which is attenuated by decreasing $\ve$ but still present even
for $\ve = 10^{-3}$ (panel (a)).  This bias is also present for the
FLA at $\ve = 10^{-1}$ but it is already removed for $\ve = 10^{-2}$
and $\ve = 10^{-3}$. Expectedly, there is no such a bias for the CLA,
even at $\ve = 10^{-1}$.  A larger scatter of points at large $\ell$
is a consequence of an insufficient statistics for rare trajectories
with large $\ell_t$.  Similar results were obtained for other simple
domains such as a circular layer, a sphere, and a spherical shell, see
\ref{app:valid}.  We only mention that our modification SIM2 yields
more accurate results in three dimensions than the original method
SIM1 (see \ref{app:Skorokhod} for further discussion).

%%% 2D disk
\begin{table}[t!]
  \centering 
\begin{tabular}{ c|c|c|c|r|c|c|c|c|r } 
\hline\hline
Method & $\varepsilon$ & CPU time (s) & $\langle \ell_\delta \rangle_\mathrm{emp}$ & Rel. Error & Method & $\varepsilon$ & CPU time (s) & $\langle \ell_\delta \rangle_\mathrm{emp}$ & Rel. Error  \\
\hline
\multirow{3}{2em}{SIM1} & $10^{-1}$ & $7.88\times10^0$ & 1.7420 & $-7.43 \%$ % sim-120
&     \multirow{3}{2em}{SIM2} & $10^{-1}$ & $7.37\times10^0$ & 1.7305 & $-8.04 \%$  % sim-77
\\
& $10^{-2}$ & $6.97\times10^1$ & 1.8989 & $-0.91 \%$ % sim-121
&     & $10^{-2}$ & $5.62\times10^1$ & 1.7943 & $-4.65 \%$  % sim-78
\\
& $10^{-3}$ & $6.86\times10^2$ & 1.9156 & $1.80 \%$ % sim-122
&     & $10^{-3}$ & $5.39\times10^2$ & 1.8006 & $-4.31 \%$  % sim-79
\\
\hline
\multirow{3}{2em}{FLA} & $10^{-1}$ & $9.09\times10^0$ & 1.7906 & $-4.84 \%$ % sim-71
&     \multirow{3}{2em}{CLA} & $10^{-1}$ & $1.09\times10^1$ & 1.8912 & $0.50 \%$  \\ % sim-74
& $10^{-2}$ & $7.87\times10^1$ & 1.8707 & $-0.59 \%$  % sim-72
&     & $10^{-2}$ & $9.56\times10^1$ & 1.8860 & $0.22 \%$  \\ % sim-75
& $10^{-3}$ & $7.28\times10^2$ & 1.8798 & $-0.10 \%$  % sim-73
&     & $10^{-3}$ & $9.18\times10^2$ & 1.8808 & $-0.05 \%$  \\ % sim-76
\hline
\hline
\end{tabular}
\caption{
Comparison of SIM, FLA and CLA with $N=10^6$ and different $\ve$ for
the unit disk (Fig. \ref{fig:illustration}a) with $p=1$ and $r_0 =
0.5$. The mean and standard deviation of $\ell_\dlt$ are $\langle
\ell_\delta \rangle = 1.8817$, and $\sigma = 2.2113$, so that the
relative statistical error is $\frac{\sigma}{\lla\ell_\dlt\rra
\sqrt{N}} \simeq 0.12 \%$.  } \label{table:2Ddisk}
\end{table}

\subsection{Random pack of disks in 2D}
\label{sec:pack}

We proceed the validation by computing the statistics of the boundary
local time $\ell_\delta$ of reflected Brownian motion outside a random
pack of circular holed (disks) of the same radius $R_0$, confined by a
large circle of radius $R$.  The positions of these holes were chosen
randomly.  In the considered example shown in
Fig. \ref{fig:illustration}c, we set $R=1$, and used 10 small holes of
radius $R_0=0.1$.  In this setting, the probability density of
$\ell_\delta$ is still given by the spectral expansion
(\ref{eq:PDF_general}); however, there is no explicit formula for the
eigenvalues and eigenfunctions of the Steklov problem.  We therefore
compute them by a finite-element method \cite{Chaigneau24}.  The
spectral expansion was truncated to 21 terms, and the maximal mesh
size was 0.005 (244614 triangles).
The accuracy of this computation was checked by changing the
truncation order and the mesh size.  In the following, we consider the
probability density $\Psi(\ell,p|\x_0)$ from
Eq. (\ref{eq:PDF_general}) as a benchmark solution, to which Monte
Carlo simulations of the boundary local time $\ell_\dlt$ spent on the
outer large circle of radius $R$ will be compared with.  In turn, the
encounters with the circular obstacles were ignored.  This choice
allows us to illustrate the flexibility of the proposed method, which
can access the statistics of encounters with either the whole
boundary, or its subsets.
 
\begin{figure}[t!]
\begin{center}
\includegraphics[width=0.45\linewidth]{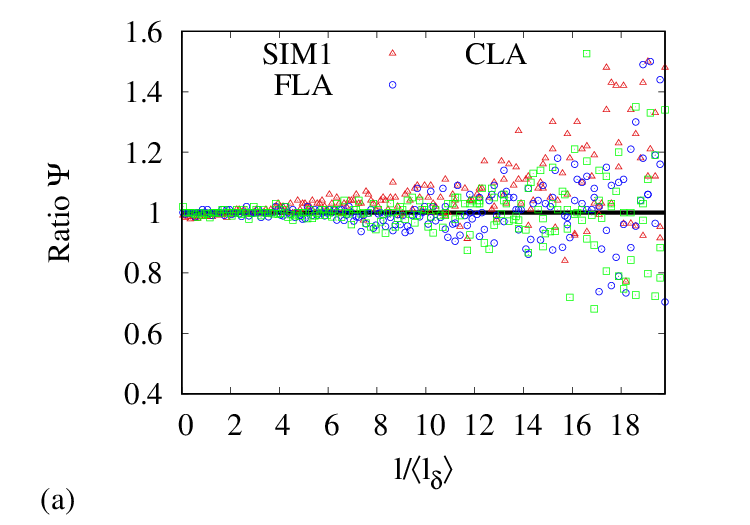}
\includegraphics[width=0.45\linewidth]{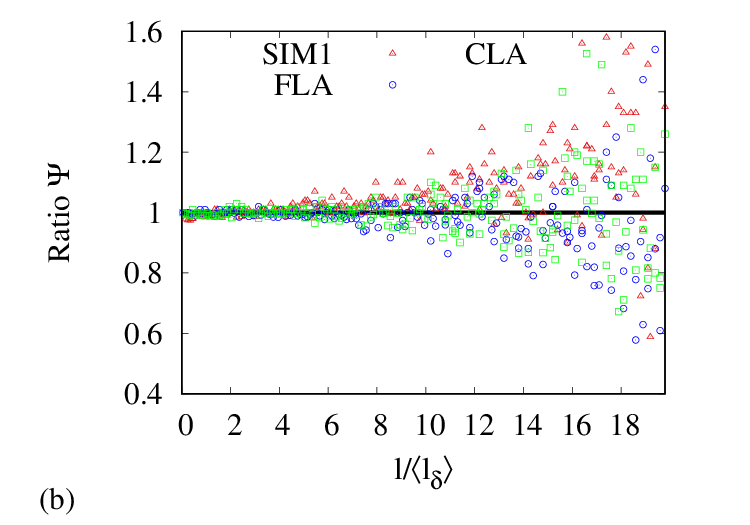}
\end{center}
\caption{
Ratio $\Psi_\mrm{emp}(\ell,p|\x_0)$ over $\Psi(\ell,p|\x_0)$ (obtained
from the finite-element method) as a function of
$\ell/\lla\ell_\dlt\rra$ for the unit disk perforated by 10 circular
holes (Fig. \ref{fig:illustration}c), with $p=1$, $R_0=0.1$, and the
boundary layer width $\ve=10^{-3}$.  The empirical PDF
$\Psi_\mrm{emp}$ was estimated from $N=10^6$ simulated values of
$\ell_\dlt$ for different starting points:
\tbf{(a)} $\x_0 = (1,0)$, 
\tbf{(b)} $\x_0 = (0,0)$, 
with different Monte Carlo methods: SIM1 (triangles), FLA (circles),
and CLA (squares).  The trajectories that never touched the boundary
and thus yielded $\ell_\delta = 0$, were excluded.}
\label{fig:2Drandom_comp}
\end{figure}

%%% 
Table \ref{table:2Dmulti} summarizes the CPU time, the empirical mean
$\lla \ell_\dlt \rra_{\rm emp}$, and its relative error for three
methods.  As the boundary layer width $\ve$ should necessarily be
(much) smaller than the radius $R_0 = 0.1$ of circular holes, we
provide results only for $\ve = 10^{-3}$.  The conclusions on the
comparison between three methods are quite similar to those for the
unit disk: while the CPUs of three methods are comparable, the FLA and
especially CLA provide much more accurate results.  Figure
\ref{fig:2Drandom_comp} shows the comparison of empirical densities
with those derived by the finite-element method, for $\ve = 10^{-3}$.
As previously, the SIM1 yields a bias (which is consistent with the $2
\%$ relative error in the mean), whereas both FLA and CLA are not
biased.

%%% 2D multi
\begin{table}[t!]
  \centering 
\begin{tabular}{ c|c|c|c|r | c|c|c|c|r } 
\hline\hline
$\x_0$ & Method & CPU time (s) & $\lla \ell_\dlt \rra_\mathrm{emp}$ & Rel. Error
& $\x_0$ & Method & CPU time (s) & $\lla \ell_\dlt \rra_\mathrm{emp}$ & Rel. Error \\
\hline
\multirow{4}{3em}{$(+1,0)$} 
& SIM1 & $4.47\times10^3$ & 2.5204 & $ 1.70 \%$   % sim-39
&     \multirow{4}{2em}{$(0,0)$} 
& SIM1 & $4.24\times10^3$ & 1.9771 & $ 1.91 \%$  \\ % sim-40
& SIM2 & $3.74\times10^3$ & 2.3648 & $ -4.58 \%$  % sim-41
&     & SIM2 & $3.61\times10^3$ & 1.8522 & $ -4.53 \%$  \\ % sim-42
& FLA & $4.18\times10^3$ & 2.4702 & $ -0.33 \%$  % sim-21
&     & FLA & $3.98\times10^3$ & 1.9319 & $ -0.42 \%$  \\ % sim-33
& CLA & $4.51\times10^3$ & 2.4746 & $ -0.15 \%$  % sim-20
&     & CLA & $4.33\times10^3$ & 1.9402 & $ 0.01 \%$  \\ % sim-32
\hline
\hline
\end{tabular}
\caption{
Comparison of SIM, FLA, CLA with $N=10^6$ and $\ve = 10^{-3}$ for the
unit disk perforated by 10 circular holes
(Fig. \ref{fig:illustration}c) with $p=1$ and $R_0=0.1$.  The
``reference'' mean value $\lla \ell_\sigma \rra$ was computed from the
spectral expansion (\ref{eq:PDF_general}) whose elements being
obtained by a finite-element method \cite{Chaigneau24}: $\lla
\ell_\delta \rra = 2.4783$ for $\x_0 = (1,0)$; $\lla \ell_\delta \rra
= 1.9400$ for $\x_0 = (0,0)$.  } 
\label{table:2Dmulti}
\end{table}

\subsection{Torus}
\label{sec:tore}

As the last example, we consider a torus in three dimensions, which
can be parameterized as:
\eq{ \llc \begin{aligned}
x(\tta,\vphi) &= (R + R_0\cos\tta) \cos\vphi \,,\\
y(\tta,\vphi) &= (R + R_0\cos\tta) \sin\vphi \,,\\
z(\tta,\vphi) &= R_0 \sin\tta \,,
\end{aligned} \rrd }
using angular coordinates $\tta, \vphi \in \lls 0, 2\pi \rrp$, where
the minor radius $R_0$ is the radius of the tube, and the major radius
$R$ is the distance between the center of the tube and the center of
the torus.  We select the torus with $R = 1.5$ and $R_0 = 0.5$
(Fig. \ref{fig:illustration}f), and set $p=1$.  For this domain, we
compute the eigenvalues and eigenfunctions of the Steklov problem by a
finite-element method (adapted for axisymmetric three-dimensional
problems) \cite{Chaigneau24}.  The spectral expansion
(\ref{eq:PDF_general}) is truncated to 4 terms, and the maximal mesh
size was 0.005 (75314 triangles).  This benchmark solution is compared
to the empirical probability density obtained by Monte Carlo
simulations.

\begin{figure}[t!]
\begin{center}
\includegraphics[width=0.45\linewidth]{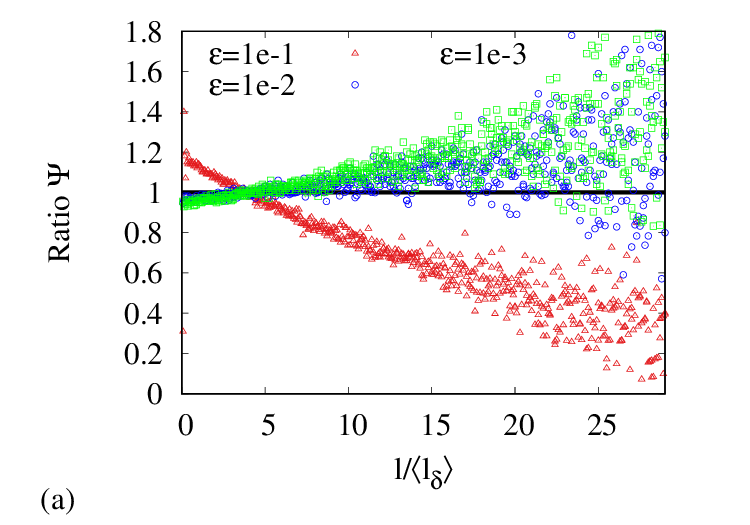}
\includegraphics[width=0.45\linewidth]{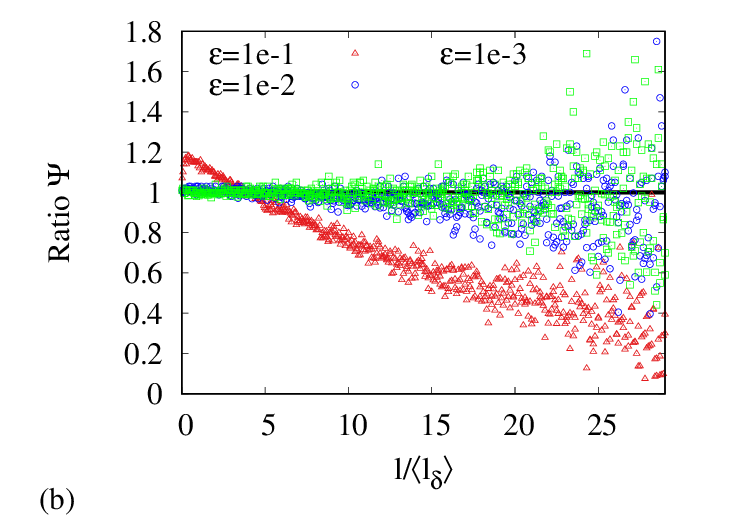}
\end{center}
\caption{
Ratio $\Psi_\mrm{emp}(\ell,p|\x_0)$ over $\Psi(\ell,p|\x_0)$ (obtained
by the finite-element method) as a function of
$\ell/\lla\ell_\dlt\rra$ for the torus (Fig. \ref{fig:illustration}f),
with $p=1$, $R_0=0.5$, $R=1.5$, and the initial point located at
$(1.5,0,0)$.  The empirical PDF $\Psi_\mrm{emp}$ was estimated from
$N=10^6$ simulated values of $\ell_\dlt$ by \tbf{(a)} SIM1, \tbf{(b)}
FLA, with different choices of the boundary layer width: $\ve=10^{-1}$
(triangles), $\ve=10^{-2}$ (circles), and $\ve=10^{-3}$ (squares).
The trajectories that never touched the boundary and thus yielded
$\ell_\delta = 0$, were excluded. }
\label{fig:torus_comp}
\end{figure}

As the boundary points on the torus have two different radii of
curvature, the CLA is not applicable.  We test therefore only the FLA
and the SIM.  Table \ref{table:3Dtorus} compares the empirical means
with the ``reference'' one.  In this three-dimensional setting, our
modification SIM2 yields more accurate results than the original
method SIM1.  Curiously, the FLA at $\ve = 10^{-1}$ yields a bigger
relative error than the SIM2.  However, as $\ve$ decreases, the
relative error of the FLA drops much faster than that of both SIM1 and
SIM2.  Figure \ref{fig:torus_comp} presents the comparison of the PDFs
of the boundary local time $\ell_\dlt$, with similar conclusions.

%%% 3D torus
\begin{table}[t!]
  \centering 
\begin{tabular}{ c|c|c|c|r | c|c|c|c|r } 
\hline\hline
Method & $\varepsilon$ & CPU time (s) & $\lla \ell_\dlt \rra_\mathrm{emp}$ & Rel. Error
& Method & $\varepsilon$ & CPU time (s) & $\lla \ell_\dlt \rra_\mathrm{emp}$ & Rel. Error \\
\hline
\multirow{3}{2em}{SIM1} & $10^{-1}$ & $1.49\times10^1$ & 
3.3024 & $ -14.77 \%$  % sim-18
&     \multirow{3}{2em}{SIM2} & $10^{-1}$ & $1.55\times10^1$ & 3.5039 & $ -9.57 \%$  \\ % sim-11
& $10^{-2}$ & $2.38\times10^2$ & 
4.0124 & $ 3.55 \%$  % sim-19
&     & $10^{-2}$ & $2.08\times10^2$ & 3.7825 & $ -2.38 \%$  \\ % sim-12
& $10^{-3}$ & $2.53\times10^3$ & 
4.1076 & $ 6.01 \%$  % sim-20
&     & $10^{-3}$ & $2.06\times10^3$ & 3.8179 & $ -1.47 \%$  \\ % sim-13
\hline
\multirow{3}{2em}{FLA} & $10^{-1}$ & $2.86\times10^1$ & 
3.2777 & $ -15.41 \%$  \\ % sim-14
& $10^{-2}$ & $3.74\times10^2$ & 
3.8034 & $ -1.84 \%$  \\ % sim-15
& $10^{-3}$ & $3.81\times10^3$ & 
3.8629 & $ -0.31 \%$  \\ % sim-16
\hline
\hline
\end{tabular}
\caption{
Comparison between SIM and FLA with $N=10^6$ and different $\ve$ for
the torus (Fig. \ref{fig:illustration}f) with $p=1$, $R_0=0.5$,
$R=1.5$, and the initial position $(1.5, 0, 0)$.  The ``reference''
mean value $\lla \ell_\sigma \rra = 3.8748$ was computed from the
spectral expansion (\ref{eq:PDF_general}) whose elements being
obtained by a finite-element method \cite{Chaigneau24}.  }
\label{table:3Dtorus}
\end{table}

\section{Discussion and conclusion}
\label{sec:discon}

In this paper, we proposed an efficient numerical tool to access the
statistics of the boundary local time $\ell_t$ of reflected Brownian
motion in Euclidean domains with smooth boundaries.  Using the
approximate solution of the escape problem for a thin boundary layer,
we managed to replace a refined simulation of the random trajectory
near the boundary by a single escape from that layer.  This step
eliminated the most time-consuming part of former Monte Carlo
techniques and allowed us to relax the requirement of a too small
layer width $\ve$ that was needed to control the accuracy of the
estimated boundary local time.  Our implementation of this step
employed several approximations: (i) the boundary layer was locally
approximated by either a flat layer (FLA), or a curved layer with a
known curvature (CLA); (ii) the conditional escape times $\tau_0$ and
$\tau_\ell$ were replaced by their mean values $\langle \tau_0\rangle
$ and $\langle \tau_\ell\rangle$; (iii) the escape position $\X'_\tau$
was replaced by its mean value $\langle
\X'_\tau \rangle$; (iv) the exact stopping condition (e.g., $t_{k+1} =
T$) was replaced by a loose condition (e.g., $t_{k+1} > T$).  Despite
these approximations, both FLA and CLA provided more accurate results
than the state-of-the-art SIM under a comparable computational time.
Moreover, the CLA was remarkably accurate even at relatively large
boundary layer width (e.g., $\ve/R = 0.1$).  Even though the
approximations (ii)-(iv) can be further relaxed (see discussions in
Sec. \ref{sec:stopcond} and \ref{app:condproden}), the accuracy of the
CLA was high enough to ignore these improvements.

At the same time, the approximation (i) was central to our approach.
While the FLA is valid for any Euclidean domain  in $\R^d$ ($d
\geq 2$) with a smooth boundary, the applicability of the CLA is more
restricted.  In fact, it is valid for any planar domain with a smooth
boundary, as well as for three-dimensional domains whose smooth
boundary can be locally approximated by a sphere (i.e., with two equal
radii of curvature).  For instance, the CLA was not applicable for a
torus that we discussed in Sec. \ref{sec:tore}.  Further
investigations of the escape problem from ellipsoidal or even more
sophisticated layers, as well as to higher dimensions, present an
interesting perspective of this work.  Another important research
direction is an extension of our approach to domains with angles (like
polygons or prefractal domains) or more sophisticated singularities
(e.g., cusps).  In this case, one can consider escape events from
triangular or rectangular regions, in analogy to efficient extensions
of the WOS algorithm \cite{Deaconu06,Zein10}.  For this purpose, one
would need to find an approximate solution of the underlying escape
problem \cite{Grebenkov23}.

As explained in Sec. \ref{sec:intro}, the boundary local time plays
the central role in the theory of stochastic processes and has
numerous applications.  First of all, our numerical approach allows
one to simulate reflected Brownian motion $\X_t$ and the associated
boundary local time $\ell_t$ in various confinements.  For instance,
random packs of monodisperse or polydisperse disks or spheres (as an
example considered in Sec. \ref{sec:pack}) are typical models of
porous media in material sciences, chemistry, and biology.  In this
way, one can compute the statistics of encounters between the
diffusing particle and the boundary or a specific region on that
boundary (e.g., we considered separately the boundary local times on
the inner and outer circles of a circular annulus in
\ref{app:annulus}).  Moreover, one can easily treat more general
problems with multiple diffusing particles or/and multiple targets,
and access the joint distribution of various boundary local times and
their mutual correlations \cite{Grebenkov22b,Grebenkov20c}. These
quantities help to characterize the competition between particles for
a target or the competition between targets for a particle.  An
extension of this numerical approach for simulating different
functionals of the boundary local time presents an interesting
perspective \cite{majumdar2007brownian,pacchiarotti1998numerical}.
From a broader perspective, one can study the boundary local
time of a general diffusive process with given drift and covariance
matrix \cite{Ito,Freidlin}.  There are numerous Monte Carlo methods
for simulating such stochastic processes as Markov chains at small
timesteps (see
\cite{Milshtein96,Milstein97,Slominski01,Milstein_book,Bernal16,Bernal19,Leimkuhler23}
and references therein).  While these methods can deal with much more
general diffusive processes, their implementation in multi-scale media
may be computationally prohibited due to small timesteps.  However, if
an applied potential and the induced drift are localized near the
boundary, a combination of the above methods with the WOS algorithm
can be beneficial.  In this way, one could investigate the effect of
the boundary potential (e.g., short-ranged electric attraction or
repulsion) onto the boundary local time and the consequent surface
reactions.

While the statistics of the boundary local time can be accessed by
different numerical methods (e.g., by a finite-element method used in
Sec. \ref{sec:pack}), Monte Carlo simulations have numerous
advantages:

\begin{enumerate}[(i)]
%(i) 
\item 
One can directly generate the boundary local time $\ell_t$ at a fixed
time $t$, whereas the methods relying on the spectral expansion
(\ref{eq:PDF_general}) are less suitable in time domain, as they would
require the inverse Laplace transform.  For this reason, we did not
look at the probability density function of $\ell_t$ that will be
reported elsewhere.

%(ii) 
\item 
Monte Carlo techniques are very flexible and can be easily implemented
in any dimension; in contrast, an implementation of a finite-element
method in dimensions $d \geq 3$ can be rather sophisticated and very
time-consuming.

%(iii) 
\item 
One can easily treat multiple targets (as the example shown in
Fig. \ref{fig:illustration}c) and compute distributions of the
boundary local times $\ell_t^i$ on these targets.  Such joint
distributions are not easily accessible via other methods (see further
discussions on multiple targets in \cite{Grebenkov20c}).
\end{enumerate}

Potential applications of the proposed method go far beyond the
statistics of encounters.  For instance, the position $\X_{\mathcal
T}$ of the process at the stopping time ${\mathcal T} =
\inf\{ t> 0~: \ell_t > \hat{\ell}\}$ when the boundary local time
$\ell_t$ exceeds a random threshold $\hat{\ell}$, determines the
position of the reaction event.  The exponential distribution of the
threshold, $\P\{ \hat{\ell} > \ell\} = e^{-q\ell}$, corresponds to the
most common case of a constant partial reactivity $\kappa = q D$ in
diffusion-controlled reactions \cite{Grebenkov23b}.  
In this case, the distribution of the position $\X_{\mathcal T}$ is
known as the spread harmonic measure
\cite{Grebenkov06c,grebenkov2015analytical}, whereas the stopping time
${\mathcal T}$ is called the first-reaction time or the first-passage
time to the reaction event
\cite{redner2001guide,schuss2015brownian,metzler2014first,masoliver2018random,lindenberg2019chemical,dagdug2024diffusion,Grebenkov_book}. 
Both quantities can be accessed by the proposed method.  However,
other choices for the threshold distribution allow one to represent
more sophisticated surface reactions such as progressive passivation
or activation of the reactive boundary by the diffusing particles
\cite{Grebenkov20}.  Our numerical method allows one to investigate
the distribution of the reaction positions and thus to identify the
most prolific subsets of the boundary with the highest reaction
probability.  This information can be further used to address various
optimization problems in chemical engineering.  From an even more
general perspective, as Monte Carlo techniques are often employed to
solve partial differential equations, the escape-from-a-layer approach
can thus be helpful in treating problems with Neumann and Robin
boundary conditions
\cite{Sabelfeld,Sabelfeld2,Milshtein,Zhou17,maire2013monte,Hsu85,Papanicolaou90,Zhou16,brosamler1976probabilistic,bencherif2009probabilistic,lions1984stochastic,morillon1997numerical}.

\section*{Acknowledgements}
D.S.G. acknowledges a partial financial support from the Alexander von
Humboldt Foundation through a Bessel Research Award.

\appendix
\vskip 10mm

\section{Escape time distribution}
\label{app:exittime}

The WOS algorithm required generation of the random escape time
$\tau_c$ from a disk or a ball.  As this is a fairly standard
procedure (see \cite{grebenkov2014efficient} and references therein),
we only sketch the main steps.  The survival probability of a particle
located in the center of the unit disk, $S_c$, reads
\begin{equation}
S_c(t) = \mathbb{P} \{ \tau_c > t \} = 2 \sum_{k=0}^\infty \frac{\exp(-\alpha_{0k}^2 t)}{\alpha_{0k} J_1(\alpha_{0k})} \,, 
\end{equation}
where $\alpha_{0k}$ refers to the $k$-th zero of the Bessel function
$J_0(x)$ of the first kind, and we set $D=1$.  In turn, for the unit
sphere with a particle located at center, the survival probability is
\begin{equation}
S_c(t) = 2 \sum_{k=1}^\infty (-1)^{k+1} e^{-\pi^2 k^2 t} .
\end{equation}
Explicit formulas are also known in higher dimensions.

In order to generate the random variable $\tau_c$, one needs to invert
the survival probability, i.e., to find the function $T_c(x)$ such
that $S_c[T_c(x)] = x$, for any $x \in (0,1)$.  If $\eta$ is uniformly
distributed on $(0,1)$, then $\tau_c = T_c(\eta)$ is the escape time.
In practice, one can discretize the interval $(0,1)$ into a sequence
of points $x_k = k/K$, such as $k = 1, 2, \cdots, K-1$ with large
enough $K$, solve the equation $S_c(t_k) = x_k$ for each $k$ (e.g. by
the bisection method), and set $T_c(x_k) = t_k$.  Since the survival
probability is a smooth, monotonously decreasing function, this
computation can be done rapidly even for large $K$.  Most importantly,
this computation has to be done only once and then the stored values
of $T_c(x_k)$ can be preloaded before running Monte Carlo simulations.
%%%
Once a uniform random variable $\eta \in (0,1)$ is generated, one
rapidly gets $\tau_c = T_c(\eta)$ by interpolation.

In order to be able to accurately generate large escape times, one can
rely on the long-time behavior:
\begin{equation}
\label{eq:sclong} S_c(t) \simeq \left\{ \begin{aligned}
\frac{2}{\al_{00} J_1(\al_{00})} e^{-\al_{00}^2 t} \quad (d=2) \,, \\
2e^{-\pi^2 t} \quad (d=3) \,.
\end{aligned} \right. 
\end{equation}
If the random variable $\eta$ is below $x_1 = 1/K$ (i.e., $\tau$ is
large), one can use Eq. (\ref{eq:sclong}) to invert $S_c(t)$ at large
$t$, yielding an extrapolation relation
\begin{equation}
\tau_c = \llc \begin{aligned}
- \invs{\al_{00}^2} \ln \lls \frac{\eta \al_{00} J_1(\al_{00})}{2} \rrs \quad (d=2) \,, \\
- \invs{\pi^2} \ln \llp \frac{\eta}{2} \rrp \quad (d=3) \,.
\end{aligned} \rrd 
\end{equation}
Replacing $\tau_c$ by $\frac{\rho^2}{D} \tau_c$ allows one to get the
escape time from a disk or a ball of radius $\rho$ and a given
diffusion coefficient $D$.

To skip the above inversion of the survival probability, it is quite
common to replace the random escape time $\tau_c$ by its mean value,
which is particularly simple for the escape from the center of a ball
of radius $\rho$ in $\mathbb{R}^d$: $\lla \tau_c \rra =
\frac{\rho^2}{2dD}$. However, this simplification can lead to
significant errors when the particle starts far from the boundary so
that the first jump to a large distance $\rho$ gives a large (fixed)
contribution $\frac{\rho^2}{2dD}$ to the time counter (see further
discussion in \ref{sec:justify}).  For this reason, we generated
random escape times $\tau_c$ in all simulations of this paper.

\section{Conditional probability density}
\label{app:condproden}

In this Appendix, we briefly discuss the difficulties in generating
conditional escape times and their possible solutions.  We focus here
on the escape from a flat layer and consider separately two
conditional escape times: without hitting the bottom surface and with
reflections on it.

The inverse Laplace transform of Eq. (\ref{eq:H0_int}) yields
\begin{equation}
H_0(t|y_0) = \frac{2\pi D}{\ve y_0} \sum\limits_{k=1}^\infty (-1)^{k-1} k \sin(\pi k y_0/\ve) e^{-Dt \pi^2 k^2/\ve^2} .
\end{equation}  % [Ht] = A_Yilin_1D_Hve_t_fig();
The survival probability of this random variable, 
\begin{equation}
S_0(t|y_0) = 
\mathbb{P}_{y_0}\{ \tau_0 > t \} = \int\limits_t^\infty \md t^\prime H_0(t^\prime|y_0) = 
\frac{2\ve}{\pi y_0} \sum\limits_{k=1}^\infty \frac{(-1)^{k-1}}{k} \sin(\pi k y_0/\ve) e^{-Dt \pi^2 k^2/\ve^2} ,%.
\end{equation}  
can be numerically inverted to generate the conditional escape time
$\tau_0$ (see \ref{app:exittime} for a similar procedure for a disk).
However, the main practical difficulty is that this survival
probability depends on $y_0/\ve$ so that its inversion may be quite
costly.

The situation is much more difficult for the second conditional escape
time $\tau_\ell$, whose moment-generating function is given by
Eq. (\ref{eq:Hcond2}).  There are two technical difficulties to
perform its inverse Laplace transform: first, there is no simple
explicit representation for this inversion (see, however, some related
representations derived in \cite{Grebenkov20c}); second, there are
three parameters, $t$, $y_0$ and $\ell$, that determine this
probability density.  As a consequence, a direct computation of this
quantity for the whole set of parameters could be problematic.

To partly overcome the second difficulty, one can employ the following
simplification.  The expression in Eq. (\ref{eq:Hcond2}) can be
written as
\begin{equation}  \label{eq:H_int}
\tilde{H}(p|y_0,\ell) = \frac{\ve \, \sinh(\sqrt{p/D}(\ve - y_0))}{(\ve - y_0)\sinh(\sqrt{p/D} \ve)}  \, \tilde{H}(p|0,\ell).
\end{equation}
The first factor is the Laplace transform of the probability density
of the escape time, conditioned to the escape through the endpoint
$0$, which is actually equal to $H_0(t|\ve - y_0)$ due to the
symmetry.  In turn, the second factor is the Laplace transform of the
conditional probability density $H(t|0,\ell)$ when the starting point
is at $0$.  The product of these factors indicates that the escape
time $\tau_\ell$ can be written as the sum of two independent random
times: the time $\tau_1$ to reach the endpoint $0$ and the time
$\tau_2$ to cross the interval $(0,\ve)$ and to escape through the
endpoint $\ve$.  The probability density of the latter time is then
determined by the inverse Laplace transform:
\begin{equation}  \label{eq:H_1D}
H(t|0,\ell) = \L^{-1} \bigl\{ \tilde{H}(p|0,\ell) \bigr\} 
 = \L^{-1} \biggl\{ 
\frac{\ve \sqrt{p/D} }{\sinh(\sqrt{p/D} \ve)}  e^{-\ell  (\sqrt{p/D} \, \ctanh(\sqrt{p/D} \ve) - 1/\ve)} \biggr\}.
\end{equation} % A_Yilin_1D_H0_t_fig();
Even though this inverse Laplace transform needs to be computed
numerically, it does not depend on $y_0$ that simplifies its practical
implementation.  In the same way, one can compute the related survival
probability:
\begin{equation}
S(t|0,\ell) = 1 - \L^{-1} \biggl\{ \frac{\tilde{H}_0(p|0,\ell)}{p} \biggr\}  
= 1 - \L^{-1} \biggl\{ 
\frac{\ve \sqrt{p/D} }{p\sinh(\sqrt{p/D} \ve)}  e^{-\ell  (\sqrt{p/D} \, \ctanh(\sqrt{p/D} \ve) - 1/\ve)} \biggr\}.
\end{equation} % A_Yilin_1D_S0_t_fig();
To avoid technical issues with Laplace transform inversion, we did not
use these exact representations and replaced the random variables
$\tau_0$ and $\tau_\ell$ by their mean values, which were computed
explicitly.

\section{Justification of the approximation by means}
\label{sec:justify}

One practically relevant approximation consists in replacing
random escape times by their mean values.  At a first thought, this
approximation may sound rather crude.  While a rigorous analysis of
the accuracy of this approximation is beyond the scope of the
manuscript, we provide some qualitative arguments and rationales.

To convey the idea, we consider a simpler, slightly different problem
of estimating the first-passage time $T$ to an absorbing boundary and
assume that a random trajectory is simulated by generating a sequence
of spheres of the same radius $\rho$.  In other words, one draws a
sphere of radius $\rho$ centered at the starting position and moves
the particle to a random point on that sphere.  The escape time from
the sphere, $\tau$, is a random variable, with explicitly known
distribution (see \ref{app:exittime}).  The process is repeated again
and again, until the newly drawn $(n+1)$th sphere crosses the
boundary; the number $n$ of generated spheres is thus also a random
variable.  The first-passage time $T$ to the boundary is then
estimated by summing independent random escape times generated until
hitting the boundary: $T = \tau_1 + \tau_2 + \ldots + \tau_n$.  In
turn, the approximation consists in replacing random variables
$\tau_1, \tau_2, \ldots, \tau_n$ by their means, $\langle \tau\rangle
= \rho^2/(6D)$, i.e., $T_{\rm app} = n \langle \tau\rangle = n
\rho^2/(6D)$.  While this approximation yields the correct mean FPT,
$\langle T \rangle = \langle n \rangle \, \langle \tau\rangle =
\langle T_{\rm app}\rangle$, the distributions of $T$ and $T_{\rm
app}$ are different.  How accurately can the distribution of $T_{\rm
app}$ approximate that of $T$?

To estimate the error of this approximation, let us compute the
variance of $T$.  A straightforward computation gives: 
\begin{equation} \label{eq:varT}
\Var\{T\} = \langle n\rangle \Var\{\tau\} + \Var\{n\} \langle\tau\rangle^2 .
\end{equation}
As the approximation replaces random variables $\tau_1, \tau_2,
\ldots, \tau_n$ by their means, the variance of the approximated
first-passage time is $\Var\{T_{\rm app}\} = \Var\{n\}
\langle\tau\rangle^2$, i.e., it ignores the variance of $\tau$ (the
first term in Eq. (\ref{eq:varT})).  We conclude that the proposed
approximation always underestimates the variance of the first-passage
time.  

However, in many practical cases, the second term in
Eq. (\ref{eq:varT}) is dominant so that $\Var\{T\} \approx
\Var\{T_{\rm app}\}$, suggesting that the approximation can be
accurate.  Indeed, the variance of the escape time from a sphere,
$\Var\{\tau\}$, can be found explicitly; but even without an exact
computation, it must be proportional to $(\rho^2/D)^2$ from
dimensional analysis, and thus to $\langle \tau\rangle^2$.  Moreover,
the variance of the FPT is also generally proportional to the square
of its mean, $\Var\{T\} \sim \langle T\rangle^2$, i.e.,
Eq. (\ref{eq:varT}) reads as $\langle n\rangle^2 \sim \langle n\rangle
+ \Var\{n\}$.  If the mean number of generated spheres, $\langle
n\rangle$, is large enough, the variance $\Var\{n\}$ scales as
$\langle n\rangle^2$, i.e., the relative contribution of the first
term to Eq. (\ref{eq:varT}) is of the order of $1/\langle n\rangle$
and thus can be neglected.

We stress that the above arguments are qualitative and may not be
valid in general.  For instance, the relation $\Var\{T\} \sim \langle
T\rangle^2$ may not hold.  Moreover, when the radii of generated
spheres are not fixed (as in the WOS algorithm), the contribution from
few large spheres can be dominant.  For this reason, we do not apply
the approximation to simulate Brownian motion in the bulk, i.e., when
using spheres of variable radii; in this part of the algorithm, we
generate random escape times (see \ref{app:exittime}).  The
approximation is only used to simulate the escape from a layer of a
fixed small width $\ve$.

\section{Skorokhod integral method}
\label{app:Skorokhod}

For completeness, we briefly recall here the Skorokhod integral method
developed in \cite{Schumm23}.  The WOS algorithm is employed until the
particle enters the boundary layer of width $\ve$.  Then, the radius
$\rho$ is fixed to be proportional to $\ve$: $\rho = 2 \veps$ in two
dimensions, and $\rho = 3 \veps$ in three dimensions.  After the jump
to the distance $\rho$, the particle can leave the boundary layer,
stay inside it, or exit the domain.  In the first case, the WOS
algorithm with variable-distance jumps is resumed; in the second case,
one makes another jumps with the radius $\rho = d \ve$. In turn, the
latter option is treated as a reflection event.  The current position
generated outside the domain is reflected back with respect to the
boundary (Fig. \ref{rflcc}a).  The time counter is increased by $\dlt
= \frac{\rho^2}{D} \tau_c$, where $\tau_c$ is the escape time from the
unit disk or the unit ball.  In turn, the Skorokhod integral
representation allows one to increment the boundary local time as
\cite{Zhou17,Schumm23}:
\begin{equation}
\ell_{t_{k+1}} = \ell_{t_k} + \sqrt{\pi/2} \sqrt{D\delta} \,.
\label{eq:ellincrement}
\end{equation}
We refer to this method as SIM1.

\begin{figure}[t!]
\centering
\begin{tikzpicture}[scale = 2]
    \begin{scope}[xshift=0cm]
    \clip  (-2.1, -2.75) rectangle (0.0, -1.6); % clip some region
	    \filldraw[color=gray!120, fill=orange!10, thick] (0, 0) circle (2.5);
	    \filldraw[color=gray!90, fill=red!0, dashed, thick] (0, 0) circle (2.3);
	    \filldraw[color=red, fill=red] (-0.8, -2.3) circle (0.02);
	    \draw[color=red!40, very thick] (-0.8, -2.3) circle (0.4);
	    
	    \begin{scope}[xshift = -0.8cm, yshift = -2.3cm]
	    	\draw[color=red!60] (0,0) -- (190: 0.4);
			\filldraw[color=green, fill=green] (190: 0.4) circle (0.02);
			\draw[color=violet, dashed, thick] (0.8, 2.3) -- (190: 0.4);
	    \end{scope}
	    
	    \draw[color=blue!60] (-1.193923, -2.369459) -- (-1.1249565, -2.2325945);
%		\filldraw[color=blue, fill=blue] (-1.05599, -2.09573) circle (0.02);
		\filldraw[color=blue, fill=blue] (-1.1249565, -2.2325945) circle (0.02);
		\filldraw[color=violet, fill=violet] (0.0, 0.0) circle (0.02);
            %\draw [fill=white] (-2, -2.65) circle (0.0cm) node () {(a)};
		\node[] at (-2, -2.65) () {(a)};
	\end{scope}
	
    \begin{scope}[xshift=4cm]
    \clip  (-2.1, -2.75) rectangle (0.0, -1.6); % clip some region
	    \filldraw[color=gray!120, fill=orange!10, thick] (0, 0) circle (2.5);
	    \filldraw[color=gray!90, fill=red!0, dashed, thick] (0, 0) circle (2.3);
	    \filldraw[color=red, fill=red] (-0.8, -2.3) circle (0.02);
	    \draw[color=red!40, very thick] (-0.8, -2.3) circle (0.4);
	    
	    \begin{scope}[xshift = -0.8cm, yshift = -2.3cm]
	    	\draw[color=red!60] (0,0) -- (190: 0.4);
			\filldraw[color=green, fill=green] (190: 0.4) circle (0.02);
			\draw[color=violet, dashed, thick] (0.8, 2.3) -- (190: 0.4);
	    \end{scope}
	    
	    \draw[color=blue!60] (-1.193923, -2.369459) -- (-1.05599, -2.09573);
		\filldraw[color=blue, fill=blue] (-1.05599, -2.09573) circle (0.02);
		\filldraw[color=violet, fill=violet] (0.0, 0.0) circle (0.02);
            %\draw [fill=white] (-2, -2.65) circle (0.0cm) node () {(b)};
		\node[] at (-2, -2.65) {(b)};
	\end{scope}
\end{tikzpicture}
\caption{
Illustration of the reflection on a circular boundary.  When a
particle (red dot) is inside the boundary layer (light orange) of
width $\veps$, one draws a circle of radius $\rho=2\veps$ and
generates a random position on it.  If this position (green dot) turns
out to be outside the domain, the final position (blue dot) is either
attached on the boundary along the normal direction (violet dashed
line) in SIM1 \tbf{(a)}, or obtained by a mirror reflection along the
normal direction (violet dashed line) in SIM2
\tbf{(b)}. 
}
\label{rflcc}
\end{figure}

Despite a number of tests realized in \cite{Zhou17,Schumm23}, we
provide an independent validation to check that our implementation of
the SIM1 yields the correct results.  In particular, we check and
validate the choice $\rho = d \ve$ for the radius $\rho$ inside the
boundary layer.  For this purpose, we compute numerically the spread
harmonic measure density $\omega_q(\x|\x_0)$ on the circle and on the
sphere, for which exact solutions are available
\cite{grebenkov2015analytical}.  For a fixed starting point $\x_0$,
$\omega_q(\x|\x_0) \md \x$ is the probability that $\X_{\mathcal T}
\in (\x , \x + \md \x)$ where ${\mathcal T} = \mathrm{inf} \{ t>0 :
\ell_t > \hat\ell \}$ is the first crossing time of a random threshold
$\hat\ell$ by the $\ell_t$.  In other words, we employ here the third
stopping condition (see Sec. \ref{sec:stopcond}) with the exponential
threshold $\hat\ell$ such that $\mathbb{P} \{ \hat\ell > \ell \} =
e^{-q \ell}$, for a given $q > 0$.  From the applicative point of
view, $\X_{\mathcal T}$ is the random position of the reaction event
on the boundary with partial reactivity, $\kappa = q D$
\cite{Grebenkov20,Grebenkov20b}.

\subsection{Test for a disk}

The spread harmonic measure density on the circle of radius $R$ reads
\cite{grebenkov2015analytical}
\begin{equation}
\omega_q (\theta|r_0,\theta_0) = \frac{1}{2\pi R} \llp 1 + 2 \sum_{j=1}^\infty \llp \frac{r_0}{R} \rrp^j 
\frac{\cos \lls j(\theta-\theta_0) \rrs}{1 + \frac{j}{qR}} \rrp \,, 
\end{equation}
where $(r_0, \tta_0)$ is the starting point in polar coordinates.  In
order to estimate this density from Monte Carlo simulations, we divide
the circle into $K$ equal arcs such that the $k$-th arc contains $\tta
\in \lls k-1, k \rrp \times \frac{2\pi}{K}$, $k = 1,2,3,...,K$.  The
probability $p_k$ of the reaction event on this arc reads then
\begin{align}
p_k &= R \int\limits_{2\pi(k-1)/K}^{2\pi k/K} \omega_q(\theta|r_0,\theta_0) \md \theta 
= \invs{K} + \sum_{j=1}^\infty  \frac{2 q R}{\pi j (j + q R)} \left(\frac{r_0}{R}\right)^j \sin 
\left(\frac{\pi  j}{K}\right)  \cos \left( j \lls \frac{2\pi}{K} \llp k-\demi \rrp - \tta_0 \rrs \right) \,. 
 \label{eq:diskSHM} 
\end{align}
In turn, the probabilities $p_k$ can be estimated by running Monte
Carlo simulations with the third stopping condition and counting the
number $N_k$ of trajectories that stopped on the $k$-th arc, so that
$N_k/N$ approximates $p_k$.

\begin{figure}[t!]
\centering
\includegraphics[width=0.32\linewidth]{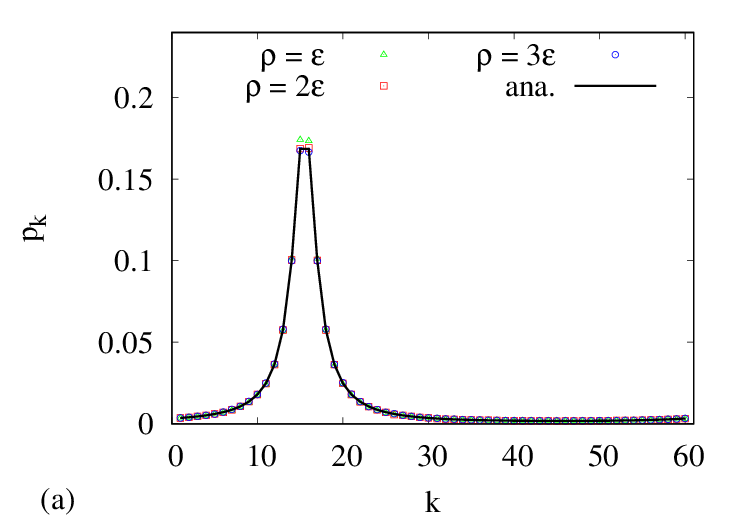}
\includegraphics[width=0.32\linewidth]{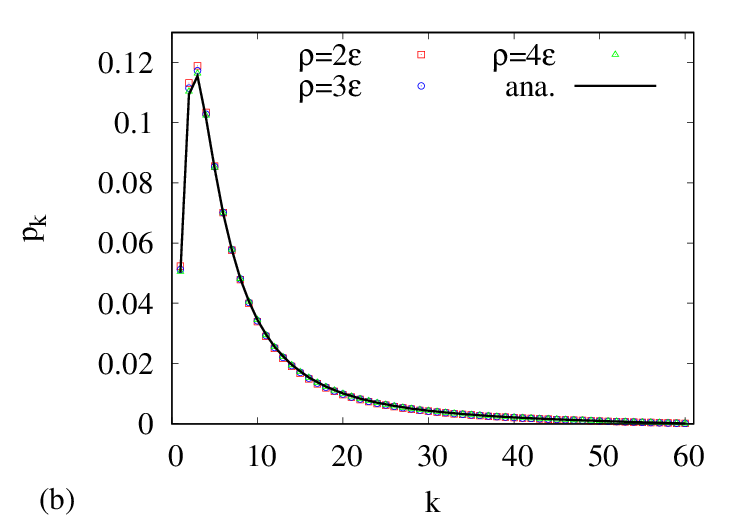}
\includegraphics[width=0.32\linewidth]{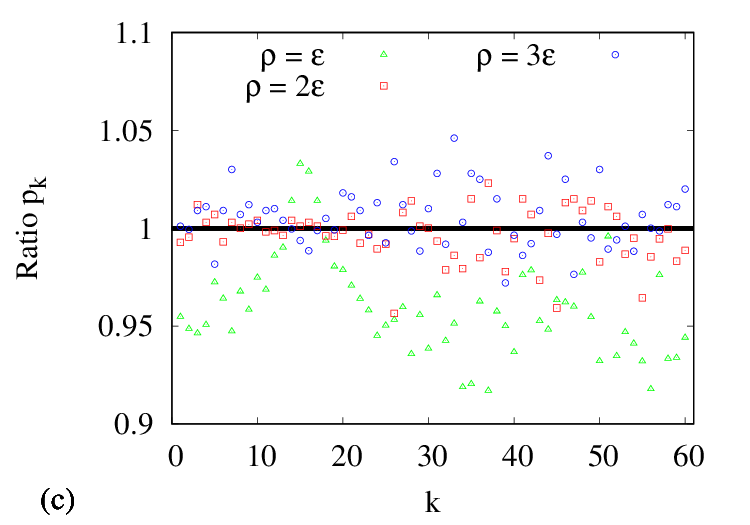}
\includegraphics[width=0.32\linewidth]{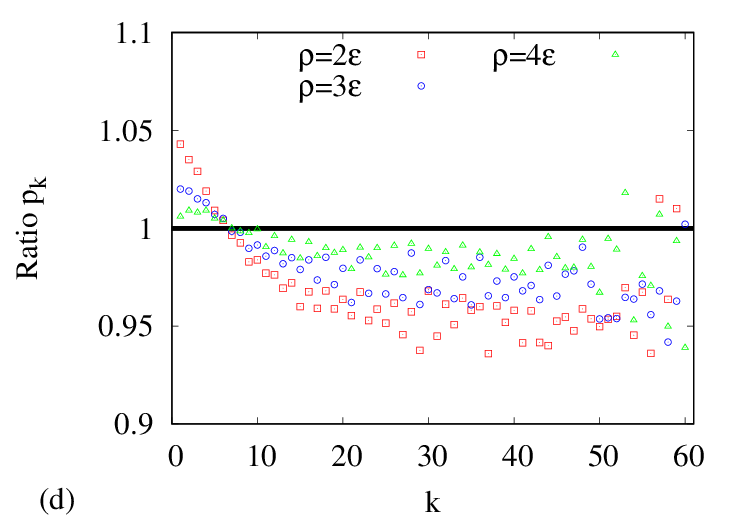}
\includegraphics[width=0.32\linewidth]{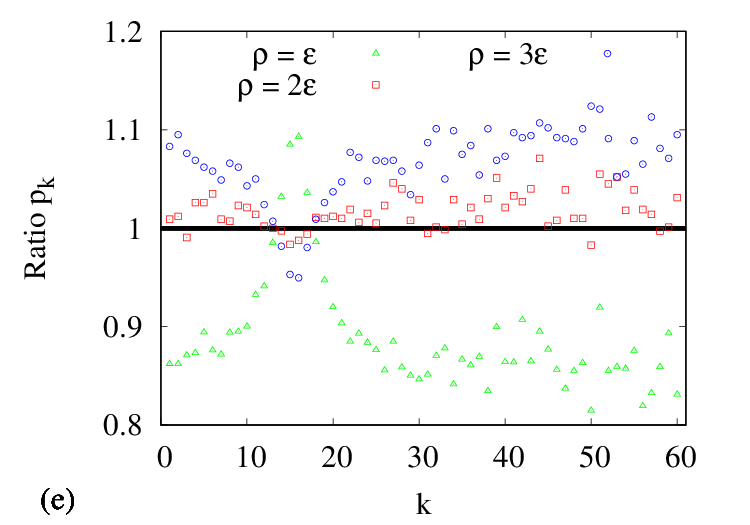}
\includegraphics[width=0.32\linewidth]{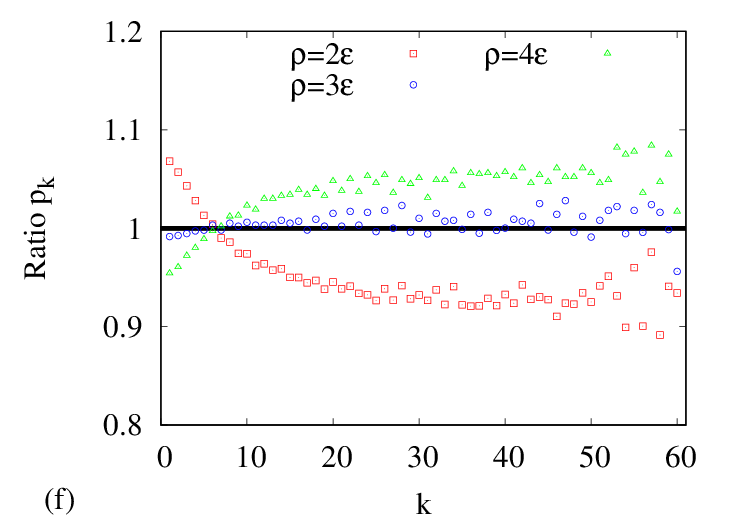}
\caption{
\tbf{(a)} 
Spread harmonic measure density on the unit circle ($R=1$): comparison
between the analytical expression (\ref{eq:diskSHM}) with $K=60$,
truncated at $j_{\max}=50$, shown by the black line, and numerical
estimates by SIM1.  Parameters are: the boundary layer width $\ve =
10^{-3}$, reactivity $q=10$, particle number $N = 10^6$, and the
initial point is located at $(0, 0.9)$, with different choices of the
ratio $\rho/\ve$: $\rho = \ve$ (triangles), $\rho = 2\ve$ (squares),
and $\rho = 3\ve$ (circles).
\tbf{(b)} 
Spread harmonic measure density on the unit sphere ($R = 1$):
comparison between the analytical expression (\ref{eq:sphereSHM}) with
$K=60$, truncated at $n_{\max}=60$, shown by black line, and numerical
ones by SIM1.  Parameters are: the boundary layer width $\ve =
10^{-3}$, reactivity $q=10$, particle number $N = 10^7$, and the
initial point is located at $(0,0,0.9)$, with different choices of the
ratio $\rho/\ve$: $\rho = 2\ve$ (squares), $\rho = 3\ve$ (circles),
and $\rho = 4\ve$ (triangles).
\tbf{(c)} 
Ratio of numerical estimates $p_k$ shown in panel (a) to the
analytical results (\ref{eq:diskSHM}), with different choices of the
ratio $\rho/\ve$: $\rho = \ve$ (triangles), $\rho = 2\ve$ (squares),
and $\rho = 3\ve$ (circles).
\tbf{(d)} 
Ratio of numerical estimates $p_k$ shown in panel (b) to the
analytical results (\ref{eq:sphereSHM}), with different choices of the
ratio $\rho/\ve$: $\rho = 2\ve$ (squares), $\rho = 3\ve$ (circles),
and $\rho = 4\ve$ (triangles).
\tbf{(e)} 
Ratio of numerical estimates $p_k$ obtained by SIM2 (not shown) to the
analytical results (\ref{eq:diskSHM}), with different choices of the
ratio $\rho/\ve$: $\rho = \ve$ (triangles), $\rho = 2\ve$ (squares),
and $\rho = 3\ve$ (circles).
\tbf{(f)} 
Ratio of numerical estimates $p_k$ obtained by SIM2 (not shown) to the
analytical results (\ref{eq:sphereSHM}), with different choices of the
ratio $\rho/\ve$: $\rho = 2\ve$ (squares), $\rho = 3\ve$ (circles),
and $\rho = 4\ve$ (triangles).  }
\label{pbb_rho}
\end{figure}

Figure \ref{pbb_rho}a compares the exact probabilities $p_k$ from
Eq. (\ref{eq:diskSHM}) to the empirical ones obtained by Monte Carlo
simulations with SIM1. We compare three choices for the radius $\rho$
inside the boundary layer and illustrate that $\rho = 2\ve$ yields the
most accurate results.  Other starting points and values of $q$ were
tested and gave similar results (not shown).

\subsection{Test for a sphere}
\label{app:testsphere}

For a sphere of radius $R$, the spread harmonic measure density reads
\cite{grebenkov2019semi}:
\begin{equation}
\omega_q(\x|\x_0) = \invs{4\pi R^2} \sum_{n=0}^\infty P_n \llp \cos\gamma \rrp \llp \frac{|\x_0|}{R} \rrp^n  \frac{2n+1}{1 + n/(qR)} \,,
\end{equation}
where $\x$ is the position on the sphere, $\cos\gamma = \frac{\llp \x
\cdot \x_0 \rrp }{|\x| |\x_0|}$, $P_n(x)$ is the Legendre
polynomial. Using the spherical coordinates, $\x = (R,\tta, \phi)$,
$\x_0 = (|\x_0|,\tta_0, \phi_0)$, it is convenient to integrate over
the angle $\phi$ to get
\begin{equation}
\bar{\omega}(\theta | \theta_0) = \frac{1}{2} \sum\limits_{n=0}^\infty P_n(\cos\theta) P_n(\cos\theta_0) \llp \frac{r_0}{R} \rrp^n  \frac{2n+1}{1 + n/(qR)} \,.
\end{equation} 
As previously, we divide the sphere into $K$ regions such that the
$k$-th layer contains $\tta \in \lls k-1, k \rrp \times
\frac{\pi}{K}$, with $k = 1, 2, \cdots, K$. The probability $p_k$ of
the reaction event on $k$-th region is thus expressed by
\begin{align}
p_k &= \int\limits_{(k-1)\pi/K}^{k\pi/K} \bar\omega(\tta|\tta_0) \sin\tta \md \tta  
= \frac{\xi_{k-1} - \xi_k}{2} - \demi \sum_{n=1}^\infty \frac{(r_0/R)^n}{1 + n/(qR)} 
\Big\{ \lls P_{n+1}(\xi_k) - P_{n-1} (\xi_k) \rrs - \lls P_{n+1}(\xi_{k-1}) - P_{n-1} (\xi_{k-1}) \rrs \Big\} \,, \label{eq:sphereSHM} 
\end{align}
with $\xi_k = \cos(k\pi/K)$.

We test different choices for the ratio $\rho/\ve$ by computing the
spread harmonic measure on a unit sphere with $R=1, \veps=10^{-3},
q=10$, and the initial point is located at $(0, 0, 0.9)$.  Figure
\ref{pbb_rho}b compares the exact probabilities $p_k$ from
Eq. (\ref{eq:sphereSHM}) to the empirical ones obtained by Monte Carlo
simulations with SIM1.
%%%
In particular, panel (d) indicates that the choice $\rho = 3\ve$
suggested in \cite{berezhkovskii2013trapping}, does not yield the most
accurate results. Moreover, the results remain biased even for a
larger value $\rho = 4\ve$.  This bias can potentially explain higher
relative errors of the SIM1 in three dimensions.

\begin{comment}
\eq{
\lla T \rra = \frac{R^2 - r_0^2}{2dD} + \frac{R}{dDq}
\label{eq:meanT}
}
\end{comment}

\subsection{A modification of SIM}

While Schumm and Bressloff designed and validated their SIM for planar
domains \cite{Schumm23}, its extension to three-dimensional domains
was straightforward. However, no validation of this extension was
performed.  Our test for a sphere in \ref{app:testsphere} revealed the
accuracy issues of this extension.  Moreover, the relative error of the
mean $\lla \ell_\dlt \rra$ was quite big for all the considered
three-dimensional domains, even at $\ve = 10^{-3}$ (see Tables
\ref{table:3Dtorus}, \ref{table:3Dsphere}, \ref{table:3Dshell}).  In an
attempt to understand this discrepancy, we tried different ways of
modeling the reflection event.  In fact, the original method suggests
to relocate the particle, which left the domain, to the boundary point
along the normal direction (Fig. \ref{rflcc}a).  To some extent, this
is an arbitrary choice. For instance, most algorithms simulating
reflected diffusion employ a mirror reflection across the boundary
(see Fig. \ref{rflcc}b), but other choices are as well possible.  It
is generally believed that the choice of reflection, occuring at a
small scale $\ve$, provides only a minor effect onto simulation
results.  However, as the boundary local time is incremented only at
such reflections, the minor errors potentially induced by the
``wrong'' choice, can accumulate and result in considerable bias.

In order to check this point, we propose a minor modification of the
SIM, which consists in two steps: (i) the relocation to the boundary
is replaced by a mirror reflection (compare Fig. \ref{rflcc}a and
\ref{rflcc}b); (ii) after the reflection, we accumulated the durations
of the consecutive jumps inside the boundary layer, until the particle
leaves it. The accumulated duration $\dlt$ is then used to increment
the boundary local time via Eq. (\ref{eq:ellincrement}).  We refer to
this method as SIM2.  Figure \ref{pbb_rho}f shows that the SIM2 with
$\rho = d \ve$ yields unbiased results for the spread harmonic measure
in both two- and three-dimensional cases.  We used this choice
throughout all the manuscript.  Tables \ref{table:3Dtorus},
\ref{table:3Dsphere}, \ref{table:3Dshell} indicate that SIM2 yields
more accurate results than SIM1 in three-dimensional examples, but
less accurate results in two-dimensional examples.  A more systematic
analysis of this observation presents an interesting problem, which
is, however, beyond the scope of this paper.

\section{Validation for other domains}
\label{app:valid}

\subsection{Annulus in 2D}
\label{app:annulus}

We validate Monte Carlo simulations for a circular annulus of
radii $R$ and $L$ (Fig. \ref{fig:illustration}b).  In this case, we
evaluate separately the statistics of the boundary local times
$\ell_\delta^{\rm i}$ and $\ell_\delta^{\rm o}$ on the inner and outer
circles.  The rotational symmetry of this domain implies again the
simple form (\ref{eq:PDF_elldelta}), with the following parameters
\cite{Grebenkov20b}:

(i) for the boundary local time $\ell_\dlt^{\rm i}$ on the inner
circle:
\begin{subequations}
\begin{align}
\pi_0 & = 1 - \frac{I_1(\alpha L) K_0(\alpha r_0) + K_1(\alpha L) I_0(\alpha r_0)}
{I_1(\alpha L) K_0(\alpha R) + K_1(\alpha L) I_0(\alpha R)} \,, \\
\mu_0^{(p)} & = \alpha\, \frac{I_1(\alpha L) K_1(\alpha R) - K_1(\alpha L) I_1(\alpha R)}
{I_1(\alpha L) K_0(\alpha R) + K_1(\alpha L) I_0(\alpha R)} \,.
\end{align}
\end{subequations}

(ii) for the boundary local time $\ell_\dlt^{\rm o}$ on the outer
circle:
\begin{subequations}
\begin{align}
\pi_0 & = 1 - \frac{K_1(\alpha R) I_0(\alpha r_0) + I_1(\alpha R) K_0(\alpha r_0)}
{K_1(\alpha R) I_0(\alpha L) + I_1(\alpha R) K_0(\alpha L)} \,, \\
\mu_0^{(p)} & = \alpha\, \frac{K_1(\alpha R) I_1(\alpha L) - I_1(\alpha R) K_1(\alpha L)}
{K_1(\alpha R) I_0(\alpha L) + I_1(\alpha R) K_0(\alpha L)} \,.
\end{align}
\end{subequations}

As previously for the case of a disk, Table \ref{table:2Dannulus}
compares the empirical mean values of $\ell_\delta^{\rm i}$ and
$\ell_\delta^{\rm o}$ found by different methods and different widths
$\ve$ with their analytical expectations, given by
Eq. (\ref{eq:ellmean}).  As in Table \ref{table:2Ddisk}, the most
accurate results by the SIM1 are achieved at an intermediate width
$\ve = 10^{-2}$, whereas its accuracy does not improve at smaller
$\ve$.  For other methods, smaller $\ve$ results in more accurate
values. The CLA outperforms two other methods for both
$\ell_\delta^{\rm i}$ and $\ell_\delta^{\rm o}$.  Accidentally, the
relative error of $\lla \ell_\delta^{\rm i} \rra_\mrm{emp}$ for the
SIM1 at $\ve = 10^{-2}$ is comparable to that of the CLA. However,
this is not true for $\lla \ell_\delta^{\rm o} \rra_\mrm{emp}$.

\begin{figure}[t!]
\begin{center}
\includegraphics[width=0.32\linewidth]{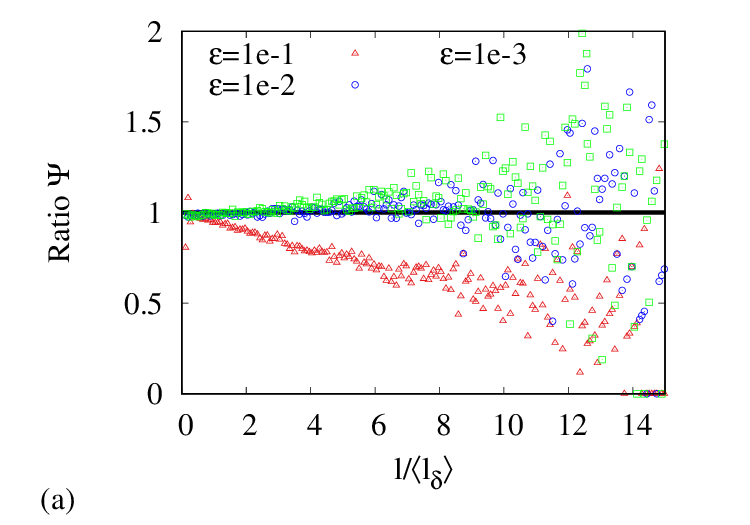}
\includegraphics[width=0.32\linewidth]{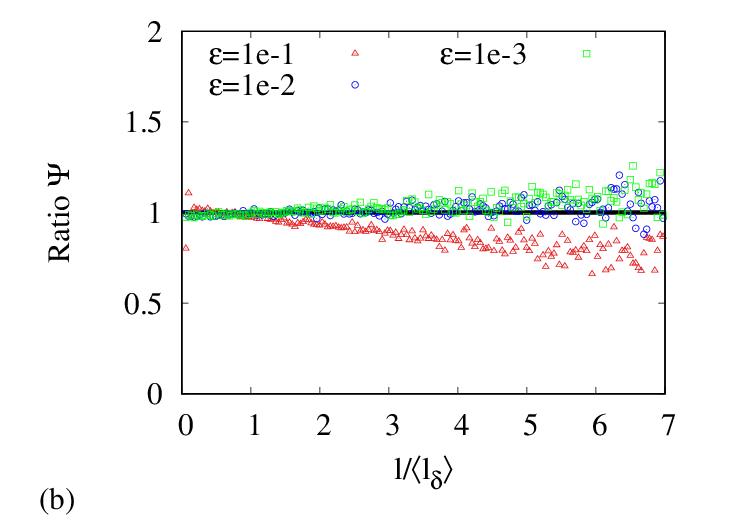}
\includegraphics[width=0.32\linewidth]{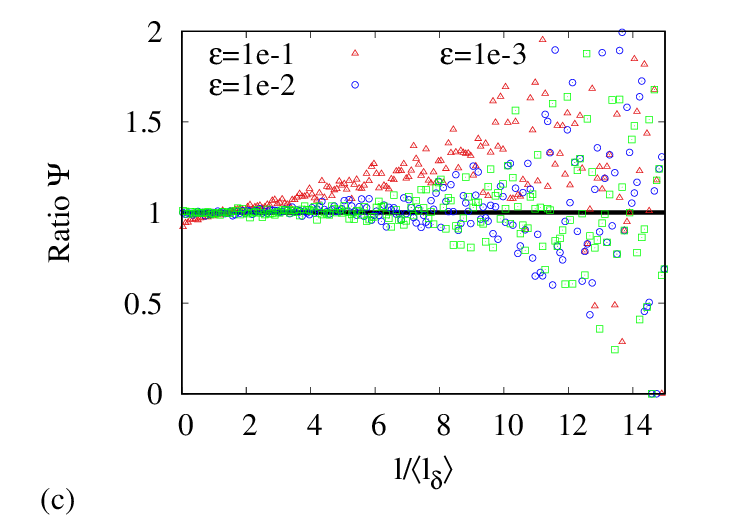}
\includegraphics[width=0.32\linewidth]{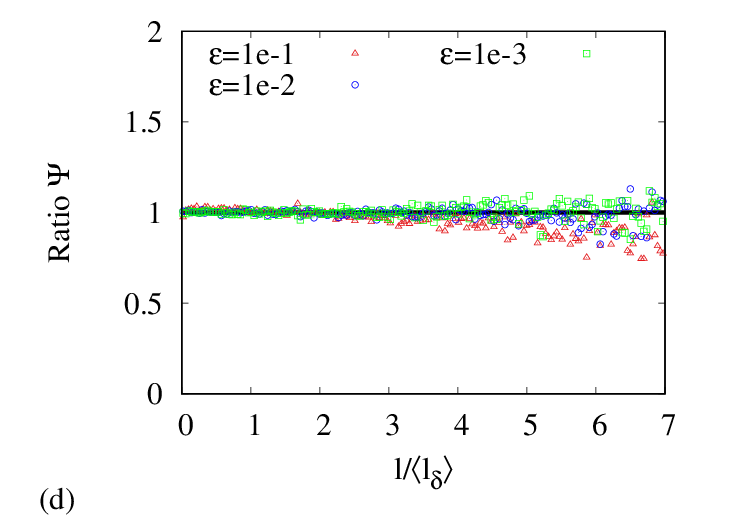}
\includegraphics[width=0.32\linewidth]{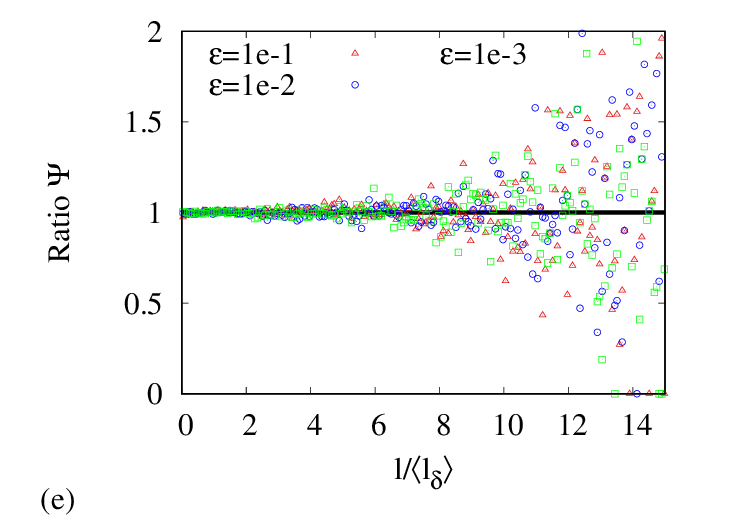}
\includegraphics[width=0.32\linewidth]{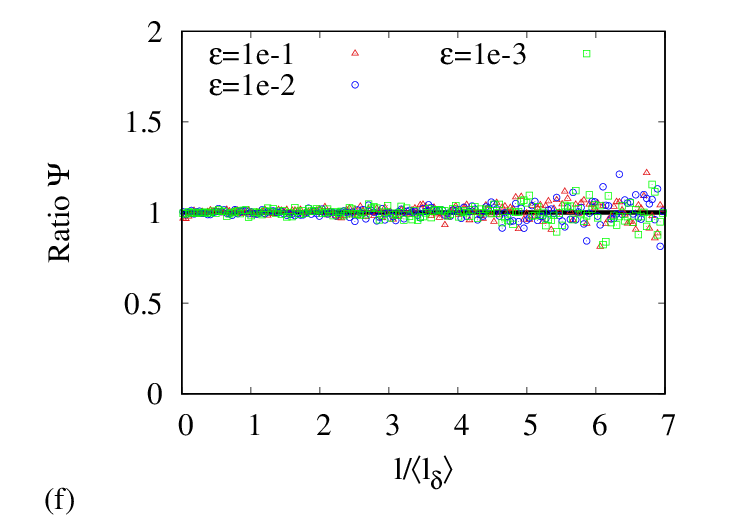}
\end{center}
\caption{
Ratio $\Psi_\mrm{emp}(\ell,p|\x_0)$ over $\Psi(\ell,p|\x_0)$ as a
function of $\ell/\lla\ell_\dlt\rra$ for the circular annulus
(Fig. \ref{fig:illustration}b), with $p=1$, $R=1$, $L=2$, and $r_0 =
1.5$.  The empirical PDF $\Psi_\mrm{emp}$ was estimated from $N=10^6$
simulated values of $\ell_\dlt$ by \tbf{(a,b)} SIM1, \tbf{(c,d)} FLA,
\tbf{(e,f)} CLA, with different choices of the boundary layer width:
$\ve=10^{-1}$ (triangles), $\ve=10^{-2}$ (circles), and $\ve=10^{-3}$
(squares).
\tbf{(a,c,e)} 
refer to $\ell^\mrm{i}_\dlt$ on the inner circle and \tbf{(b,d,f)}
refer to $\ell^\mrm{o}_\dlt$ on the outer circle.  The trajectories
that never touched the boundary and thus yielded $\ell_\delta = 0$,
were excluded.}
\label{fig:annulus_comp}
\end{figure}

Figure \ref{fig:annulus_comp} shows the comparison of the statistics
of two boundary local times, $\ell_\delta^{\rm i}$ and
$\ell_\delta^{\rm o}$, obtained by different Monte Carlo methods, with
the exact solution (\ref{eq:PDF_elldelta}), for $p = 1$ and several
values of $\ve = 10^{-1}, 10^{-2}, 10^{-3}$.  Curiously, the FLA
presents opposite trends for $\ell_\delta^{\rm i}$ and
$\ell_\delta^{\rm o}$ as $\ell$ increases.  The CLA yields accurate
results even for $\ve=10^{-1}$.

%%% 2D annulus
\begin{table}[t!]
  \centering 
\begin{tabular}{ c|c|c|c|r|c|r } 
\hline\hline
Method & $\varepsilon$ & CPU time (s) & 
$\langle \ell^\mathrm{i}_\delta \rangle_\mathrm{emp}$ & Rel. Error &  
$\langle \ell^\mathrm{o}_\delta \rangle_\mathrm{emp}$ & Rel. Error  \\
\hline
\multirow{3}{2em}{SIM1} & $10^{-1}$ & $9.57\times10^0$ & 
0.5516 & $-15.07 \%$ & 
1.1524 & $-8.98 \%$  \\ % sim-95
& $10^{-2}$ & $7.63\times10^1$ & 
0.6504 & $0.15 \%$ & 
1.2769 & $0.86 \%$  \\  % sim-96
& $10^{-3}$ & $7.11\times10^2$ & 
0.6621 & $1.95 \%$ & 
1.2914 & $2.02 \%$  \\  % sim-97
\hline
\multirow{3}{2em}{SIM2} & $10^{-1}$ & $8.87\times10^0$ & 
0.6386 & $-1.66 \%$ & 
1.1820 & $-6.63 \%$  \\ % sim-76
& $10^{-2}$ & $6.32\times10^1$ & 
0.6217 & $-4.27 \%$ &
1.2077 & $-4.61 \%$  \\ % sim-77
& $10^{-3}$ & $5.73\times10^2$ & 
0.6185 & $-4.75 \%$ &
1.2091 & $-4.49 \%$  \\ % sim-78
\hline
\multirow{3}{2em}{FLA} & $10^{-1}$ & $1.31\times10^1$ & 
0.6883 & $6.00 \%$ & 
1.2344 & $-2.50 \%$  \\ % sim-91
& $10^{-2}$ & $9.21\times10^1$ & 
0.6532 & $0.58 \%$ & 
1.2619 & $-0.33 \%$  \\  % sim-92
& $10^{-3}$ & $8.02\times10^2$ & 
0.6508 & $0.22 \%$ & 
1.2665 & $-0.04 \%$  \\  % sim-93
\hline
\multirow{3}{2em}{CLA} & $10^{-1}$ & $1.31\times10^1$ & 
0.6522 & $0.43 \%$ &
1.2694 & $0.27 \%$ \\ % sim-88
& $10^{-2}$ & $1.04\times10^2$ & 
0.6483 & $-0.17 \%$ & 
1.2645 & $-0.12 \%$  \\ % sim-89
& $10^{-3}$ & $9.74\times10^2$ & 
0.6483 & $-0.02 \%$ & 
1.2668 & $-0.06 \%$ \\ % sim-90
\hline
\hline
\end{tabular}
\caption{
Comparison of the SIM, FLA, CLA with $N=10^6$ and different $\ve$ for
the annulus (Fig. \ref{fig:illustration}b) with $p=1$, $R=1$, $L=2$,
and $r_0 = 1.5$. The mean and standard deviation are: $\langle
\ell^\mrm{i}_\delta \rangle = 0.6494$, $\sigma^{\rm i} = 0.9112$, and
$\langle \ell_\delta^{\rm o} \rangle = 1.2660$, $\sigma^{\rm o} =
1.5931$, so that the relative statistical errors are
$\frac{\sigma^{\rm i}}{\langle \ell^\mrm{i}_\delta \rangle \sqrt{N}}
\simeq 0.14 \%$, and $\frac{\sigma^{\rm o}}{\langle
\ell^\mrm{o}_\delta \rangle \sqrt{N}} \simeq 0.13 \%$.  }
\label{table:2Dannulus}
\end{table}

\subsection{Sphere and spherical shell in 3D}
\label{app:3dvalid}

The expression (\ref{eq:PDF_elldelta}) is also valid for the sphere of
radius $R$, with
\begin{equation}
\mu_0^{(p)} = \alpha \, \ctanh(\alpha R) - \frac{1}{R} \,, \qquad
\pi_0 = 1 - \frac{R \sinh(\alpha r_0)}{r_0 \sinh(\alpha R)} \,,
\end{equation}
where $r_0$ refers to the distance between the initial point and the
sphere center, and $\al=\sqrt{p/D}$.

\begin{figure}[t!]
\begin{center}
\includegraphics[width=0.32\linewidth]{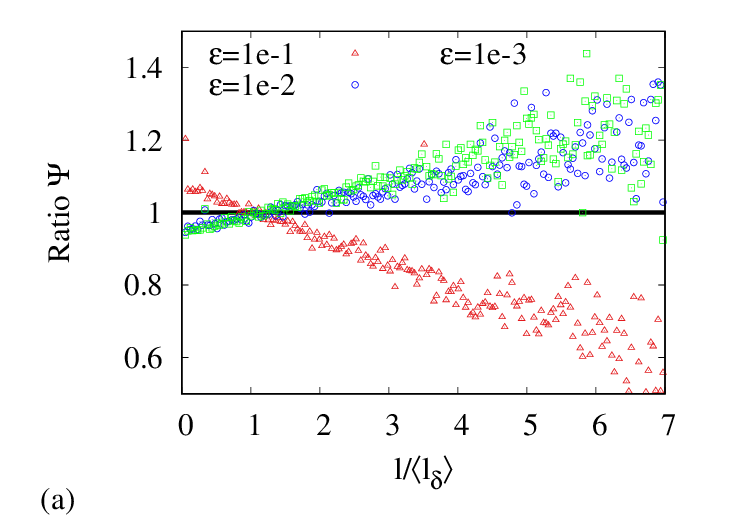}
\includegraphics[width=0.32\linewidth]{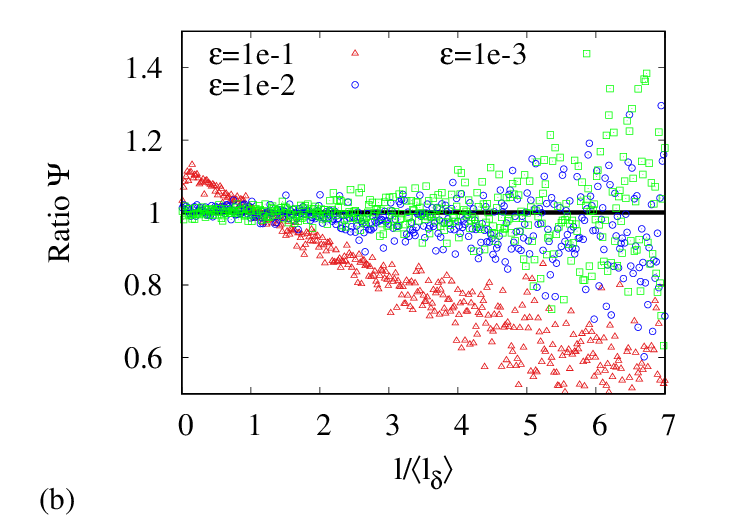}
\includegraphics[width=0.32\linewidth]{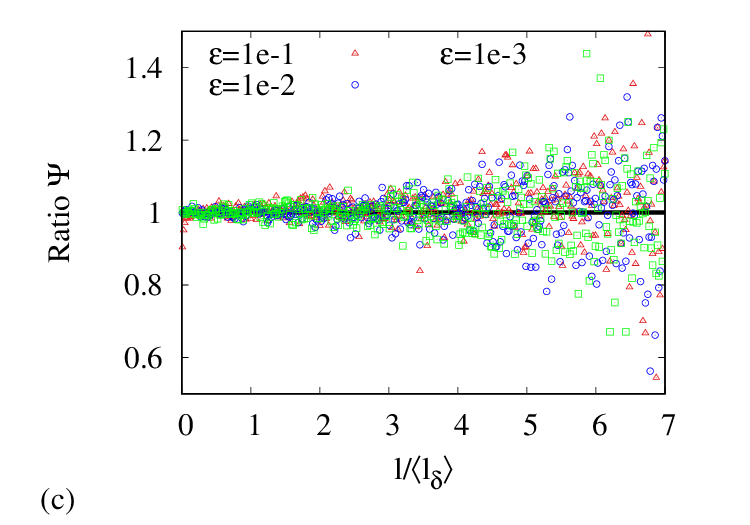}
\end{center}
\caption{
Ratio $\Psi_\mrm{emp}(\ell,p|\x_0)$ over $\Psi(\ell,p|\x_0)$ as a
function of $\ell/\lla\ell_\dlt\rra$ for the unit sphere
(Fig. \ref{fig:illustration}d), with $p=1$ and $r_0 = 0.5$.  The
empirical PDF $\Psi_\mrm{emp}$ was estimated from $N=10^6$ simulated
values of $\ell_\dlt$ by \tbf{(a)} SIM1, \tbf{(b)} FLA, \tbf{(c)} CLA,
with different choices of the boundary layer width: $\ve=10^{-1}$
(triangles), $\ve=10^{-2}$ (circles), and $\ve=10^{-3}$ (squares).
The trajectories that never touched the boundary and thus yielded
$\ell_\delta = 0$, were excluded.}
\label{fig:sphere_comp}
\end{figure}

Figure \ref{fig:sphere_comp} presents the comparison of the exact PDF
in Eq. (\ref{eq:PDF_elldelta}) with the empirical ones obtained by
different Monte Carlo methods.  For $\ve = 10^{-1}$, both SIM and FLA
present a bias, which is removed at smaller $\ve$.  In turn, the CLA
shows again excellent agreement for all cases.

Table \ref{table:3Dsphere} compares three methods for evaluating the
mean value $\lla \ell_\dlt \rra$. As for the disk, the SIM is
outperformed by the FLA (especially for small $\ve$), whereas the CLA
yields the most accurate results. Curiously, the accuracies are higher
in three dimensions than in two dimensions.  Note that the SIM2 yields
more accurate results than SIM1.

%%% 3D sphere
\begin{table}[t!]
  \centering 
\begin{tabular}{ c|c|c|c|r | c|c|c|c|r } 
\hline\hline
Method & $\varepsilon$ & CPU time (s) & $\lla \ell_\dlt \rra_\mathrm{emp}$ & Rel. Error
& Method & $\varepsilon$ & CPU time (s) & $\lla \ell_\dlt \rra_\mathrm{emp}$ & Rel. Error \\
\hline
\multirow{3}{2em}{SIM1} & $10^{-1}$ & $8.23\times10^0$ & 
2.5995 & $ -8.24 \%$ % sim-64
&     \multirow{3}{2em}{SIM2} & $10^{-1}$ & $8.24\times10^0$ & 2.5394 & $ - 10.36 \%$  \\ % sim-18
& $10^{-2}$ & $1.22\times10^2$ & 
2.9557 & $ 4.33 \%$ % sim-65
&     & $10^{-2}$ & $1.04\times10^2$ & 2.7688 & $ -2.27 \%$  \\ % sim-19
& $10^{-3}$ & $1.33\times10^3$ & 
3.0006 & $ 5.91 \%$  % sim-66
&     & $10^{-3}$ & $1.08\times10^3$ & 2.7893 & $ -1.54 \%$  \\ % sim-20
\hline
\multirow{3}{2em}{FLA} & $10^{-1}$ & $1.54\times10^1$ & 2.5353 & $ - 10.51 \%$ % sim-24
%&     \multirow{3}{2em}{CLA} & $10^{-1}$ & $1.49\times10^1$ & 2.8568 & $ 0.84 \%$  \\ % sim-21
&     \multirow{3}{2em}{CLA} & $10^{-1}$ & $1.53\times10^1$ & 2.8476 & $ 0.52 \%$  \\ % sim-67
& $10^{-2}$ & $1.82\times10^2$ & 2.7988 & $ -1.21 \%$ % sim-25
%&     & $10^{-2}$ & $1.78\times10^2$ & 2.8358 & $ 0.10 \%$  \\ % sim-22
&     & $10^{-2}$ & $1.90\times10^2$ & 2.8353 & $ 0.08 \%$  \\ % sim-70
& $10^{-3}$ & $1.91\times10^3$ & 2.8351 & $ 0.07 \%$  % sim-26
%&     & $10^{-3}$ & $1.92\times10^3$ & 2.8330 & $ -0.0003 \%  \\ % sim-23
&     & $10^{-3}$ & $2.01\times10^3$ & 2.8362 & $ 0.11 \%$  \\ % sim-23
\hline \hline
\end{tabular}
\caption{
Comparison of the SIM, FLA, CLA with $N=10^6$ and different $\ve$ for
the unit sphere (Fig. \ref{fig:illustration}d) with $p=1$ and $r_0 =
0.5$. The mean and standard deviation of $\ell_\dlt$ are $\langle
\ell_\delta \rangle = 2.8330$ and $\sigma = 3.1740$, so that the
relative statistical error is $\frac{\sigma}{\lla\ell_\dlt\rra
\sqrt{N}} \simeq 0.11 \%$.  } 
\label{table:3Dsphere}
\end{table}

Furthermore, we also validate Monte Carlo simulations for a spherical
shell between concentric spheres of radii $R$ and $L$ by evaluating
$\ell_\delta^{\rm i}$ and $\ell_\delta^{\rm o}$.  Their PDFs are still
given by Eq. (\ref{eq:PDF_elldelta}), with

(i) for the boundary local time $\ell_\delta^{\rm i}$ on the inner
sphere:
\begin{subequations}
\begin{align}
\pi_0 & = 1 - \frac{R}{r_0} \, \frac{\alpha L \cosh(\alpha(L-r_0)) - \sinh(\alpha (L-r_0))}{\alpha L \cosh(\alpha(L-R)) - \sinh(\alpha (L-R))} \,, \\
\mu_0^{(p)} & = \frac{(\alpha^2 LR-1) \sinh(\alpha(L-R)) + \alpha(L-R) \cosh(\alpha (L-R))}
{R(\alpha L \cosh(\alpha(L-R)) - \sinh(\alpha (L-R)))}\,.
\end{align}
\end{subequations}

(ii) for the boundary local time $\ell_\delta^{\rm o}$ on the outer
sphere:
\begin{subequations}
\begin{align}
\pi_0 & = 1 - \frac{L}{r_0} \, \frac{\alpha R \cosh(\alpha(r_0-R)) + \sinh(\alpha (r_0-R))}{\alpha R \cosh(\alpha(L-R)) + \sinh(\alpha (L-R))} \,, \\
\mu_0^{(p)} & = \frac{(\alpha^2 LR-1) \sinh(\alpha(L-R)) + \alpha(L-R) \cosh(\alpha (L-R))}
{L(\alpha R \cosh(\alpha(L-R)) + \sinh(\alpha (L-R)))}\,.
\end{align}
\end{subequations}

\begin{figure}[t!]
\begin{center}
\includegraphics[width=0.32\linewidth]{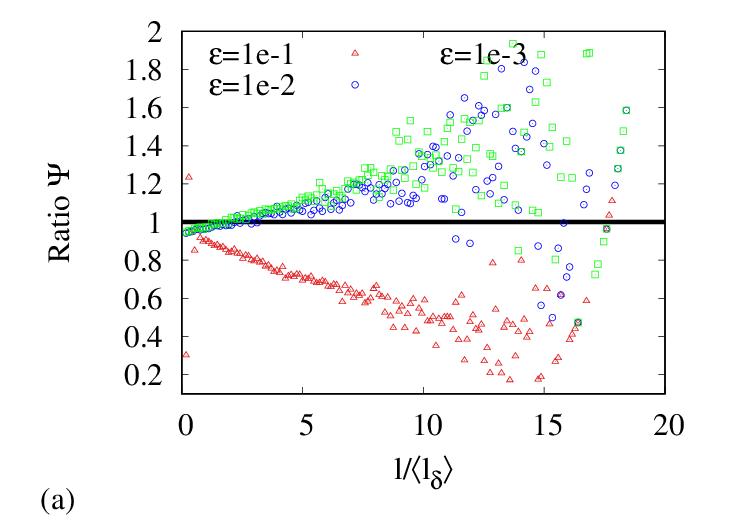}
\includegraphics[width=0.32\linewidth]{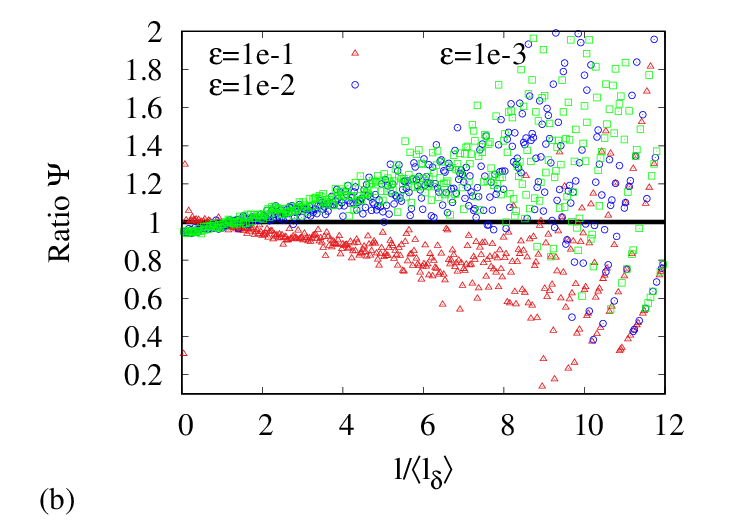}
\includegraphics[width=0.32\linewidth]{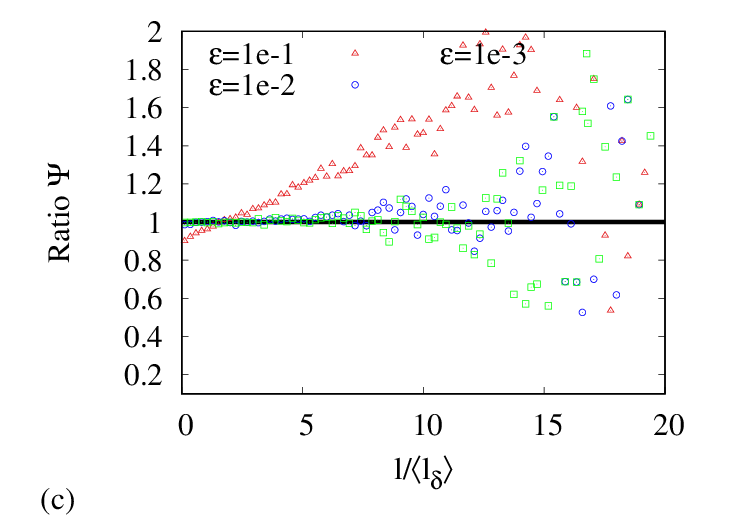}
\includegraphics[width=0.32\linewidth]{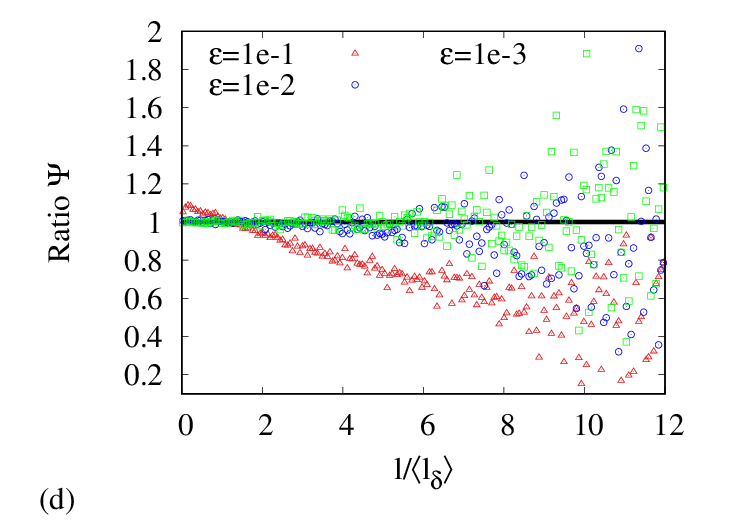}
\includegraphics[width=0.32\linewidth]{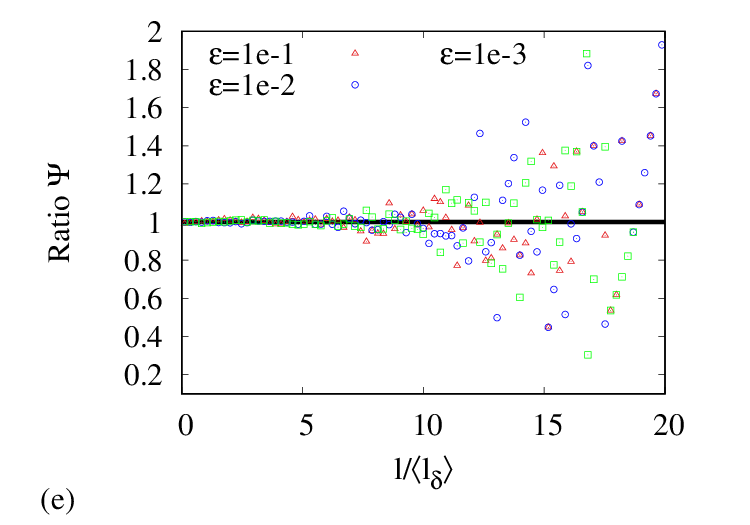}
\includegraphics[width=0.32\linewidth]{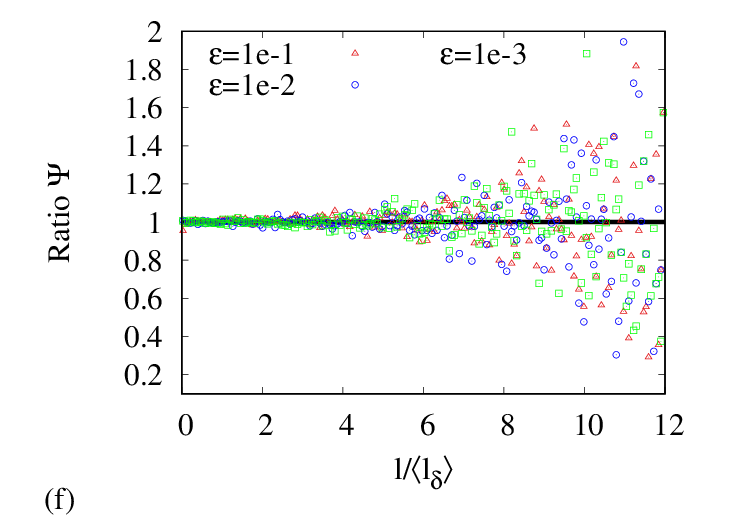}
\end{center}
\caption{
Ratio $\Psi_\mrm{emp}(\ell,p|\x_0)$ over $\Psi(\ell,p|\x_0)$ as a
function of $\ell/\lla\ell_\dlt\rra$ for the spherical shell
(Fig. \ref{fig:illustration}e), with $p=1$, $R=1$, $L=2$, and $r_0 =
1.5$.  The empirical PDF $\Psi_\mrm{emp}$ was estimated from $N=10^6$
simulated values of $\ell_\dlt$ by \tbf{(a,b)} SIM1, \tbf{(c,d)} FLA,
\tbf{(e,f)} CLA, with different choices of the boundary layer width:
$\ve=10^{-1}$ (triangles), $\ve=10^{-2}$ (circles), and $\ve=10^{-3}$
(squares).
\tbf{(a,c,e)} 
refer to $\ell^\mrm{i}_\dlt$ on the inner sphere and \tbf{(b,d,f)}
refer to $\ell^\mrm{o}_\dlt$ on the outer sphere.  The trajectories
that never touched the boundary and thus yielded $\ell_\delta = 0$,
were excluded.}
\label{fig:shell_comp}
\end{figure}

Table \ref{table:3Dshell} and Figure \ref{fig:shell_comp} present the
comparison between three methods, with the same trends as for the
circular annulus.

%%% 3D shell
\begin{table}[t!]
  \centering 
\begin{tabular}{ c|c|c|c|r|c|r } 
\hline\hline
% \multicolumn{7}{c}{$N=10^6, p=1, r_0 = 1.5, R = 1, L = 2$} \\
% \hline
Method & $\varepsilon$ & CPU time (s) & 
$\langle \ell^\mathrm{i}_\delta \rangle_\mathrm{emp}$ & Rel. Error & 
$\langle \ell^\mathrm{o}_\delta \rangle_\mathrm{emp}$ & Rel. Error  \\
\hline
\multirow{3}{3em}{SIM1} & $10^{-1}$ & $1.26\times10^1$ & 
0.3346 & $-21.33 \%$ &
1.5072 & $-6.81 \%$  \\ % sim-66
& $10^{-2}$ & $1.39\times10^2$ & 
0.4385 & $3.09 \%$ &
1.6888 & $4.41 \%$  \\  % sim-67
& $10^{-3}$ & $1.34\times10^2$ & 
0.4496 & $5.71 \%$ &
1.7145 & $6.00 \%$  \\  % sim-68
\hline
\multirow{3}{3em}{SIM2} & $10^{-1}$ & $1.06\times10^1$ & 
0.4536 & $6.65 \%$ &
1.4994 & $-7.30 \%$  \\ % sim-10
& $10^{-2}$ & $1.03\times10^2$ & 
0.4221 & $-0.75 \%$ &
1.5825 & $-2.16 \%$  \\ % sim-11
& $10^{-3}$ & $1.01\times10^3$ & 
0.4196 & $-1.34 \%$ &
1.5961 & $-1.32 \%$  \\ % sim-12
\hline
\multirow{3}{3em}{FLA} & $10^{-1}$ & $1.80\times10^1$ & 
0.4656 & $9.46 \%$ &
1.4762 & $-8.73 \%$  \\ % sim-16
& $10^{-2}$ & $1.88\times10^2$ & 
0.4282 & $0.69 \%$ &
1.6011 & $-1.01 \%$  \\  % sim-17
& $10^{-3}$ & $1.78\times10^2$ & 
0.4259 & $0.15 \%$ &
1.6159 & $-0.10 \%$  \\  % sim-18
\hline
\multirow{3}{3em}{CLA} & $10^{-1}$ & $1.90\times10^1$ & 
0.4263 & $0.24 \%$ &
1.6258 & $0.52 \%$  \\ % sim-13
& $10^{-2}$ & $1.86\times10^2$ & 
0.4253 & $-0.0002 \%$ &
1.6160 & $-0.09 \%$  \\ % sim-14
& $10^{-3}$ & $1.73\times10^3$ & 
0.4252 & $-0.03 \%$ &
1.6174 & $-0.004 \%$  \\ % sim-15
\hline
\hline
\end{tabular}
\caption{
Comparison of the SIM, FLA, CLA with $N=10^6$ and different $\ve$ for
the spherical shell (Fig. \ref{fig:illustration}e) with $p=1$, $R=1$,
$L=2$, and $r_0 = 1.5$.  The mean and standard deviation are: $\langle
\ell^\mrm{i}_\delta \rangle = 0.4253$, $\sigma^{\rm i} = 0.6458$, and
$\langle \ell_\delta^{\rm o} \rangle = 1.6174$, $\sigma^{\rm o} =
1.9631$, so that the relative statistical errors are
$\frac{\sigma^{\rm i}}{\langle \ell^\mrm{i}_\delta \rangle \sqrt{N}}
\simeq 0.15 \%$, and $\frac{\sigma^{\rm o}}{\langle
\ell^\mrm{o}_\delta \rangle \sqrt{N}} \simeq 0.12 \%$.  }
\label{table:3Dshell}
\end{table}

%%%%% %%%%% %%%%% %%%%% %%%%% %%%%% %%%%% %%%%% %%%%% %%%%% %%%%% %%%%% 
%%%%% %%%%% %%%%% %%%%% %%%%% %%%%% %%%%% %%%%% %%%%% %%%%% %%%%% %%%%% 
%%%%% %%%%% %%%%% %%%%% %%%%% %%%%% %%%%% %%%%% %%%%% %%%%% %%%%% %%%%% 

\bibliographystyle{elsarticle-num} % numeric-style citation call-outs

\begin{thebibliography}{99}


\bibitem{Levy}				P. L\'evy,
					{\it Processus Stochastiques et Mouvement Brownien}
					(Paris, Gauthier-Villard, 1965).

\bibitem{Ito}				K. Ito and H. P. McKean,
					{\it Diffusion Processes and Their Sample Paths}
					(Springer-Verlag, Berlin, 1965).

\bibitem{Freidlin}			M. Freidlin,
					{\it Functional Integration and Partial Differential Equations}
					(Annals of Mathematics Studies, Princeton University Press, Princeton, New Jersey, 1985).



\bibitem{Grebenkov20}			D. S. Grebenkov, 
					Paradigm Shift in Diffusion-Mediated Surface Phenomena, 
					Phys. Rev. Lett. {\bf 125}, 078102 (2020).


%## Added 05/04/24
\bibitem{mckean1975brownian} 		H. P. McKean, 
					Brownian local time, 
					Adv. Math. {\bf 15}, 91-111 (1975). 

\bibitem{borodin2015handbook} 		A. N. Borodin and P. Salminen, 
					{\it Handbook of Brownian Motion: Facts and Formulae} 
					(Birkhauser Verlag, Basel-Boston-Berlin, 1996).

\bibitem{majumdar2007brownian}		S. N. Majumdar, 
					Brownian functionals in physics and computer science, 
					Curr. Sci. {\bf 89}, 2076 (2007).

%##


\bibitem{Grebenkov21}			D. S. Grebenkov,
					Statistics of boundary encounters by a particle diffusing outside a compact planar domain,
					J. Phys. A.: Math. Theor. {\bf 54}, 015003 (2021).
					
					

\bibitem{Grebenkov22a}			D. S. Grebenkov,  
					An encounter-based approach for restricted diffusion with a gradient drift,
					J. Phys. A: Math. Theor. {\bf 55}, 045203 (2022).

\bibitem{Bressloff22}			P. C. Bressloff, 
					Narrow capture problem: An encounter-based approach to partially reactive targets,
					Phys. Rev. E {\bf 105}, 034141 (2022).

\bibitem{Grebenkov22}			D. S. Grebenkov, 
					Statistics of diffusive encounters with a small target: Three complementary approaches,
					J. Stat. Mech. 083205 (2022).  


\bibitem{Bressloff22d}			P. C. Bressloff,
					Diffusion-mediated surface reactions and stochastic resetting,
					J. Phys. A: Math. Theor. {\bf 55}, 275002 (2022).

\bibitem{Benkhadaj22}			Z. Benkhadaj and D. S. Grebenkov, 
					Encounter-based approach to diffusion with resetting,
					Phys. Rev. E {\bf 106}, 044121 (2022).

\bibitem{Grebenkov22b}			D. S. Grebenkov,  
					Depletion of Resources by a Population of Diffusing Species, 
					Phys. Rev. E {\bf 105}, 054402 (2022). 

\bibitem{Grebenkov23b}			D. S. Grebenkov, 
					Diffusion-controlled reactions with non-Markovian binding/unbinding kinetics, 
					J. Chem. Phys. {\bf 158}, 214111 (2023). 

\bibitem{Grebenkov24f}			D. S. Grebenkov,
					Adsorption and permeation events in molecular diffusion, 
					Molecules {\bf 29}, 5012 (2024).




\bibitem{Bressloff22c}			P. C. Bressloff, 
					A probabilistic model of diffusion through a semipermeable barrier,
					Proc. Roy. Soc. A {\bf 478}, 20220615 (2022).

\bibitem{Bressloff23a}			P. C. Bressloff, 
					Renewal equation for single-particle diffusion through a semipermeable interface,
					Phys. Rev. E. {\bf 107}, 014110 (2023).

\bibitem{Bressloff23b}			P. C. Bressloff,
					Renewal equations for single-particle diffusion in multilayered media,
					SIAM J. Appl. Math. {\bf 83}, 1518--1545 (2023).

%%% Added 05/04/24
\bibitem{fieremans2010monte} 		E. Fieremans, D. S. Novikov, J. H. Jensen, and J. A. Helpern, 
					Monte Carlo study of a two-compartment exchange model of diffusion, 
					NMR Biomed. {\bf 23}, 711-724 (2010). 
%%% 


%Ref1:
\bibitem{Lejay16}			A. Lejay, 
					The snapping out Brownian motion, 
					Ann. Appl. Probab. {\bf 26}, 1727-1742 (2016).
%%% sec56_ref1

%%% Added 05/04/24
\bibitem{lejay2018monte} 		A. Lejay,
					Monte Carlo estimation of the mean residence time in cells 
					surrounded by thin layers, 
					Math. Comput. Simul. {\bf 143}, 65-77 (2018) .


\bibitem{Grebenkov19}			D. S. Grebenkov, 
					Probability distribution of the boundary local time of 
					reflected Brownian motion in Euclidean domains, 
					Phys. Rev. E {\bf 100}, 062110 (2019). 

\bibitem{Grebenkov20b}			D. S. Grebenkov, 
					Surface Hopping Propagator: An Alternative Approach to Diffusion-Influenced Reactions, 
					Phys. Rev. E {\bf 102}, 032125 (2020).









\bibitem{Sabelfeld}			K. K. Sabelfeld,
					{\it Monte Carlo Methods in Boundary Value Problems}
					(Springer-Verlag: New York - Heidelberg, Berlin, 1991).


\bibitem{Sabelfeld2}			K. K. Sabelfeld and N. A. Simonov,
					{\it Random Walks on Boundary for Solving PDEs}
					(Utrecht, The Netherlands, 1994).

\bibitem{Milshtein}			G. N. Milshtein,
					{\it Numerical Integration of Stochastic Differential Equations}
					(Kluwer, Dordrecht, the Netherlands, 1995).




%%%%%

\bibitem{litwin1980monte} 		S. Litwin, 
					Monte Carlo simulation of particle adsorption rates at high cell concentration,
					Biophys. J. {\bf 31}, 271-277 (1980).

\bibitem{Torquato89}			S. Torquato and I. C. Kim, 
					Efficient simulation technique to compute effective properties of heterogeneous media,
					Appl. Phys. Lett. {\bf 55}, 1847 (1989).

\bibitem{Zheng89}			L. H. Zheng and Y. C. Chiew,
					Computer simulation of diffusion-controlled reactions in dispersions of spherical sinks,
					J. Chem. Phys. {\bf 90}, 322-327 (1989).

\bibitem{Lee89}				S. B. Lee, I. C. Kim, C. A. Miller, and S. Torquato,
					Random-walk simulation of diffusion-controlled processes among static traps,
					Phys. Rev. B {\bf 39}, 11833-11839 (1989).

\bibitem{Ossadnik91}			P. Ossadnik, 
                        		Multiscaling Analysis of Large-Scale Off-Lattice DLA,
					Physica A {\bf 176}, 454 (1991).

\bibitem{batsilas2003stochastic}	L. Batsilas, A. M. Berezhkovskii, and S. Y. Shvartsman,
					Stochastic model of autocrine and paracrine signals in cell culture assays,
					Biophys. J. {\bf 85}, 3659-3665 (2003).

\bibitem{dagdug2003diffusion} 		L. Dagdug, A. Berezhkovskii, S. M. Bezrukov, and G. H. Weiss,
					Diffusion-controlled reactions with a binding site hidden in a channel,
					J. Chem. Phys. {\bf 118}, 2367-2373 (2003).


\bibitem{Grebenkov05a}			D. S. Grebenkov,
                        		What Makes a Boundary Less Accessible,
					Phys. Rev. Lett. {\bf 95}, 200602 (2005).

\bibitem{Grebenkov05b}			D. S. Grebenkov, A. A. Lebedev, M. Filoche and B. Sapoval,   
					Multifractal Properties of the Harmonic Measure on Koch Boundaries in Two and Three Dimensions,
					Phys. Rev. E {\bf 71}, 056121 (2005).

\bibitem{Grebenkov06c}			D. S. Grebenkov,
					Scaling Properties of the Spread Harmonic Measures,
					Fractals {\bf 14}, 231-243 (2006).

\bibitem{Levitz06}			P. Levitz, D. S. Grebenkov, M. Zinsmeister, K. Kolwankar  and B. Sapoval, 
					Brownian flights over a fractal nest and first passage statistics on irregular surfaces,
					Phys. Rev. Lett. {\bf 96}, 180601 (2006). 


%Ref2
\bibitem{Deaconu06}			M. Deaconu and A. Lejay, 
					A Random Walk on Rectangles Algorithm, 
					Method. Comput. Appl. Probab. {\bf 8}, 135-151 (2006).

\bibitem{opplestrup2006first} 		T. Opplestrup, V. V. Bulatov, G. H. Gilmer, M. H. Kalos, and B. Sadigh, 
					First-passage Monte Carlo algorithm: Diffusion without all the hops, 
					Phys. Rev. Lett. {\bf 97}, 230602 (2006).

\bibitem{hall2009convergence} 		M. Hall and D. C. Alexander,
					Convergence and Parameter Choice for Monte-Carlo Simulations of Diffusion MRI, 
					IEEE Trans. Med. Imag. {\bf 28}, 1354 (2009).

\bibitem{Zein10}			S. Zein, A. Lejay, and M. Deaconu, 
					An efficient algorithm to simulate a Brownian motion 
					over irregular domains, 
					Commun. Comput. Phys. {\bf 8}, 901-916 (2010). 
%%% sec56_ref2

\bibitem{berezhkovskii2013trapping}	A. M. Berezhkovskii, L. Dagdug, M.-V. Vazquez, V. A. Lizunov, J. Zimmerberg, and S. M. Bezrukov, 
					Trapping of diffusing particles by clusters of absorbing disks on a reflecting wall 
					with disk centers on sites of a square lattice, 
					J. Chem. Phys. {\bf 138}, 064105 (2013).

\bibitem{ghosh2015non}			S. K. Ghosh, A. G. Cherstvy, and R. Metzler,
					Non-universal tracer diffusion in crowded media of non-inert obstacles,
					Phys. Chem. Chem. Phys. {\bf 17}, 1847-1858 (2015).

\bibitem{ghosh2015anomalous}		S. K. Ghosh, A. G. Cherstvy, D. S. Grebenkov, and R. Metzler,
					Anomalous, non-Gaussian tracer diffusion in heterogeneously crowded environments,
					New J. Phys. {\bf 18}, 013027 (2016).

\bibitem{bernoff2018boundary}		A. J. Bernoff, A. E. Lindsay, and D. D. Schmidt,
					Boundary Homogenization and Capture Time Distributions of Semipermeable Membranes 
					with Periodic Patterns of Reactive Sites,
					Multiscale Model. Simul. {\bf 16}, 1411-1447 (2018).

\bibitem{palombo2019generative}		M. Palombo, D. C. Alexander, and H. Zhang,
					A generative model of realistic brain cells with application to 
					numerical simulation of the diffusion-weighted MR signal,
					NeuroImage {\bf 188}, 391-402 (2019).

\bibitem{ianus2021mapping}		A. Ianus, D. C. Alexander, H. Zhang, and M. Palombo,
					Mapping complex cell morphology in the grey matter with double diffusion 
					encoding MR: A simulation study,
					NeuroImage {\bf 241}, 118424 (2021).

\bibitem{le2022first}			F. Le Vot, S. B. Yuste, E. Abad, and D. S. Grebenkov,
					First-encounter time of two diffusing particles in two- and 
					three-dimensional confinement,
					Phys. Rev. E {\bf 105}, 044119 (2022).

%%% Added 05/04/24
\bibitem{cherry2022trapping}		J. Cherry, A.E. Lindsay, A. Navarro Hern\'andez, and B. Quaife, 
					Trapping of planar Brownian motion: full first passage time distributions 
					by kinetic Monte Carlo, asymptotic, and boundary integral methods, 
					Multiscale Model. Simul. {\bf 20}, 1284-1314 (2022).
%%%

%%%%%

%## Added 05/04/24
\bibitem{bossy2004symmetrized}		M. Bossy, E. Gobet, and D. Talay, 
					A symmetrized Euler scheme for an efficient approximation of reflected diffusions, 
					J. Appl. Probab. {\bf 41}, 877-889 (2004). 
%##


\bibitem{Grebenkov07}			D. S. Grebenkov, 
					Residence times and other functionals of reflected Brownian motion, 
					Phys. Rev. E 76, 041139 (2007).

\bibitem{Muller56}			M. E. Muller,
					Some Continuous Monte Carlo Methods for the Dirichlet Problem,
					Ann. Math. Statist. {\bf 27}, 569 (1956).


\bibitem{Zhou17}			Y. Zhou, W. Cai, and E. Hsu,
					Computation of the local time of reflecting Brownian motion and 
					the probabilistic representation of the Neumann problem,
					Comm. Math. Sci. {\bf 15}, 237-259 (2017).

\bibitem{Schumm23}			R. D. Schumm and P. C. Bressloff,
					A numerical method for solving snapping out Brownian motion in 2D bounded domains,
					J. Comput. Phys. {\bf 493}, 112479 (2023).

\bibitem{Scher23}			Y. Scher, S. Reuveni, and D. S. Grebenkov, 
					Escape of a sticky particle, 
					Phys. Rev. Research {\bf 5}, 043196 (2023).
	
\bibitem{Scher24}			Y. Scher, S. Reuveni, and D. S. Grebenkov, 
					Escape from textured adsorbing surfaces, 
					J. Chem. Phys. {\bf 160}, 184105 (2024).				
					
\bibitem{Grebenkov23}			D. S. Grebenkov, 
					Encounter-based approach to the escape problem, 
					Phys. Rev. E {\bf 107}, 044105 (2023).




\bibitem{Binder12}			I. Binder and M. Braverman,
					The rate of convergence of the walk of sphere algorithm,
					Geom. Func. Anal. {\bf 22}, 558-587 (2012).
%%39
\bibitem{Levitin}			M. Levitin, D. Mangoubi, and I. Polterovich,
					{\em Topics in Spectral Geometry} (Vol. 237), American Mathematical Society (2023)
					%(Preliminary version, May 29, 2023; \url{https://www.michaellevitin.net/Book/TSG230529.pdf})
					
%%46				
\bibitem{Carlsson10}			T. Carlsson, T. Ekholm, and C. Elvingson,
					Algorithm for generating a Brownian motion on a sphere,
					J. Phys. A: Math. Theor. {\bf 43}, 505001 (2010).
%%%47
\bibitem{Burrage22}			K. Burrage, P. M. Burrage, and G. Lythe, 
					Effective numerical methods for simulating diffusion on a spherical surface in three dimensions,
					Numer. Algor. {\bf 91}, 1577-1596 (2022).


%%% 43
\bibitem{Yuste13}			S. B. Yuste, E. Abad, and K. Lindenberg, 
					Exploration and trapping of mortal random walkers, 
					Phys. Rev. Lett. {\bf 110}, 220603 (2013).

\bibitem{Meerson15}			B. Meerson and S. Redner, 
					Mortality, redundancy, and diversity in stochastic search, 
					Phys. Rev. Lett. {\bf 114}, 198101 (2015).
%% 45
\bibitem{Grebenkov17d}			D. S. Grebenkov and J.-F. Rupprecht, 
					The escape problem for mortal walkers, 
					J. Chem. Phys. {\bf 146}, 084106 (2017).

%%% Added 05/04/24
\bibitem{maire2013monte}		S. Maire and E. Tanr\'e,
					Monte Carlo approximation of the Neumann problem, 
					Monte Carlo Methods Appl. {\bf 19}, 201–236 (2013).
%%% 

%%%42
\bibitem{Grebenkov11}			D. S. Grebenkov, 
					A fast random walk algorithm for computing the pulsed-gradient spin-echo 
					signal in multiscale porous media, 
					J. Magn. Reson. {\bf 208}, 243-255 (2011).


%%%41
\bibitem{Chaigneau24}			A. Chaigneau and D. S. Grebenkov,
					A numerical study of the Dirichlet-to-Neumann operator in planar domains,
					J. Phys. A: Math. Theor. {\bf 57}, 445201 (2024). 



%%%40
\bibitem{Grebenkov20c}			D. S. Grebenkov, 
					Joint distribution of multiple boundary local times and 
					related first-passage time problems with multiple targets, 
					J. Stat. Mech. 103205 (2020).


%## Added 05/04/24
\bibitem{pacchiarotti1998numerical} 	C. Costantini, B. Pacchiarotti, and F. Sartoretto, 
					Numerical approximation for functionals of reflecting diffusion processes, 
					SIAM J. Appl. Math. {\bf 58}, 73-102 (1998).
%##


\bibitem{Milshtein96}			G. N. Mil'shtein, 
					Application of the numerical integration of stochastic equations for 
					the solution of boundary-value problems with Neumann's boundary conditions,
					Theory Probab. Appl. {\bf 41}, 170-177 (1996).

\bibitem{Milstein97}			G. N. Milstein, 
					Weak approximation of a diffusion process in a bounded domain,
					Stoch. Stoch. Rep. {\bf 62}, 147-200 (1997).

\bibitem{Slominski01}			L. S{\l}omi\'nski,
					Euler's approximations of solutions of SDEs with reflecting boundary,
					Stoch. Proc. Appl. {\bf 94}, 317-337 (2001).

\bibitem{Milstein_book}			G. N. Milstein and M. V. Tretyakov, 
					{\it Stochastic Numerics for Mathematical Physics. Scientific Computation}
					(Springer, Berlin, 2004).

\bibitem{Bernal16}			F. Bernal and J. A. Acebr\'on, 
					A comparison of higher-order weak numerical schemes for stochastic 
					differential equations in bounded domains,
					Commun. Comput. Phys. {\bf 20}, 703-732 (2016).

\bibitem{Bernal19}			F. Bernal, 
					An implementation of Milstein's method for general bounded diffusions,
					J. Sci. Comp. {\bf 79}, 867-890 (2019).

\bibitem{Leimkuhler23}			B. Leimkuhler, A. Sharma, and M. V. Tretyakov, 
					Simplest random walk for approximating Robin boundary value problems and 
					ergodic limits of reflected diffusions,
					Ann. Appl. Probab. {\bf 33}, 1904-1960 (2023).



%%% Moved 
\bibitem{grebenkov2015analytical} 	D. S. Grebenkov, 
					Analytical representations of the spread harmonic measure density, 
					Phys. Rev. E. {\bf 91}(5), 052108 (2015).
%%%


%## Added 05/04/24
\bibitem{redner2001guide} 		S. Redner, 
					{\it A Guide to First Passage Processes} 
					(Cambridge University press, 2001).

\bibitem{schuss2015brownian} 		Z. Schuss, 
					{\it Brownian Dynamics at Boundaries and Interfaces in Physics, Chemistry and Biology} 
					(Springer, New York, 2013).

\bibitem{metzler2014first} 		R. Metzler, G. Oshanin, and S. Redner (Eds), 
					{\it First-Passage Phenomena and Their Applications} 
					(World Scientific Press, Singapore, 2014).
 
\bibitem{masoliver2018random} 		J. Masoliver, 
					{\it Random Processes: First-passage and Escape} 
					(World Scientific, 2018).

\bibitem{lindenberg2019chemical} 	K. Lindenberg, R. Metzler, and G. Oshanin (Eds), 
					{\it Chemical Kinetics: Beyond the Textbook} 
					(World Scientific, New Jersey, 2019).
 
\bibitem{dagdug2024diffusion} 		L. Dagdug , J. Pe\~{n}a , and I. Pompa-Garc\'{\i}a, 
					{\it Diffusion Under Confinement: A Journey Through Counterintuition} 
					(Springer, 2024).
%##

\bibitem{Grebenkov_book}		D. S. Grebenkov, R. Metzler, and G. Oshanin (Eds),
					{\it Target Search Problems}
					(Springer, Cham, 2024).


%Ref4:
\bibitem{Hsu85}				E. Hsu, 
					Probabilistic approach to the neumann problem, 
					Commun. Pure Appl. Math. {\bf 38}, 445 (1985).

\bibitem{Papanicolaou90}		V. G. Papanicolaou, 
					The probabilistic solution of the third boundary value problem for 
					second order elliptic equations, 
					Probab. Theory Relat. Fields {\bf 87}, 27 (1990).

\bibitem{Zhou16}			Y. Zhou and W. Cai, 
					Numerical Solution of the Robin Problem of Laplace Equations 
					with a Feynman-Kac Formula and Reflecting Brownian Motions, 
					J. Scient. Comp. {\bf 69}, 107-121 (2016). 
%%% sec56_ref4


%%% Added 05/04/24
\bibitem{brosamler1976probabilistic} 	G. A. Brosamler, 
					A probabilistic solution of the Neumann problem, 
					Mathematica Scandinavica, {\bf 38}, 137-147 (1976).

\bibitem{bencherif2009probabilistic} 	A. Bench\'erif-Madani and \'E. Pardoux, 
					A probabilistic formula for a Poisson equation with Neumann boundary condition, 
					Stoch. Anal. Appl. {\bf 27}, 739-746 (2009).

\bibitem{lions1984stochastic} 		P. L. Lions and A. S. Sznitman, 
					Stochastic differential equations with reflecting boundary conditions,
					Comm. Pure Appl. Math. {\bf 37}, 511-537 (1984).

%\bibitem{maire2013monte} S. Maire and E. Tanr´e, Monte Carlo approximations of the Neumann problem, Monte Carlo Meth. Appl. 19, 201-236 (2013). %%% Repeated with ref above. 

\bibitem{morillon1997numerical}		J. P. Morillon, 
					Numerical solutions of linear mixed boundary value problems using stochastic representations,
					Int. J. Numer. Meth. Engng. {\bf 40}, 387-405 (1997).
%%%




\bibitem{grebenkov2014efficient} 	D. S. Grebenkov, 
					Efficient Monte Carlo methods for simulating diffusion-reaction processes in complex systems, 
					in "First-Passage Phenomena and Their Applications", Eds. R. Metzler, G. Oshanin, S. Redner (World Scientific Press, Singapore, 2014).



\bibitem{grebenkov2019semi} 		D. S. Grebenkov and S. D. Traytak, 
					Semi-analytical computation of Laplacian Green functions in three-dimensional domains with disconnected spherical boundaries, 
					J. Comput. Phys. {\bf 379}, 91-117 (2017). 









\end{thebibliography}

\end{document}